\documentclass[a4paper,12pt]{article}
\pdfoutput=1
\usepackage{graphicx}
\usepackage{subcaption}
\usepackage[margin=1in]{geometry}
\usepackage{afterpage}
\usepackage{youngtab}
\usepackage{amsmath}
\usepackage{color}

\begin{document}
\makeatletter
\@addtoreset{equation}{section}
\makeatother
\renewcommand{\theequation}{\thesection.\arabic{equation}}
\vspace{1.8truecm}

{\LARGE{ \centerline{\bf Absorption of closed strings by giant gravitons}  }}  

\vskip.5cm 

\thispagestyle{empty} 
\centerline{ {\large\bf Gaoli Chen$^{b,}$\footnote{{\tt galic.chen@gmail.com}}, 
Robert de Mello Koch$^{a,b,}$\footnote{{\tt robert@neo.phys.wits.ac.za}}, 
Minkyoo Kim$^{b,}$\footnote{{\tt minkyoo.kim@wits.ac.za}} }}\par
\vspace{.2cm}
\centerline{{\large\bf and
Hendrik J.R. Van Zyl${}^{b,}$\footnote{ {\tt hjrvanzyl@gmail.com}} }}

\vspace{.4cm}
\centerline{{\it ${}^{a}$ School of Physics and Telecommunication Engineering},}
\centerline{{ \it South China Normal University, Guangzhou 510006, China}}

\vspace{.4cm}
\centerline{{\it ${}^{b}$ National Institute for Theoretical Physics,}}
\centerline{{\it School of Physics and Mandelstam Institute for Theoretical Physics,}}
\centerline{{\it University of the Witwatersrand, Wits, 2050, } }
\centerline{{\it South Africa } }

\vspace{1truecm}

\thispagestyle{empty}

\centerline{\bf ABSTRACT}

\vskip.2cm 
A new approach to the computation of correlation functions involving two determinant operators as well as one 
non-protected single trace operator has recently been developed by Jiang, Komatsu and Vescovi.
This correlation function provides the holographic description of the absorption of a closed string by a giant graviton. 
The analysis has a natural interpretation in the framework of group representation theory, which admits a generalization to
general Schur polynomials and restricted Schur polynomials.
This generalizes the holographic description to any giant or dual giant gravitons which carry more than one angular momentum
on the sphere.
For a restricted Schur polynomial labeled by a column with $N$ boxes (dual to a maximal giant graviton) 
we find evidence in favor of integrability.

\setcounter{page}{0}
\setcounter{tocdepth}{2}
\newpage
\tableofcontents
\setcounter{footnote}{0}
\linespread{1.1}
\parskip 4pt

{}~
{}~

\section{Introduction}

The AdS/CFT correspondence\cite{Maldacena:1997re,Gubser:1998bc,Witten:1998qj} is a remarkable duality between a 
quantum theory of gravity on an asymptotically Anti-de Sitter (AdS) background and an ordinary quantum conformal field 
theory (CFT) living in fewer dimensions. 
A well studied example involves type IIB string theory on AdS$_5\times$S$^5$ with $N$ units of RR five form flux and 
${\cal N} = 4$ super Yang-Mills theory in four dimensions.
Physical observables of the theory can be expanded in a double expansion on both sides.
In the CFT for example, the expansion is in terms of $1/N$ and the 't Hooft coupling $\lambda$.
Computations performed on both sides will agree at both the perturbative and non-perturbative levels.
We expect non-perturbative effects both in $1/N$ and in $\lambda$.
One well understood source of non-perturbative effects in $1/N$ comes from 
correlators \cite{Balasubramanian:2001nh,Berenstein:2003ah,Corley:2001zk,Aharony:2002nd,Brown:2006zk,Hirano:2018xmh} 
of operators dual to giant gravitons\cite{McGreevy:2000cw} and dual giant gravitons\cite{Grisaru:2000zn,Hashimoto:2000zp}.
The operators dual to both giant and dual giant gravitons are given by Schur polynomials\cite{Corley:2001zk} which can 
be generalized to restricted Schur polynomials to describe excited giant 
gravitons\cite{Balasubramanian:2004nb,deMelloKoch:2007rqf,deMelloKoch:2007nbd,Bekker:2007ea}
(see also \cite{Balasubramanian:2002sa,Berenstein:2005fa,Berenstein:2006qk}).
The operator corresponding to the maximal giant graviton is particularly simple: it corresponds to a 
determinant.
Methods that sum all orders in the $1/N$ expansion, in the free CFT, have been developed in a series of articles \cite{Corley:2001zk,Corley:2002mj,Kimura:2007wy,Brown:2007xh,Bhattacharyya:2008rb,Bhattacharyya:2008xy,Brown:2008ij}.
For the case of ${1\over 2}$-BPS correlators there are no 't Hooft couping corrections \cite{Baggio:2012rr} and the resulting 
expressions are in fact exact.
The $1/N$ expansion of these correlators exhibits the Stokes phenomenon (a hallmark of non-perturbative contributions) 
but are Borel summable\cite{deMelloKoch:2019dda}.

Significant recent progress\cite{Jiang:2019xdz} has been made by considering correlators involving determinant operators 
as well as one non-protected single trace operator.
This particular correlator provides the holographic description of the absorption of a closed string by a giant graviton, and 
has been considered previously in \cite{Bak:2011yy,Bissi:2011dc,Caputa:2012yj,Lin:2012ey}.
These previous computations focused on protected correlators so that there are no $\lambda$ corrections.
The key advance reported in \cite{Jiang:2019xdz} is a determination of the exact 't Hooft coupling dependence at large $N$.
The maximal giant gravitons are non-perturbative objects of string theory, and therefore are heavy. 
In the weak coupling limit they should be understood as boundary states of the string world sheet CFT.
In the large radius limit they become geometric objects since they are the locus where strings can end.
Making use of results for defect one-point functions \cite{deLeeuw:2015hxa}, the paper \cite{Jiang:2019xdz} argues that
the correlator corresponds to overlaps on the string worldsheet between an integrable boundary state and a state dual to the
single trace operator.
The analysis is capable of providing non-perturbative expressions using the framework of the thermodynamic Bethe 
ansatz (TBA)\cite{Zamolodchikov:1991et}.
This is a remarkable result and warrants further study.

The goal of this paper is to reconsider and extend results reported in \cite{Jiang:2019xdz}.
A powerful approach to the problem of computing correlators of Schur polynomials in general and determinant operators
in particular, which makes use of group representation theory, was developed in \cite{Corley:2001zk}.
Each Schur polynomial can be expressed as the trace of a projection operator times the tensor product of the fields
appearing in the polynomial.
Then the two point correlation function, for example, is simply given by the trace of a product of projectors, which
is easily computed.
This interpretation can be extended to restricted Schur polynomials\cite{Bhattacharyya:2008rb}, which are constructed
using more than one matrix. 
The results of \cite{Jiang:2019xdz} have a simple interpretation within this framework.
In particular, to obtain an effective action for the determinant operators, the authors of \cite{Jiang:2019xdz} carried
out a sequence of rewritings which they then suggested have a natural interpretation from string theory.
The first step in the sequence of rewritings is precisely building the projection operator needed to construct the maximal
giant graviton.
This new interpretation of the results of \cite{Jiang:2019xdz} not only contributes towards making the discussion of the 
physics richer by providing some mathematical depth and context, but it also allows significant generalizations.
The generalizations we will pursue include
\begin{itemize}
\item[1.] Generalizing from maximal giant gravitons to any giant gravitons.
\item[2.] Generalizing from giant gravitons to dual giant gravitons.
\item[3.] Generalizing from giant (and dual giant) gravitons constructed  from a single matrix (and hence described by Schur polynomials) to giant (and dual giant) gravitons built from many matrices (and hence described by restricted Schur polynomials).
\end{itemize}
The effective action description provides a useful approach to evaluating the correlator:
by extremizing the effective action for the determinant operators, one obtains an emergent classical background and the
computation of the correlator of determinants and one single-trace operator amounts to evaluating the single trace operator
on the emergent classical background.
We will see that this also generalizes nicely for the three cases listed above. 

A fascinating observation of \cite{Jiang:2019xdz} is that already at the level of the free theory there are convincing hints 
of integrability.
As an example, if the single trace operator belongs to the $SU(2)$ sector, 
the correlator is only non-zero when (i) both the length of 
the spin chain $L$ and the number of magnons $M$ are both even and (ii) the rapidities of the magnons are parity symmetric,
namely ${\bf u}=\{u_1,-u_1,u_2,-u_2,\cdots,u_{M\over 2},-u_{M\over 2}\}$.
These selection rules have a natural interpretation: they arise as overlaps between an integrable boundary state and a state
corresponding to the single trace operator.
The resulting overlaps are generalizations of $g$-functions of integrable quantum field theory and concrete proposals
for the all orders in $\lambda$ can then be obtained using the TBA formalism.
The proposal passes a number of non-trivial tests, including comparison to direct one loop computations and comparison to the
superconformal block expansion of one-loop four-point functions of two determinants and two single trace operators.
How much of this remarkable picture generalizes? 
It turns out that the selection rule is not obeyed for less than maximal giants or for dual giants, but it is obeyed for the maximal
giant and operators close to the maximal giant.
For restricted Schur polynomials there is one case in which we find the selection rules are obeyed: restricted Schur polynomials
corresponding to maximal giant gravitons.

In Section \ref{Sec:Giants} we explain the connection between projection operators and Schur polynomials and then the
connection between projection operators and the rewriting performed in \cite{Jiang:2019xdz}.
This immediately allows us to generalize the effective field theory of \cite{Jiang:2019xdz} to less than maximal giant gravitons.
The section ends by developing the Feynman diagrams of the effective field theory (see also \cite{Brezin:2007aa,Brezin:2007wv}) and showing some of the details of the graph duality \cite{Rajesh} underlying the description. 
In Section \ref{Sec:DualGiants} we explain how a very similar construction can be given for dual giants.
One again obtains an effective theory and there is again an underlying graph duality.  
In Section \ref{Sec:GiantCorrelators} we discuss the free theory for correlation functions involving two giant gravitons
and one non-protected single trace operator. 
We specifically look for evidence of the selection rules exhibited in \cite{Jiang:2019xdz}. 
Our conclusion is that the selection rules are only obeyed for the maximal giant and operators close to the maximal giant.
In Section \ref{Sec:DualGiantCorrelators} we extend the analysis to correlation functions involving dual giant gravitons, 
again concluding that the selection rules are not obeyed.
The giant gravitons are Schur polynomials labeled with a Young diagram that has a single column, while dual giant
gravitons are labeled by giant gravitons with a single row.
The general Schur polynomial has many rows and is interpreted as a bound state of giant gravitons (or of dual giant
gravitons).
There are formulas that express the general Schur polynomial in terms of either giant gravitons or in terms of dual giant
gravitons.
We use these formulas in Section \ref{Sec:BoundStates} to construct an effective field theory for general bound states of
giant or dual giant gravitons. 
Another possible generalization is to consider determinants (and more generally other Schur polynomials) constructed using
more than one matrix.
This generalization is worked out in detail in Section \ref{Sec:RestrictedSchur}.
We find that in this case the maximal giant graviton again exhibits the selection rules of \cite{Jiang:2019xdz}, suggesting that this 
case may indeed be integrable.
We collect our conclusions in Section \ref{Sec:Conclusions} and also suggest further directions for study.

\section{Giants}\label{Sec:Giants}

A giant graviton\cite{McGreevy:2000cw} with momentum $k$ is dual to a Schur polynomial labeled with a Young diagram 
that has a single column with $k$ boxes. 
Denote a Young diagram that has a single column with $k$ boxes by $(1^k)$.
The Schur polynomial can be written as
\begin{equation}
\chi_{(1^k)}(Z)={1\over k!}\sum_{\sigma\in S_k}\chi_{(1^k)}(\sigma)Z^{i_1}_{i_{\sigma (1)}}\cdots Z^{i_k}_{i_{\sigma (k)}}
\end{equation}
where $\chi_{(1^k)}(\sigma)$ is a character of the symmetric group $S_k$.
A few examples of the Schur polynomials, which will be useful to check formulas below, are
\begin{equation}
\chi_{\tiny\yng(1)}(Z)={\rm Tr}(Z)
\end{equation}
\begin{equation}
\chi_{\tiny\yng(1,1)}(Z)={1\over 2!}\left({\rm Tr}(Z)^2-{\rm Tr}(Z^2)\right)  \label{2Schur}
\end{equation}
\begin{equation}
\chi_{\tiny\yng(1,1,1)}(Z)={1\over 3!}\left({\rm Tr}(Z)^3-3{\rm Tr}(Z){\rm Tr}(Z^2)+2{\rm Tr}(Z^3)\right)
\end{equation}
\begin{equation}
\chi_{(1^N)}(Z)=\det Z
\end{equation}

The giant two point function is \cite{Corley:2001zk}
\begin{equation}
   \langle \chi_{(1^k)}(x_1) \chi_{(1^k)}^\dagger (x_2)\rangle = {N!\over (N-k)!}
\left( {g_{YM}^2 \over 4\pi^2}\right)^k {1\over |x_1-x_2|^{2k}}
\end{equation}
Three and higher point functions of the operators dual to giants are also easily computed \cite{Corley:2001zk}.
For example
\begin{equation}
\langle \chi_{(1^{J_1})}(x_1)\cdots \chi_{(1^{J_{Q-1}})}(x_{Q-1})
\chi_{(1^{J_{Q}})}^\dagger (x_{Q})\rangle
={N!\over (N-J_Q)!}\left({g_{YM}^2\over 4\pi^2}\right)^{J_Q}\,\,
\prod_{i=1}^{Q-1}{1\over |x_{i Q}|^{2J_i}}
\end{equation}
where $J_Q=J_1+\cdots+J_{Q-1}$.
The Schur polynomials located at $x_1,x_2,...,x_{Q-1}$ are constructed from $Z$'s while the operator at $x_Q$ is 
constructed from $Z^\dagger$.

\subsection{Projectors}

In this section we explain how the Schur polynomial can be written in terms of projection operators.
To do this we need a little more notation.
The matrix $Z^i_j$ transforms the states of an $N$ dimensional vector space $V$.
We employ the multi-index notation (first introduced in \cite{Corley:2001zk,Corley:2002mj}) to write
\begin{equation}
(Z^{\otimes k})^I_J=Z^{i_1}_{j_1}\cdots Z^{i_k}_{j_k}
\end{equation}
\begin{equation}
(\sigma)^I_J=\delta^{i_1}_{j_{\sigma(1)}}\cdots \delta^{i_k}_{j_{\sigma(k)}}
\end{equation}
$Z^{\otimes k}$ transforms the states of $V^{\otimes k}$.
The object
\begin{equation}
\left(P_{(1^k)}\right)^I_J={1\over k!}\sum_{\sigma\in S_k}\chi_{(1^k)}(\sigma)\sigma^I_J
\end{equation}
is a projection operator, acting on $V^{\otimes k}$.
By Schur Weyl duality the space formed by taking the tensor product of $k$ copies of $V$ can be decomposed into
a direct sum, with each component in the sum carrying an irreducible representation of $U(N)$ and an irreducible representation
of $S_k$
\begin{equation}\label{DSdecomp}
V^{\otimes k}=\bigoplus_{\Lambda\vdash k}V^{(S_k)}_{\Lambda}\otimes V^{(U(N))}_{\Lambda}
\end{equation}
$\left(P_{(1^k)}\right)^I_J$ projects onto the $\Lambda =(1^k)$ component in this direct sum decomposition.
In terms of this projection operator, the Schur polynomial can be written as
\begin{equation}
\chi_{(1^k)}(Z)={\rm Tr}_{V^{\otimes k}} (P_{(1^k)} Z^{\otimes k})
\end{equation}

To motivate the discussion below, note that it is possible to write $\chi_{(1^k)}(Z)$ in terms of $\chi_{(1^N)}(Z)$ as follows
(this formula was given in \cite{deMelloKoch:2004crq} and proved in \cite{deMelloKoch:2007rqf})\footnote{The general form of the formula says that if you act with ${\rm Tr}\left({d\over dZ}\right)$ on a Schur polynomial labeled by Young diagram $R$, the result is a sum of Schur polynomials. The labels of the Schur polynomials which appear in the sum, are all possible Young diagrams that can be obtained from $R$ by dropping a box. The coefficient of a given term is the factor of the box that is removed. Recall that a box in row $i$ and column $j$ has a factor $N-i+j$. It is simple to test this formula using the explicit formulas for the Schur polynomials given above.}
\begin{equation}
\chi_{(1^p)}(Z)={1\over (N-p)!}{\rm Tr}\left({d\over dZ}\right)^{N-p}\chi_{(1^N)}(Z)
\end{equation}
Using this formula we have
\begin{eqnarray}
\chi_{(1^p)}(Z)&=&{\rm Tr}_{V^{\otimes p}}(P_{(1^p)}Z^{\otimes p})\cr\cr
&=&{1\over (N-p)!}{N!\over p!}{\rm Tr}_{V^{\otimes N}}(P_{(1^N)}Z^{\otimes p}\otimes{\bf 1}^{\otimes N-p})
\end{eqnarray}
which gives another representation for the projector
\begin{equation}\label{genform}
P_{(1^p)}={1\over (N-p)!}{N!\over p!}{\rm Tr}_{V^{\otimes N-p}}(P_{(1^N)}{\bf 1}^{\otimes N-p})
\end{equation}

We now state an integral representation for the projector $P_{(1^k)}$, which connects to and generalizes formulas written in \cite{Jiang:2019xdz}.
The integral representation uses a zero dimensional complex Grassman vector variable with $N$ components, $\chi_i,\bar\chi^j$ where $i,j=1,...,N$. 
Our normalization conventions are as follows\footnote{This is a rather standard choice.
The only disadvantage is that it is in tension with the conventions of \cite{Jiang:2019xdz} as spelled out in their eqn (3.4).}
\begin{equation}
\int d\bar\chi^N\cdots d\bar\chi^1 \int d\chi_1\cdots d\chi_N \,\,\bar\chi^1 \chi_1\cdots \bar\chi^N\chi_N=(-1)^N
\equiv \int d^N\bar\chi\int d^N\chi \,\,\bar\chi^1 \chi_1\cdots \bar\chi^N \chi_N
\end{equation}
The integral representation for the projector then takes the following form
\begin{equation}
(P_{(1^N)})^I_J={(-1)^N\over N!}\int d\bar\chi^N\cdots d\bar\chi^1 \int d\chi_1\cdots d\chi_N
 \,\,\bar\chi^{i_1} \chi_{j_1}\cdots \bar\chi^{i_N}\chi_{j_N}
\end{equation}
Consequently, we have
\begin{equation}
{\rm Tr}(P_{(1^N)} Z)={(-1)^N\over N!}\int d\bar\chi^N\cdots d\bar\chi^1 \int d\chi_1\cdots d\chi_N
 \,\,\bar\chi^{i_1} Z^{j_1}_{i_1}\chi_{j_1}\cdots \bar\chi^{i_N}Z^{j_N}_{i_N}\chi_{j_N}
\end{equation}
which can also be written as
\begin{eqnarray}
{\rm Tr}(P_{(1^N)} Z)
&=&{(-1)^N\over N!}\int d\bar\chi^N\cdots d\bar\chi^1 \int d\chi_1\cdots d\chi_N \,\, (\bar\chi^{i} Z^{j}_{i}\chi_{j})^N\cr
&=&\int d\bar\chi^N\cdots d\bar\chi^1 \int d\chi_1\cdots d\chi_N \,\, e^{-\bar\chi^{i} Z^{j}_{i}\chi_{j}}
\end{eqnarray}
This is the formula written in \cite{Jiang:2019xdz}, apart from the change in convention for the fermion measure.
Now, using (\ref{genform}) we can write a more general integral representation
\begin{eqnarray}
{\rm Tr}(P_{(1^p)} Z)
&=&{(-1)^N\over (N-p)!p!}\int d\bar\chi^N\cdots d\bar\chi^1 \int d\chi_1\cdots d\chi_N \,\, (\bar\chi^{i} Z^{j}_{i}\chi_{j})^p
(\bar\chi^{i} \chi_{i})^{N-p}\cr\cr
&=&{(-1)^{N-p}\over (N-p)!}\int d\bar\chi^N\cdots 
d\bar\chi^1 \int d\chi_1\cdots d\chi_N \,\, e^{-\bar\chi^{i} Z^{j}_{i}\chi_{j}}
(\bar\chi^{i} \chi_{i})^{N-p}\label{GenRep}
\end{eqnarray}
Finally, we can collect the above results to define the generating function
\begin{equation}
\sum_{k=0}^N t^k {\rm Tr}(P_{(1^k)} Z)
=\int d\bar\chi^N\cdots d\bar\chi^1 \int d\chi_1\cdots d\chi_N \,\, e^{-t\bar\chi^{i} Z^{j}_{i}\chi_{j}-\bar\chi^{i} \chi_{i}}
\end{equation}
Of course, the integral over the Grassman numbers can be carried out and this leads to the well known identity
\begin{equation}
\det ({\bf 1}+tZ)=\sum_{k=0}^N t^k {\rm Tr}(P_{(1^k)} Z)
\end{equation}
The reader may find it useful to check this answer for $N=3$. 
Work in the basis in which $Z$ is diagonal.
In this case
\begin{equation}
\det ({\bf 1}+tZ)=\det \left[\begin{matrix} t+z_1 &0 &0\cr 0 &t+z_2 &0\cr 0 &0 &t+z_3\end{matrix}\right]
= z_1 z_2 z_3 t^3 +(z_1 z_2 + z_1 z_3 + z_2 z_3)t^2 + (z_1 + z_2 + z_3)t+ 1
\end{equation}
These are indeed the correct answers since
\begin{equation}
\chi_{\tiny\yng(1)}(Z)={\rm Tr}(Z)=z_1+z_2+z_3
\end{equation}
\begin{equation}
\chi_{\tiny\yng(1,1)}(Z)={1\over 2!}\left({\rm Tr}(Z)^2-{\rm Tr}(Z^2)\right)
=z_1z_2+z_1 z_3+z_2z_3   
\end{equation}
\begin{eqnarray}
\chi_{\tiny\yng(1,1,1)}(Z)&=&{1\over 3!}\left({\rm Tr}(Z)^3-3{\rm Tr}(Z){\rm Tr}(Z^2)+2{\rm Tr}(Z^3)\right)
=z_1 z_2 z_3
\end{eqnarray}

\subsection{Large $N$ Effective Theory}

Following \cite{Jiang:2019xdz} we start with the free theory.
Consider correlation functions of giant gravitons at positions $x_A$ for $A=1,2,...,Q$.
The operator at $x_A$ is given by the Schur polynomial $\chi_{(1^{p_A})}(x_A)$.
The Schur polynomial located at $x_A=(0,a_A,0,0)$ is constructed using the field
\begin{equation}
{\cal Z}(x_A)=Z+\kappa^2 a_A^2 Z^\dagger +\kappa a_A (Y-Y^\dagger)\equiv\vec{\cal Y}\cdot\vec\phi\label{ForY}
\end{equation}
where
\begin{equation}
Z={1\over\sqrt{2}}(\phi^1+i\phi^2)\qquad
Y={1\over\sqrt{2}}(\phi^3+i\phi^4)
\end{equation}
One can read off ${\cal Y}^I$ in moving from the second to third expression in (\ref{ForY}).
There are no UV divergences that need to be considered to define our operator because
\begin{equation}
\left\langle {\cal Z}^i_j(x_K){\cal Z}^r_s(x_K)\right\rangle =0
\end{equation}
In an effort to carry out a general discussion, we consider the generating function
\begin{eqnarray}
&&\sum_{i_1,i_2,\cdots i_Q=1}^N t_1^{i_1}t_2^{i_2}\cdots t_Q^{i_Q}
\left\langle \chi_{(1^{i_1})}(x_1)\chi_{(1^{i_2})}(x_2)\cdots \chi_{(1^{i_Q})}(x_Q)\,{\cal O}(x)\right\rangle
=\int [d\phi^I] \cr\cr
&&\int \prod_{K=1}^Q
[d\bar\chi_K d\chi_K]e^{-{1\over g_{YM}^2}\int d^4 x\left( \sum_{I=1}^6 {1\over 2}
{\rm Tr}(\partial_\mu \phi^I\partial^\mu \phi^I)
+g_{YM}^2\sum_{K=1}^Q \delta (x-x_K)\bar \chi_K^i(\delta^j_i+t_K {\cal Z}^j_i)\chi_{Kj}\right)}
{\cal O}(\phi^I)\cr\cr
&&
\end{eqnarray}
where ${\cal O}$ is a general single trace operator and the measure is normalized so that
\begin{equation}
\int [d\phi^I] e^{-{1\over g_{YM}^2}\int d^4 x \sum_{I=1}^6{1\over 2}{\rm Tr}(\partial_\mu \phi^I \partial^\mu \phi^I)}=1\label{NormPhi}
\end{equation}
In what follows, ${\cal O}$ is constructed as a normal ordered operator, so that contractions of fields in ${\cal O}$ vanish.
The integral over the $(\phi^I)^i_j$ field is Gaussian and hence is easily performed, after the integral identity for the Schur polynomials has been employed.
Completing the square as usual, we find
\begin{eqnarray}
&&-\int d^4 x\left( {1\over 2}{\rm Tr}(\partial_\mu \phi^I \partial^\mu \phi^I)
+g_{YM}^2\sum_{K=1}^Q \delta (x-x_K)\bar \chi_K^i(\delta^j_i+t_K {\cal Z}^j_i)\chi_{Kj}\right)\cr
&=&\!\!\!
-\int d^4 x \left(-{1\over 2} {\rm Tr}\left[(\phi^I-S^I)\partial^\mu\partial_\mu (\phi^I-S^I) 
- S^I\partial^\mu\partial_\mu S^I\right]
+g_{YM}^2\sum_{K=1}^Q \delta (x-x_K)\bar \chi_K^i\chi_{Ki}\right)\cr\cr
&&
\end{eqnarray}
where
\begin{equation}
(S^I(x))^i_j=-{g_{YM}^2\over 4\pi^2}\sum_{K=1}^Q{ t_K {\cal Y}_K^I\bar\chi_K^i \chi_{Kj}\over |x-x_K|^2}
\end{equation}
%
%
%
Changing integration variables $\phi^I\to \tilde\phi^I=\phi^I-S^I$ we find
\begin{eqnarray}
&&\sum_{i_1,i_2,\cdots i_Q=1}^N t_1^{i_1}t_2^{i_2}\cdots t_Q^{i_Q}
\left\langle \chi_{(1^{i_1})}(x_1)\chi_{(1^{i_2})}(x_2)\cdots \chi_{(1^{i_Q})}(x_Q)\,{\cal O}(x)\right\rangle\cr\cr
&=&\int [d\tilde\phi^I]\quad e^{-{1\over g_{YM}^2}\int d^4 x
\sum_{I=1}^6{1\over 2}{\rm Tr}(\partial_\mu \tilde \phi^I\partial^\mu \tilde \phi^I)} \int \prod_{K=1}^Q
[d\bar\chi_K d\chi_K]\cr
&&\quad e^{-{g_{YM}^2\over 8\pi^2}\sum_{K\ne J=1}^Q 
{t_K t_J\vec{\cal Y}_K\cdot\vec{\cal Y}_J\over x_{KJ}^2}\bar \chi_K^a \chi_{Ja}
\bar \chi_J^b \chi_{Kb}-\sum_{K=1}^Q\bar\chi_K^a\chi_{Ka}}
{\cal O}(\tilde\phi^I(x)+S^I(x))
\end{eqnarray}
The integral over $\tilde\phi^I$ will Wick contract the $\tilde\phi^I$s appearing in ${\cal O}$. 
Since ${\cal O}$ is built to be normal ordered, these contributions vanish and hence
\begin{eqnarray}
&&\sum_{i_1,i_2,\cdots i_Q=1}^N t_1^{i_1}t_2^{i_2}\cdots t_Q^{i_Q}
\left\langle \chi_{(1^{i_1})}(x_1)\chi_{(1^{i_2})}(x_2)\cdots \chi_{(1^{i_Q})}(x_Q)\,{\cal O}(x)\right\rangle\cr\cr
&=&\int [d\tilde\phi^I] e^{-{1\over g_{YM}^2}\int d^4 x
\sum_{I=1}^6{1\over 2}{\rm Tr}(\partial_\mu \tilde \phi^I\partial^\mu \tilde \phi^I)}
\int \prod_{K=1}^Q [d\bar\chi_K d\chi_K]\cr\cr
&&\qquad e^{-{g_{YM}^2\over 8\pi^2}\sum_{K\ne J=1}^Q 
{t_Kt_J\vec{\cal Y}_K\cdot\vec{\cal Y}_J\over x_{KJ}^2}\bar \chi_K^a \chi_{Ja}
\bar \chi_J^b \chi_{Kb}-\sum_{K=1}^Q\bar\chi_K^a\chi_{Ka}}
{\cal O}(S^I(x))\cr\cr
&=& \int \prod_{K=1}^Q
[d\bar\chi_K d\chi_K]e^{-{g_{YM}^2\over 8\pi^2}
\sum_{K\ne J=1}^Q {t_J t_K \vec{\cal Y}_K\cdot\vec{\cal Y}_J\over x_{KJ}^2}\bar \chi_K^a \chi_{Ja}
\bar \chi_J^b \chi_{Kb}-\sum_{K=1}^Q\bar\chi_K^a\chi_{Ka}}
{\cal O}(S^I(x))\cr
&&
\end{eqnarray}
The last step in the construction of the effective action entails performing the Hubbard-Stratanovich (H-S) transformation, 
which rewrites the quartic fermion interaction as a quadratic interaction with an auxiliary bosonic field.
The auxiliary field is an anti-hermitian matrix $\rho_{JK}$ with diagonal elements set to zero.  
In what follows normalize the measure for the H-S field so that
\begin{equation}
\int [d\rho]e^{{8\pi^2\over g_{YM}^2}{\rm Tr}(\rho^2)}=1\label{Msre}
\end{equation}
The above integral is perfectly convergent because ${\rm Tr}(\rho^2)<0$.
After the H-S transformation we have
\begin{eqnarray}
&&\sum_{i_1,i_2,\cdots i_Q=1}^N t_1^{i_1}t_2^{i_2}\cdots t_Q^{i_Q}
\left\langle \chi_{(1^{i_1})}(x_1)\chi_{(1^{i_2})}(x_2)\cdots \chi_{(1^{i_Q})}(x_Q)\,{\cal O}(x)\right\rangle\cr\cr
&=& \int [d\rho]\int \prod_{K=1}^Q
[d\bar\chi_K d\chi_K]e^{{8\pi^2\over g_{YM}^2}{\rm Tr}(\rho^2)+2
\sum_{K\ne J=1}^Q \sqrt{{t_K t_J\vec{\cal Y}_K\cdot\vec{\cal Y}_J\over x_{KJ}^2}}\rho_{KJ}\bar \chi_K^a \chi_{Ja}
-\sum_{K=1}^Q\bar\chi_K^a\chi_{Ka}}
{\cal O}(S^I(x))\cr
&&
\end{eqnarray}
where $\rho$ is a $Q\times Q$ matrix variable.
The integral over the fermions is now simple and leads to
\begin{eqnarray}
&&\sum_{i_1,i_2,\cdots i_Q=1}^N t_1^{i_1}t_2^{i_2}\cdots t_Q^{i_Q}
\left\langle \chi_{(1^{i_1})}(x_1)\chi_{(1^{i_2})}(x_2)\cdots \chi_{(1^{i_Q})}(x_Q)\,{\cal O}(x)\right\rangle\cr\cr
&=& \int [d\rho] e^{{8\pi^2 N\over\lambda}{\rm Tr}(\rho^2)+N {\rm Tr}\log\left[\delta_{JK}-2
\sqrt{t_Jt_K{\vec{\cal Y}_K\cdot\vec{\cal Y}_J\over x_{KJ}^2}}\rho_{KJ}\right]}
\langle {\cal O}(S^I(x))\rangle_\chi\label{FinalSEff}
\end{eqnarray}
and where $\langle {\cal O}^I(S^I(x))\rangle_\chi$ is defined by Wick contracting all pairs of $\chi,\bar\chi$ fields according
to Wick's theorem, with the basic contraction given by
\begin{equation}
\langle\bar\chi_J^a\chi_{Kb}\rangle =-\delta^a_b M^{-1}_{JK}
\end{equation}
where
\begin{equation}
M_{JK}=\delta_{JK}-2\sqrt{{t_J t_K\vec{\cal Y}_J\cdot\vec{\cal Y}_K\over x_{JK}^2}}\rho_{JK}
\end{equation}
Notice that the $\chi,\bar\chi$ fields transform as vectors ($\chi$ transforms in the fundamental and $\bar\chi$ in the
anti-fundamental) under $U(N)$.
The large $N$ limit of vector models is always given by contracting the pairs of fields that share the same color index.
Since ${\cal O}$ is a single trace operator, it is a sum of terms of the form
\begin{equation}
{\cal O}={\rm Tr}\left(\phi^{I_1}\phi^{I_2}\phi^{I_3}\cdots\phi^{I_J}\right)
\end{equation}
After integrating over $\phi^I$ $\phi$s become $S$s
\begin{eqnarray}
{\cal O}&=&{\rm Tr}\left(S^{I_1}S^{I_2}\cdots S^{I_J}\right)\cr\cr
&=&\left(-{g_{YM}^2\over 4\pi^2}\right)^J\sum_{K_1,\cdots,K_J=1}^Q t_{K_1}\cdots t_{K_J}
{\cal Y}^{I_1}_{K_1}\cdots {\cal Y}^{I_J}_{K_J}
{\bar\chi^{j_J}_{K_1}\chi_{K_1 j_1}\bar\chi^{j_1}_{K_2}\chi_{K_2 j_2}\cdots \bar\chi^{j_{J-1}}_{K_{J}}\chi_{K_{J} j_J}
\over |x-x_{K_1}|^2\, |x-x_{K_2}|^2\,\cdots\, |x-x_{K_J}|^2}\cr
&&
\end{eqnarray}
Thus, at the leading order in $N$ we have
\begin{eqnarray}
\langle {\cal O}(S^I(x))\rangle_\chi&=&-
\left(-{\lambda\over 4\pi^2}\right)^J\sum_{K_1,\cdots,K_J=1}^Q t_{K_1}\cdots t_{K_J}
{\cal Y}^{I_1}_{K_1}\cdots {\cal Y}^{I_J}_{K_J}\cr
&&\times {(-M^{-1})_{K_1K_2}(-M^{-1})_{K_2K_3}\cdots (-M^{-1})_{K_J K_1}\over 
|x-x_{K_1}|^2\, |x-x_{K_2}|^2\,\cdots\, |x-x_{K_J}|^2}
\end{eqnarray}
If we now define
\begin{equation}
\Phi^I(x)\equiv {\lambda\over 4\pi^2}{\rm diag}\left({t_1{\cal Y}^I_1\over |x-x_1|^2},
{t_2{\cal Y}^I_2\over |x-x_2|^2},\cdots,{t_Q{\cal Y}^I_Q\over |x-x_Q|^2}\right)\, M^{-1}
\end{equation}
we have
\begin{equation}
\langle {\cal O}(S^I(x))\rangle_\chi=-{\rm Tr}_Q (\Phi^{I_1}(x)\Phi^{I_2}(x)\cdots\Phi^{I_J}(x))
\end{equation}
It is simple to rewrite this as an overlap between a matrix product state
\begin{equation}
|\Phi\rangle=-\sum_{I_1,I_2,\cdots I_J=1}^6 {\rm Tr}_Q (\Phi^{I_1}(x)\Phi^{I_2}(x)\cdots\Phi^{I_J}(x))
|\phi^{I_1}\cdots\phi^{I_J}\rangle
\end{equation}
and a state corresponding to ${\cal O}$
\begin{equation}
|{\cal O}\rangle =|\phi^{I_1}\cdots\phi^{I_J}\rangle
\end{equation}
as
\begin{equation}
\langle {\cal O}(S^I(x))\rangle_\chi=\langle \Phi |{\cal O}\rangle
\end{equation}
This last step is simply a change in notation. 
However, as pointed out in \cite{Jiang:2019xdz} it is a useful change in notation because, following \cite{deLeeuw:2015hxa} matrix product states (MPS) have proved to be a useful language to make contact with integrability when studying one point functions in defect CFT.
The same will be true in our discussion below.
 
\subsection{Simple Tests}

To start, we will consider the large $N$ limit of the effective field theory.
Consider the case where ${\cal O}=1$ and $Q=2$.
In this case we are considering correlators of the form
\begin{equation}
\langle \chi_{(1^J)}(x_1)\chi_{(1^J)}^\dagger (x_2)\rangle ={N!\over (N-J)!}
\left({g_{YM}^2\over 4\pi^2}\right)^J
{1\over |x_1-x_2|^{2J}}
\end{equation}
where $J$ is order $N$.
Expanding at large $N$ the leading behavior is
\begin{equation}
\langle \chi_{(1^J)}(x_1)\chi_{(1^J)}^\dagger (x_2)\rangle =\left({\lambda\over 4\pi^2}\right)^J
{e^{-J}\over |x_1-x_2|^{2J}}\label{N2pnt}
\end{equation}
Does the effective field theory reproduce this correlation function?
Using the effective field theory, we have
\begin{eqnarray}
\sum_{i_1,i_2=1}^N t_1^{i_1}t_2^{i_2}
\left\langle \chi_{(1^{i_1})}(x_1)\chi_{(1^{i_2})}^\dagger (x_2)\right\rangle
= \int [d\rho] e^{{8\pi^2 N\over\lambda}{\rm Tr}(\rho^2)
+N {\rm Tr}\log\left[\delta_{KJ}-2\sqrt{t_1 t_2\over x_{12}^2}\rho_{KJ}\right]}
\end{eqnarray}
Our goal is to verify that a saddle point evaluation of the above integral reproduces the large $N$ limit of the two point
correlation function given in (\ref{N2pnt}).
Since $\rho$ is antihermitian, we make the ansatz
\begin{equation}
\rho=\left[
\begin{matrix}
0            &\rho\\
-\rho^*  &0
\end{matrix}\right]
\end{equation}
Plugging this into the ``effective action''
\begin{equation}
S_{\rm eff}=-{8\pi^2 N\over\lambda}{\rm Tr}(\rho^2)
-N{\rm Tr}\log\left[\delta_{KJ}-2\sqrt{t_1 t_2\over x_{12}^2}\rho_{KJ}\right]
\end{equation}
and extremizing we find
\begin{equation}
|\rho|^2 ={\lambda \over 16\pi^2}-{|x_{12}|^2\over 4t_1 t_2}\label{rhosaddle}
\end{equation}
Simply evaluating the integrand at this extremum (i.e. we will not perform the Gaussian integral about the saddle point and for this reason the nontrivial contribution to the measure produced by implementing (\ref{Msre}) does not contribute)
we find (the first equality is only true at the leading order of the large $N$ expansion)
\begin{eqnarray}
e^{-S_{\rm eff}}=\sum_{i_1,i_2=1}^N t_1^{i_1}t_2^{i_2}
\left\langle \chi_{(1^{i_1})}(x_1)\chi_{(1^{i_2})}(x_2)\right\rangle
= e^{-N}\sum_{k=0}^\infty {\lambda^{N-k}N^k\over k!}{t_1^{N-k}t_2^{N-k}\over (4\pi^2 |x_{12}|^2)^{N-k}}
\end{eqnarray}
Note that the spacetime dependence is correct.
Further, if we set $N-k=J$ and take ${J\over N}\ll 1$ we can write
\begin{equation}
k!=k^k e^{-k}=(N-J)^{N-J}e^{-N+J}\approx N^{N-J}e^{-N+J}=N^{k}e^{-N+J}
\end{equation}
Thus,
\begin{eqnarray}
\sum_{i_1,i_2=1}^N t_1^{i_1}t_2^{i_2}
\left\langle \chi_{(1^{i_1})}(x_1)\chi_{(1^{i_2})}(x_2)\right\rangle
= \sum_{J} t_1^Jt_2^J\,\, e^{-J}{\lambda^{J}\over (4\pi^2 |x_{12}|^2)^J}
\end{eqnarray}
which is the correct answer.

In fact, with a little extra effort we can evaluate the integral for the two point function exactly.
The effective action is
\begin{equation}
S_{\rm eff}={16\pi^2 N\over\lambda}|\rho|^2-N \log \left(1+4{t_1t_2\over x_{12}^2}|\rho|^2\right)
\end{equation}
If we set $\rho=\sqrt{r}e^{i\theta}$ and then $\tilde r={16\pi^2 N\over\lambda} r$, the effective action becomes
%
%
\begin{equation}
S_{\rm eff}=\tilde r-N \log \left(1+{g_{YM}^2 t_1t_2\over 4\pi^2 x_{12}^2}\tilde r\right)
\end{equation}
and (the normalization of the measure from (\ref{Msre}) is included here) 
\begin{equation}
\int [d\rho] e^{-S_{\rm eff}}=\int_0^\infty e^{-\tilde r}\left(1+{g_{YM}^2 t_1 t_2\over 4\pi^2 x_{12}^2}\tilde r\right)^N
d\tilde r
\end{equation}
Expanding we have
\begin{equation}
\int [d\rho] e^{-S_{\rm eff}}=\sum_{J=0}^N {N!\over (N-J)! J!}\left({g_{YM}^2 t_1 t_2\over 4\pi^2 x_{12}^2}\right)^J
\int_0^\infty e^{-\tilde r}\tilde r^J d\tilde r
=\sum_{J=0}^N {N!\over (N-J)!}\left({g_{YM}^2 t_1 t_2\over 4\pi^2 x_{12}^2}\right)^J
\end{equation}
which is the exact answer.
We can be more general: a natural correlator to consider is the extremal $Q$-point function, whose large $N$ limit is given by
\begin{equation}
\langle \chi_{(1^{J_1})}(x_1)\cdots \chi_{(1^{J_{Q-1}})}(x_{Q-1})
\chi_{(1^{J_{Q}})}^\dagger (x_{Q})\rangle
=e^{-J_Q}\left({\lambda\over 4\pi^2}\right)^{J_Q}\,\,
\prod_{i=1}^{Q-1}{1\over |x_{i Q}|^{2J_i}}\label{ExtremalSphere}
\end{equation}
where $J_Q=J_1+\cdots+J_{Q-1}$.
The Schur polynomials at $x_1,x_2,...,x_{Q-1}$ are constructed from $Z$'s while the polynomial at $x_Q$ is constructed from $Z^\dagger$.
Consequently
\begin{equation}
{\cal Y}_i\cdot{\cal Y}_Q=1\qquad i=1,2,\cdots,Q-1
\end{equation}
and all other inner products vanish. 
We want to evaluate
\begin{equation}
I_Q= \int [d\rho] e^{{8\pi^2 N\over\lambda}{\rm Tr}(\rho^2)
+N {\rm Tr}\log\left[\delta_{KJ}-2\sqrt{t_1 t_2\over x_{12}^2}\rho_{KJ}\right]}
\end{equation}
Set
\begin{equation}
\rho=\left[
\begin{matrix}
0 & \rho_{12} &\rho_{13} &\cdots &\rho_{1\,Q-1} &\rho_{1 Q}\cr
-\rho_{12}^* &0 &\rho_{23} &\cdots &\rho_{2\,Q-1} &\rho_{2\,Q}\cr
-\rho_{13}^* &\rho_{23}^* &0 &\cdots &\rho_{3\,Q-1} &\rho_{3\,Q}\cr
\vdots &\vdots &\vdots &\cdots &\vdots &\vdots\cr
-\rho_{1\,Q-1}^* &-\rho_{2\,Q-1}^* &-\rho_{3\,Q-1}^* &\cdots &0 &\rho_{Q-1\, Q}\cr
-\rho_{1\,Q}^*    &-\rho_{2\,Q}^*    &-\rho_{3\,Q}^*    &\cdots &-\rho_{Q-1\, Q}^* &0
\end{matrix}
\right]
\end{equation}
A straight forward computation shows that the effective action
\begin{equation}
S_{\rm eff}=-{8\pi^2 N\over\lambda}{\rm Tr}(\rho^2)
-N{\rm Tr}\log\left[\delta_{KJ}-2\sqrt{t_J t_K{\cal Y}_J\cdot{\cal Y}_K\over x_{JK}^2}\rho_{KJ}\right]
\end{equation}
is
\begin{equation}
S_{\rm eff}={16\pi^2 N\over\lambda}\sum_{i<j=1}^Q |\rho_{ij}|^2
-N\log\left[1+4\sum_{i=1}^{Q-1}{t_i t_Q\over x_{iQ}^2}|\rho_{iQ}|^2\right]
\end{equation}
This is rather simple: a single variable enters non-linearly.
We integrate over most of the integration variables which are basically spectators.
The variables which play a role are
\begin{equation}
\rho_i \equiv \rho_{iQ}
\end{equation}
After integrating over all other variables, the action is
\begin{equation}
S_{\rm eff}={16\pi^2 N\over\lambda}\sum_{i=1}^{Q-1} |\rho_i|^2
-N\log\left[1+4\sum_{i=1}^{Q-1}{t_i t_Q\over x_{iQ}^2}|\rho_i|^2\right]
\end{equation}
Now use the following change of variables
\begin{eqnarray}
r_i&=&|\rho_i|^2 \qquad i=1,2,\cdots,Q-2\cr
r_{Q-1}&=&\sum_{i=1}^{Q-1}{t_i t_Q\over x_{iQ}^2}|\rho_i|^2
\end{eqnarray}
which leads to the effective action
\begin{equation}
S_{\rm eff}={16\pi^2 N\over\lambda}\sum_{i=1}^{Q-2} r_i
+{16\pi^2 N\over\lambda}{x^2_{Q-1\, Q}\over t_{Q-1}t_Q}r_{Q-1}
-{16\pi^2 N\over\lambda}\sum_{i=1}^{Q-2}{x^2_{Q-1\, Q}t_i\over x^2_{iQ} t_{Q-1}}r_i
-N\log\left[1+4r_{Q-1}\right]
\end{equation}
Split the computation into two parts.
The integrations over $r_1,r_2,\cdots r_{Q-2}$ are simple to do exactly and they give
\begin{equation}
I_Q=\prod_{i=1}^{Q-2}{1\over 1-{x^2_{Q-1\, Q}t_i\over x^2_{iQ} t_{Q-1}}}
\int [d r_{Q-1}]e^{-S_{Q-1}}
\end{equation}
\begin{equation}
S_{Q-1}={16\pi^2 N\over\lambda}{x^2_{Q-1\, Q}\over t_{Q-1}t_Q}r_{Q-1}
-N\log\left[1+4r_{Q-1}\right]
\end{equation}
To get the large $N$ answer we will do the integral over $r_{Q-1}$ using a saddle point evaluation\footnote{It is also simple to
perform this integral exactly.}.
The saddle point is at
\begin{equation}
r_{Q-1}^*={1\over 4}\left[{\lambda\over 4\pi^2}{t_{Q-1}t_Q\over x^2_{Q-1\, Q}}-1\right]
\end{equation}
Evaluating $e^{-S_{Q-1}}$ at its saddle point we have
\begin{equation}
I_Q\approx \prod_{i=1}^{Q-2}{1\over 1-{x^2_{Q-1\, Q}t_i\over x^2_{iQ} t_{Q-1}}}
e^{-S_{Q-1}(r_{Q-1}^*)}=\prod_{i=1}^{Q-2}{1\over 1-{x^2_{Q-1\, Q}t_i\over x^2_{iQ} t_{Q-1}}}
e^{-N}\left({\lambda\over 4\pi^2}{t_{Q-1}t_Q\over x^2_{Q-1\, Q}}\right)^N 
e^{{4\pi^2 N\over\lambda}{x^2_{Q-1\, Q}\over t_{Q-1}t_Q}}
\end{equation}
Performing manipulations just like we did for the two point function leads to
\begin{equation}
I_Q\approx \sum_J \prod_{i=1}^{Q-2}{1\over 1-{x^2_{Q-1\, Q}t_i\over x^2_{iQ} t_{Q-1}}}
e^{-J}\left({\lambda\over 4\pi^2}{t_{Q-1}t_Q\over x^2_{Q-1\, Q}}\right)^J
\end{equation}
Expanding this answer in a power series in the $t_i$, we find complete agreement with (\ref{ExtremalSphere}) showing that we reproduce the correct large $N$ values of any extremal correlator.

It is also instructive to consider examples with a non-trivial single trace operator included in the correlator.
Towards this end consider the correlation function
\begin{equation}
\langle \chi^\dagger_{(1^K)}(x_1)\chi_{(1^{K-J})}(x_2){\rm Tr}(Z^J)(0)\rangle = (-1)^{J-1}{N!\over (N-K)!}
\left({g_{YM}^2\over 4\pi^2}\right)^K {1\over x_{12}^{2K-2J}}{1\over x_{1}^{2J}}
\end{equation}
The single trace operator is automatically normal ordered because the contraction of $Z$ with itself vanishes.
This is the exact result - we will obtain the leading large $N$ approximation to this answer, which reads
\begin{equation}
\langle \chi^\dagger_{(1^K)}(x_1)\chi_{(1^{K-J})}(x_2){\rm Tr}(Z^J)(0)\rangle = (-1)^{J-1} N^K e^{-K}
\left({g_{YM}^2\over 4\pi^2}\right)^K {1\over x_{12}^{2K-2J}}{1\over x_{1}^{2J}}
\end{equation}
The spatial dependence is easily understood: $Z$s must be contracted with $Z^\dagger$s.
Thus all the fields at $x_2$ must be contracted with fields with fields at $x_1$, and
all the fields at $0$ must be contracted with fields at $x_1$.
Using the effective field theory description we find
\begin{eqnarray}
\sum_{i_1,i_2}^N t_1^{i_1}t_2^{i_2}
\left\langle \chi_{(1^{i_1})}(x_1)\chi_{(1^{i_2})}(x_2)\,{\cal O}(0)\right\rangle
&=&\int [d\rho] e^{{8\pi^2 N\over\lambda}{\rm Tr}(\rho^2)+N {\rm Tr}\log\left[\delta_{JK}-2
\sqrt{t_1 t_2\over x_{12}^2}\rho_{KJ}\right]}
\langle {\cal O}(S^I(0))\rangle_\chi\cr\cr
&=&\sum_{L=0}^N \left( {g_{YM}^2t_1t_2\over 4\pi^2 x_{12}^{2}}\right)^L N^L e^{-L}
\langle {\cal O}(S^I(0))\rangle_\chi
\end{eqnarray}
where ${\cal O}(0)={\rm Tr}(Z^J)(0)$.
Recall that $\langle {\cal O}(S^I(x))\rangle_\chi$ is defined by Wick contracting all pairs of $\chi,\bar\chi$ fields according
to Wick's theorem, with the basic contraction given by
\begin{equation}
\langle\bar\chi_J^a\chi_{Kb}\rangle =-\delta^a_b M^{-1}_{JK}
\end{equation}
where
\begin{equation}
M_{JK}=\delta_{JK}-2\sqrt{t_1 t_2\over x_{12}^2}\rho_{JK}\qquad\Rightarrow\qquad
M^{-1}={4\pi^2 x_{12}^2\over \lambda t_1 t_2}\left[
\begin{matrix}
1 &2\sqrt{t_1 t_2\over x_{12}^2}\rho\cr
-2\sqrt{t_1 t_2\over x_{12}^2}\rho^* &1
\end{matrix}\right]
\end{equation}
To obtain this result we have used the large $N$ equation
\begin{equation}
|\rho|^2 = {\lambda\over 16\pi^2}-{|x_{12}|^2\over 4t_1t_2}
\end{equation}
Using the explicit form ${\cal O}={\rm Tr}\left(Z^J\right)$, after integrating over $\phi^I$ so that $\phi$s become $S$s 
(as above), we have
\begin{eqnarray}
{\cal O}
=\left({g_{YM}^2\over 4\pi^2}\right)^J t_1^J
{\bar\chi^{j_J}_1 \chi_{1 j_1}\bar\chi^{j_1}_{1}\chi_{1 j_2}\cdots \bar\chi^{j_{J-1}}_{1}\chi_{1 j_J}
\over |x_1|^{2J}}
\end{eqnarray}
Thus, at the leading order in $N$ we have
\begin{eqnarray}
\langle {\cal O}(S^I(x))\rangle_\chi=-
\left({\lambda\over 4\pi^2}\right)^J t_1^J 
{((-M^{-1})_{11})^J\over |x_1|^{2J}}=(-1)^{J-1}{1\over t_2^J}{x_{12}^{2J}\over x_1^J}
\end{eqnarray}
Thus, we find
\begin{eqnarray}
\sum_{i_1,i_2}^N t_1^{i_1}t_2^{i_2}
\left\langle \chi_{(1^{i_1})}(x_1)\chi_{(1^{i_2})}(x_2)\,{\cal O}(x)\right\rangle
=\sum_{L=0}^N \left( {g_{YM}^2t_1t_2\over 4\pi^2 x_{12}^{2}}\right)^L N^L e^{-L}
(-1)^{J-1}{1\over t_2^J}{x_{12}^{2J}\over x_1^{2J}}
\end{eqnarray}
Selecting the term with $L=K$ we find
\begin{equation}
\left( {g_{YM}^2t_1t_2\over 4\pi^2 x_{12}^{2}}\right)^K N^L e^{-K}
(-1)^{J-1}{1\over t_2^J}{x_{12}^{2J}\over x_1^J}
=(-1)^{J-1} N^K e^{-K}
\left({g_{YM}^2\over 4\pi^2}\right)^K {1\over x_{12}^{2K-2J}}{1\over x_{1}^{2J}}
\end{equation}
which is the correct result.
The contribution coming from $\langle {\cal O}(S^I(x))\rangle_\chi$ was a single term. 
This is not the generic case: in general we have multiple terms contributing. 
This feature will be important when we discuss selection rules obeyed by the correlator, so it is helpful to consider
one more example in which multiple terms do indeed contribute.
The example we consider is the correlator
\begin{equation}
\langle \chi^\dagger_{(1^K)}(x_1)\chi_{(1^{K})}(x_2):{\rm Tr}(ZZ^\dagger):(0)\rangle = K{N!\over (N-K)!}
\left({g_{YM}^2\over 4\pi^2}\right)^{K+1} {1\over x_{12}^{2K-2}}{1\over x_{1}^{2}}{1\over x_{2}^{2}}
\end{equation}
so that now ${\cal O}(x)=:{\rm Tr}(ZZ^\dagger):(x)$.
The contraction between $Z$ and $Z^\dagger$ is non-zero, so the normal ordering above is needed, and it indicates
that we should not contract the fields within the trace.
The above result is again exact.
For this example, the effective field theory computation involves both diagonal and off diagonal terms in $M^{-1}$.
We have
\begin{equation}
M_Z=-{\lambda\over 4\pi^2 \left(1+{4t_1t_2 |\rho|^2\over x_{12}^2}\right)}{t_1\over x_1^2}
\left[\begin{matrix} 1 &2\sqrt{t_1 t_2\over x_{12}^2}\rho\cr 0 &0\end{matrix}\right]
\end{equation}
\begin{equation}
M_{Z^\dagger}=-{\lambda\over 4\pi^2 \left(1+{4t_1t_2 |\rho|^2\over x_{12}^2}\right)}{t_2\over x_2^2}
\left[\begin{matrix} 0 &0 \cr -2\sqrt{t_1 t_2\over x_{12}^2}\rho^* &1\end{matrix}\right]
\end{equation}
where we should evaluate the above solution on the large $N$ saddle given in (\ref{rhosaddle}).
It is simple to find
\begin{equation}
\langle {\cal O}(S^I(x))\rangle_\chi=
-{\rm Tr}(M_Z M_{Z^\dagger})=
{\lambda\over 4\pi^2x_{12}^2}{x_{12}^4 \over x_1^2 x_2^2}-{1\over t_1t_2}{x_{12}^4\over x_1^2 x_2^2}
%
\end{equation}
It now follows that
\begin{eqnarray}
\sum_{i_1,i_2}^N t_1^{i_1}t_2^{i_2}
\left\langle \chi_{(1^{i_1})}(x_1)\chi_{(1^{i_2})}(x_2)\,{\cal O}(x)\right\rangle
&=&\sum_{L=0}^N \left( {g_{YM}^2t_1t_2\over 4\pi^2 x_{12}^{2}}\right)^L {N!\over (N-L)!}
\left({\lambda\over 4\pi^2x_{12}^2}{x_{12}^4 \over x_1^2 x_2^2}
-{1\over t_1t_2}{x_{12}^4\over x_1^2 x_2^2}\right)\cr
&&\label{NEFG}
\end{eqnarray}
From the first term above, $L=K$ contributes and for the second term $L=K+1$ contributes.
Summing these two terms we find
\begin{equation}
\left( {g_{YM}^2\over 4\pi^2}\right)^{K+1} {1\over x_{12}^{2K-2}}{1\over x_1^2}{1\over x_2^2}
{N!\over (N-K)!}\left( N- (N-K)\right)
=K\left( {g_{YM}^2\over 4\pi^2}\right)^{K+1} {1\over x_{12}^{2K-2}}{1\over x_1^2}{1\over x_2^2}
{N!\over (N-K)!}
\end{equation}
which is indeed the exact result\footnote{The reader may be wondering how we managed to obtain the exact result.
We have used the exact expression from the $\rho$ integration obtained earlier in this section. Thus, the only large $N$
approximation was in computing $\langle {\cal O}(S^I(x))\rangle_\chi$. This is exact because there are no 
subleading terms.}. Notice that for the maximal giant ($K=N$) the second term in (\ref{NEFG}) vanishes.
Further, for any giant for which $(N-K)/N\to 0$ as $N\to\infty$ it is a subleading contribution at large $N$, so that the
second term does not contribute at the leading order in large $N$ for giants which are nearly maximal. 

\subsection{Graph Duality}

In this subsection we will explore the $\rho$ description further. Specifically, we will explain how the graphs of the $\rho$
theory are related to the graphs of the original matrix model. In \cite{Jiang:2019xdz} a close connection between
Gopakumar's ``open-closed-open'' duality\cite{Rajesh} 
and the $\rho$ theory was pointed out. This implies that there is a one-to-one
correspondence between the graphs of the original matrix model and the graphs of the $\rho$ theory. In this section we are simply working some of these details out.
 
\begin{figure}[ht]%
\begin{center}
\includegraphics[width=0.7\columnwidth]{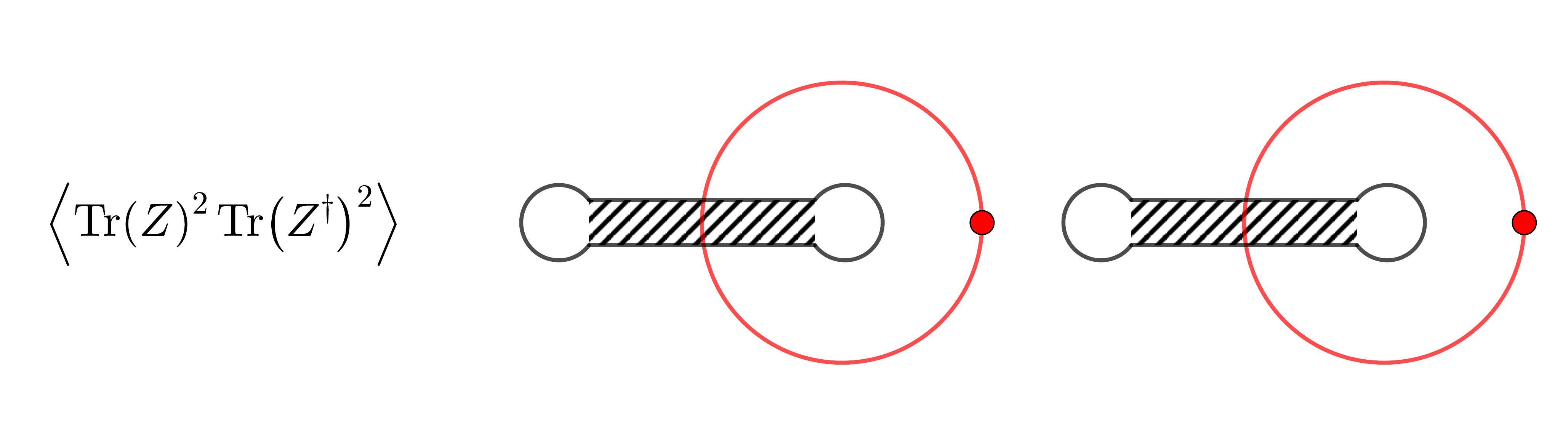}%
\caption{The original ribbon graph is in black. The $\rho$ graph is in red.}%
\label{FDiag1}%
\end{center}
\end{figure}

For simplicity, we will focus on $Q=2$.
The effective action is
\begin{eqnarray}
S_{\rm eff}&=&{16\pi^2 N\over\lambda}|\rho|^2-N\log \left(1+4{t_1 t_2\over |x_{12}|^2}|\rho|^2\right)\cr
&=&{16\pi^2 N\over\lambda}|\rho|^2-N{4t_1t_2\over x_{12}^2}|\rho|^2
+{N\over 2}\left({4t_1t_2\over x_{12}^2}\right)^2|\rho|^4
-{N\over 3}\left({4t_1t_2\over x_{12}^2}\right)^3|\rho|^6
+\cdots\cr
&&
\end{eqnarray}
From this effective action we can see that the propagator for the $\rho$ field is ${\lambda\over 16\pi^2 N}$.
There are also $2n$ point vertices with $n=1,2,3,...$.
Lets consider the simplest possible correlation function: the two point function of two giant gravitons which each are 
constructed using only 2 $Z$s. 
The relevant correlation function is
\begin{equation}
\langle ({\rm Tr}(Z)^2-{\rm Tr}(Z^2))({\rm Tr}(Z^\dagger)^2-{\rm Tr}(Z^{\dagger\,2}))\rangle
\end{equation}
The graph duality exchanges edges with edges (so for each $Z$ propagator there is a $\rho$ propagator) and it exchanges
vertices and faces.
Consider first $\langle{\rm Tr}(Z)^2{\rm Tr}(Z^\dagger)^2\rangle$.
The relevant ribbon graph with the ribbon colored in is given in Figure \ref{FDiag1}.
Each single ribbon has a single face (drawing the diagram on a sphere only cuts the sphere - it does not split it into two) so
the $\rho$ Feynman diagram has two disconnected vertices. 
Here we are making use of the fact that there is a one-to-one correspondence between the original ribbon graphs and the new $\rho$ Feynman diagrams.

The $\rho$ diagram is shown in red.
The value of the diagram is
\begin{equation}
\left({\lambda\over 16\pi^2 N}\right)^2 \left(N{4t_1t_2\over x_{12}^2}\right)^2
=\left({\lambda\over 4\pi^2}\right)^2 {t_1^2 t_2^2 \over |x_{12}|^4}
\end{equation}
This is the leading contribution in the ribbon graph language - it would have been the above expression multiplied by $N^2$.
The sign of the diagram is correct.

\begin{figure}[ht]%
\begin{center}
\includegraphics[width=0.75\columnwidth]{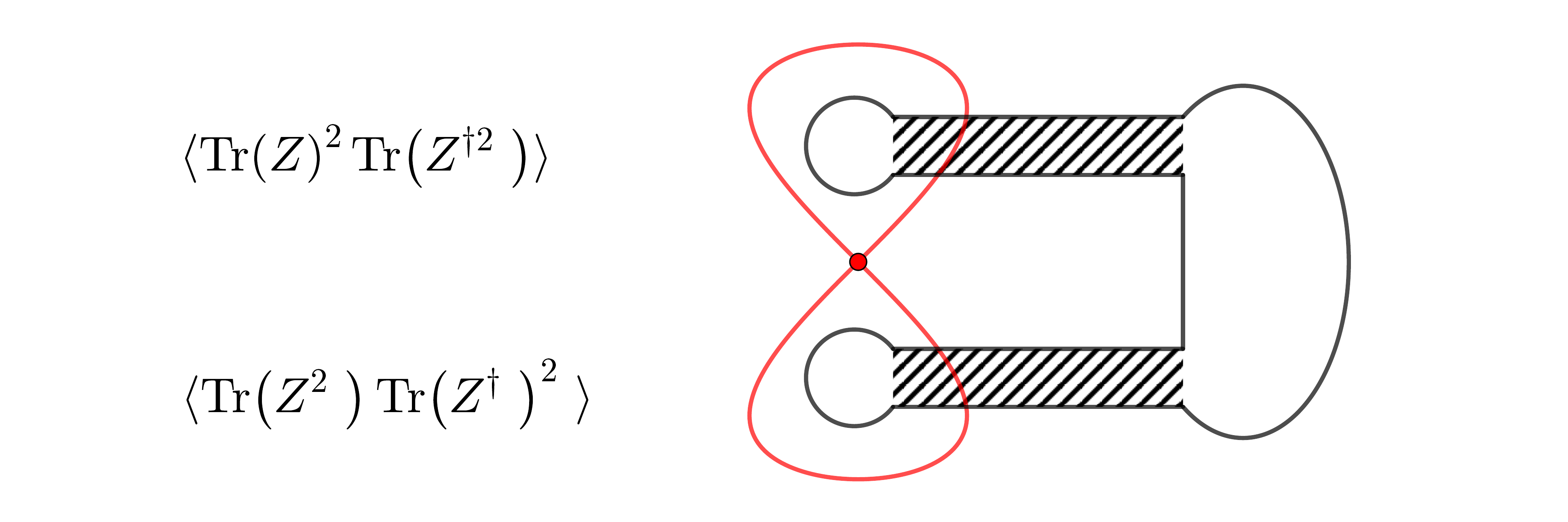}%
\caption{The original ribbon graph is in black. The $\rho$ graph is in red.}%
\label{FDiag2}%
\end{center}
\end{figure}

Now consider $\langle{\rm Tr}(Z^2){\rm Tr}(Z^\dagger)^2\rangle$ or $\langle{\rm Tr}(Z)^2{\rm Tr}(Z^{\dagger\, 2} )
\rangle$.
The relevant ribbon graph with the ribbon colored in is given in Figure \ref{FDiag2}.
The ribbon graph has a single face (again drawing the diagram on a sphere only cuts the sphere - it does not split it into two) so
the $\rho$ Feynman diagram again has one vertex. 

The $\rho$ diagram is shown in red.
The value of the diagram is
\begin{equation}
2\left({\lambda\over 16\pi^2 N}\right)^2 \left(-{N\over 2} \left({4t_1t_2\over x_{12}^2}\right)^2\right)
=-{1\over N}\left({\lambda\over 4\pi^2}\right)^2 {t_1^2 t_2^2 \over |x_{12}|^4}
\end{equation}
This is a subleading contribution in the ribbon graph language - it is down by one factor of $N$ compared to the leading term,
exactly as we see here.
Note in addition that the sign of this contribution has come out correctly.

\begin{figure}[ht]%
\begin{center}
\includegraphics[width=0.9\columnwidth]{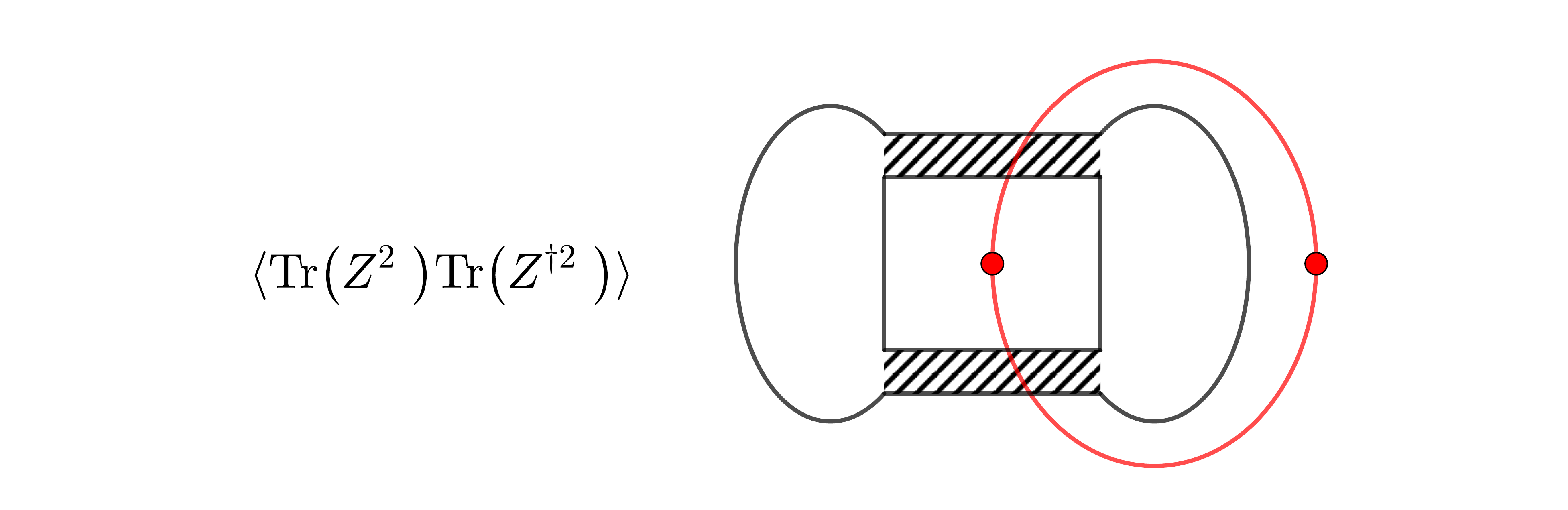}%
\caption{The original ribbon graph is in black. The $\rho$ graph is in red.}%
\label{FDiag3}%
\end{center}
\end{figure}

Now consider $\langle{\rm Tr}(Z^2){\rm Tr}(Z^{\dagger\, 2})\rangle$.
The relevant ribbon graph with the ribbon colored in is given in Figure \ref{FDiag3}.
The ribbon graph has two faces (drawing the diagram on a sphere cuts the sphere into two) so
the $\rho$ Feynman diagram has two vertices. 

The $\rho$ diagram is shown in red.
The value of the diagram is
\begin{equation}
\left({\lambda\over 16\pi^2 N}\right)^2 \left(N{4t_1t_2\over x_{12}^2}\right)^2
=\left({\lambda\over 4\pi^2}\right)^2 {t_1^2 t_2^2 \over |x_{12}|^4}
\end{equation}
The ribbon graph is a leading contribution so this is the correct $N$ dependence.
Also the sign and exact form has come out correct.

\section{Dual Giants}\label{Sec:DualGiants}

A dual giant graviton\cite{Grisaru:2000zn,Hashimoto:2000zp} with momentum $k$ is given by a Schur polynomial labeled with a Young diagram that has a single row with $k$ boxes. 
Denote a Young diagram that has a single row with $k$ boxes by $(k)$.
The Schur polynomial can be written as
\begin{equation}
\chi_{(k)}(Z)={1\over k!}\sum_{\sigma\in S_k}\chi_{(k)}(\sigma)Z^{i_1}_{i_{\sigma (1)}}\cdots Z^{i_k}_{i_{\sigma (k)}}
\end{equation}
where $\chi_{(k)}(\sigma)$ is again a character of the symmetric group $S_k$.
Lets write down a few examples which will be useful below
\begin{equation}
\chi_{\tiny\yng(1)}(Z)={\rm Tr}(Z)
\end{equation}
\begin{equation}
\chi_{\tiny\yng(2)}(Z)={1\over 2!}\left({\rm Tr}(Z)^2+{\rm Tr}(Z^2)\right)
\end{equation}
\begin{equation}
\chi_{\tiny\yng(3)}(Z)={1\over 3!}\left({\rm Tr}(Z)^3+3{\rm Tr}(Z){\rm Tr}(Z^2)+2{\rm Tr}(Z^3)\right)
\end{equation}

\subsection{Projectors}

The sum
\begin{equation}
\left(P_{(k)}\right)^I_J={1\over k!}\sum_{\sigma\in S_k}\chi_{(k)}(\sigma)\sigma^I_J
\end{equation}
is again a projection operator, acting on $V^{\otimes k}$.
$\left(P_{(k)}\right)^I_J$ projects onto the $\Lambda =(k)$ component in the direct sum decomposition (\ref{DSdecomp}).
In terms of this projection operator, the Schur polynomial can be written as
\begin{equation}
\chi_{(k)}(Z)={\rm Tr} (P_{(k)} Z^{\otimes k})
\end{equation}

There is again an integral representation for the projector $P_{(k)}$, which will allow us to further generalize the 
formulas written in \cite{Jiang:2019xdz}.
The integral representation makes use of zero dimensional complex vector variables with $N$ components, 
$\varphi_i,\bar\varphi^j$ where $i,j=1,...,N$. 
Our normalization conventions are spelled out in
\begin{equation}
\int d^N\bar\varphi\int d^N\varphi \,\,e^{-\bar\varphi^i \varphi_i}=1
\end{equation}
\begin{equation}
\int d^N\bar\varphi\int d^N\varphi \,\,e^{-\bar\varphi^i \varphi_i}\bar\varphi^k\varphi_l=\delta^k_l
\end{equation}
Using Wick's theorem we have
\begin{equation}
\int d^N\bar\varphi\int d^N\varphi \,\,e^{-\bar\varphi^i \varphi_i}\bar\varphi^{k_1}\cdots \bar\varphi^{k_q}
\varphi_{l_1}\cdots\varphi_{l_q}=\sum_{\sigma\in S_q}\sigma^K_L=k!(P_{(k)})^K_L
\end{equation}
which provides the integral representation for the projector.
Consequently, we have
\begin{equation}
{\rm Tr}(P_{(k)} Z)
={1\over k!}\int d^N\bar\varphi\int d^N \varphi\,\, e^{-\bar\varphi^i\varphi_i}(\bar\varphi^{i} Z^{j}_{i}\varphi_{j})^k
\label{DGntsProject}
\end{equation}
Thus we can write the generating function for Schur polynomials with a single row as
\begin{equation}
\sum_{k=0}^\infty t^k \chi_{(k)}(Z)=
\int d^N\bar\varphi\int d^N \varphi\,\, e^{-\bar\varphi^i\varphi_i+t\bar\varphi^{i} Z^{j}_{i}\varphi_{j}}
\label{DGnts}
\end{equation}
An important point for us is that this is again an ``action'' that is quadratic in the fields (here $\varphi,\bar\varphi$) 
so we will be able to repeat much of the analysis we carried out for the giants.
Notice that after integrating over the $\varphi$ fields we have
\begin{equation}
{1\over\det ({\bf 1}-tZ)}=\sum_{k=0}^\infty t^k \chi_{(k)}(Z)
\end{equation}
It is instructive to check this identity for the simple case that $N=2$.
A straight forwards computation gives
\begin{equation}
{1\over (1-tz_1)(1-tz_2)}=1+t(z_1+z_2)+t^2(z_1^2+z_1 z_2+z_2^2)+t^3(z_1^3+z_1^2 z_2+z_1 z_2^2+z_2^3)+\cdots
\end{equation}
To see that this is indeed correct, note that
\begin{equation}
\chi_{\tiny\yng(1)}(Z)={\rm Tr}(Z)=z_1+z_2
\end{equation}
\begin{equation}
\chi_{\tiny\yng(2)}(Z)={1\over 2!}\left({\rm Tr}(Z)^2+{\rm Tr}(Z^2)\right)
={1\over 2}\left((z_1+z_2)^2+(z_1^2+z_2^2)\right)=z_1^2+z_1z_2+z_2^2
\end{equation}
\begin{eqnarray}
\chi_{\tiny\yng(3)}(Z)={1\over 3!}\left({\rm Tr}(Z)^3+3{\rm Tr}(Z){\rm Tr}(Z^2)+2{\rm Tr}(Z^3)\right)=
z_1^3+z_1^2 z_2+z_1 z_2^2+z_2^3
\end{eqnarray}

\subsection{Large $N$ Effective Theory}

We again consider the free theory.
There are dual giant gravitons at position $x_A$ for $A=1,2,...,Q$.
The dual giant graviton at $x_A$ corresponds to the Schur polynomial $\chi_{(p_A)}(x_A)$.
Again, the Schur polynomial located at $x_A=(0,a_A,0,0)$ is constructed using the field ${\cal Z}(x_A)$.
In order to have a general discussion, we again consider a generating function
\begin{eqnarray}
&&\sum_{i_1,i_2,\cdots i_Q=1}^\infty t_1^{i_1}t_2^{i_2}\cdots t_Q^{i_Q}
\left\langle \chi_{(i_1)}(x_1)\chi_{(i_2)}(x_2)\cdots \chi_{(i_Q)}(x_Q)\,{\cal O}\right\rangle=\int [d\phi^I]\cr\cr
&&\int \prod_{K=1}^Q[d\bar\varphi_K d\varphi_K]
e^{-{1\over g_{YM}^2}\int d^4 x\left( \sum_{I=1}^6 {1\over 2}{\rm Tr}(\partial_\mu \phi^I\partial^\mu \phi^I)
+g_{YM}^2\sum_{K=1}^Q \delta (x-x_K)\bar \varphi_K^i(\delta^j_i-t_K {\cal Z}^j_i)\varphi_{Kj}\right)}
{\cal O}(\phi^I)\cr\cr
&&
\end{eqnarray}
where ${\cal O}$ is once more a general single trace operator and the measure is normalized as above (see (\ref{NormPhi})).
In what follows, ${\cal O}$ is again chosen normal ordered so contractions of fields in ${\cal O}$ will vanish, as above.
The integration over the $(\phi^I)^i_j$ field is a Gaussian integral which is easily performed to find
\begin{eqnarray}
&&-\int d^4 x\left( {1\over 2}{\rm Tr}(\partial_\mu \phi^I \partial^\mu \phi^I)
+g_{YM}^2\sum_{K=1}^Q \delta (x-x_K)\bar \varphi_K^i(\delta^j_i-t_K {\cal Z}^j_i)\varphi_{Kj}\right)\cr
&=&\int d^4 x \left({1\over 2}{\rm Tr}\left[ (\phi^I-S^I)\partial^\mu\partial_\mu (\phi^I-S^I) - S^I\partial^\mu\partial_\mu S^I\right]
-g_{YM}^2\sum_{K=1}^Q \delta (x-x_K)\bar \varphi_K^i\varphi_{Ki}\right)\cr\cr
&&
\end{eqnarray}
where
\begin{equation}
(S^I(x))^i_j= {g_{YM}^2\over 4\pi^2}\sum_{K=1}^Q{ t_K {\cal Y}_K^I\bar\varphi_K^i \varphi_{Kj}\over |x-x_K|^2}
\end{equation}
Changing integration variables $\phi^I\to \tilde\phi^I=\phi^I-S^I$ we find
\begin{eqnarray}
&&\sum_{i_1,i_2,\cdots i_Q=1}^\infty t_1^{i_1}t_2^{i_2}\cdots t_Q^{i_Q}
\left\langle \chi_{(i_1)}(x_1)\chi_{(i_2)}(x_2)\cdots \chi_{(i_Q)}(x_Q)\,{\cal O}(x)\right\rangle\cr\cr
&=&\int [d\tilde\phi^I] e^{-{1\over g_{YM}^2}\int d^4 x
\sum_{I=1}^6{1\over 2}{\rm Tr}(\partial_\mu \tilde \phi^I\partial^\mu \tilde \phi^I)} \int \prod_{K=1}^Q
[d\bar\varphi_K d\varphi_K]\cr
&&\quad e^{{g_{YM}^2\over 8\pi^2}\sum_{K\ne J=1}^Q 
{t_K t_J\vec{\cal Y}_K\cdot\vec{\cal Y}_J\over x_{KJ}^2}\bar \varphi_K^a \varphi_{Ja}
\bar \varphi_J^b \varphi_{Kb}-\sum_{K=1}^K\bar\varphi_K^a\varphi_{Ka}}
{\cal O}(\tilde\phi^I(x)+S^I(x))
\end{eqnarray}
The integral over $\tilde\phi^I$ will again Wick contract the $\tilde\phi^I$s appearing in ${\cal O}$ and, again, since 
${\cal O}$ is built normal ordered, these contributions vanish.
Hence
\begin{eqnarray}
&&\sum_{i_1,i_2,\cdots i_Q=1}^\infty  t_1^{i_1}t_2^{i_2}\cdots t_Q^{i_Q}
\left\langle \chi_{(i_1)}(x_1)\chi_{(i_2)}(x_2)\cdots \chi_{(i_Q)}(x_Q)\,{\cal O}(x)\right\rangle\cr\cr
&=&\int [d\tilde\phi^I] \int \prod_{K=1}^Q
[d\bar\varphi_K d\varphi_K]e^{-{1\over g_{YM}^2}\int d^4 x
\sum_{I=1}^6{1\over 2}{\rm Tr}(\partial_\mu \tilde \phi^I\partial^\mu \tilde \phi^I)}\cr\cr
&&\qquad\times e^{{g_{YM}^2\over 8\pi^2}\sum_{K\ne J=1}^Q 
{t_Kt_J\vec{\cal Y}_K\cdot\vec{\cal Y}_J\over x_{KJ}^2}\bar \varphi_K^a \varphi_{Ja}
\bar \varphi_J^b \varphi_{Kb}-\sum_{K=1}^Q\bar\varphi_K^a\varphi_{Ka}}
{\cal O}(S^I(x))\cr\cr
&=& \int \prod_{K=1}^Q
[d\bar\varphi_K d\varphi_K]e^{{g_{YM}^2\over 8\pi^2}
\sum_{K\ne J=1}^Q {t_J t_K \vec{\cal Y}_K\cdot\vec{\cal Y}_J\over x_{KJ}^2}\bar\varphi_K^a \varphi_{Ja}
\bar \varphi_J^b \varphi_{Kb}-\sum_{K=1}^Q\bar\varphi_K^a\varphi_{Ka}}
{\cal O}(S^I(x))\cr\cr
&&
\end{eqnarray}
In this last expression the integral does not converge. 
How do we understand this expression?
We should expand the integral in a power series in the $t_k$'s, choose a coefficient of some monomial in the $t_k$'s and then
do the integral.
For helpful background, see \cite{Eynard:2006bs}.
The last step in the construction of the effective action entails performing the H-S transformation, which rewrites
the quartic $\varphi$ interaction as a quadratic interaction with an auxiliary bosonic field.
After the H-S transformation we have
\begin{eqnarray}
&&\sum_{i_1,i_2,\cdots i_Q=1}^\infty t_1^{i_1}t_2^{i_2}\cdots t_Q^{i_Q}
\left\langle \chi_{(i_1)}(x_1)\chi_{(i_2)}(x_2)\cdots \chi_{(i_Q)}(x_Q)\,{\cal O}(x)\right\rangle\cr\cr
&=& \int [d\rho]\int \prod_{K=1}^Q
[d\bar\varphi_K d\varphi_K]e^{-{8\pi^2\over g_{YM}^2}{\rm Tr}(\rho^2)-2
\sum_{K\ne J=1}^Q \sqrt{{t_K t_J\vec{\cal Y}_K\cdot\vec{\cal Y}_J\over x_{KJ}^2}}\rho_{KJ}\bar \varphi_K^a \varphi_{Ja}
-\sum_{K=1}^Q\bar\varphi_K^a\varphi_{Ka}}
{\cal O}(S^I(x))\cr\cr
&&
\end{eqnarray}
where $\rho$ is a $Q\times Q$ hermitian matrix variable with vanishing diagonal elements.
The integral over the $\varphi,\bar\varphi$ variables is now simple and leads to 
%
\begin{eqnarray}
&&\sum_{i_1,i_2,\cdots i_Q=1}^\infty t_1^{i_1}t_2^{i_2}\cdots t_Q^{i_Q}
\left\langle \chi_{(i_1)}(x_1)\chi_{(i_2)}(x_2)\cdots \chi_{(i_Q)}(x_Q)\,{\cal O}(x)\right\rangle\cr\cr
&=& \int [d\rho] e^{-{8\pi^2 N\over\lambda}{\rm Tr}(\rho^2)-N {\rm Tr}\log\left[\delta_{KJ} + 2
\sum_{K\ne J=1}^Q \sqrt{t_Jt_K{\vec{\cal Y}_K\cdot\vec{\cal Y}_J\over x_{KJ}^2}}\rho_{KJ}\right]}
\langle {\cal O}(S^I(x))\rangle_\varphi
\end{eqnarray}
and where $\langle {\cal O}^I(S^I(x))\rangle_\varphi$ is defined by Wick contracting all pairs of $\varphi,\bar\varphi$ fields
according to Wick's theorem, with the basic contraction given by
\begin{equation}
\langle\bar\varphi_J^a\varphi_{Kb}\rangle =\delta^a_b M^{-1}_{JK}
\end{equation}
where
\begin{equation}
M_{JK}=\delta_{JK}+2\sqrt{{t_J t_K\vec{\cal Y}_J\cdot\vec{\cal Y}_K\over x_{JK}^2}}\rho_{JK}
\end{equation}
Notice that the $\varphi,\bar\varphi$ fields transform as vectors under $U(N)$.
We can now repeat the discussion we carried out above: the large $N$ limit of vector models is always given by contracting the 
pairs of fields that share the same color index.
Since ${\cal O}$ is a single trace operator, it is a sum of terms of the form
\begin{equation}
{\cal O}={\rm Tr}\left(\phi^{I_1}\phi^{I_2}\phi^{I_3}\cdots\phi^{I_J}\right)
\end{equation}
After integrating over $\phi^I$ the $\phi$s become $S$s
\begin{eqnarray}
{\cal O}&=&{\rm Tr}\left(S^{I_1}S^{I_2}\cdots S^{I_J}\right)\cr\cr
&=&\left({g_{YM}^2\over 4\pi^2}\right)^J\sum_{K_1,\cdots,K_J=1}^Q t_{K_1}\cdots t_{K_J}
{\cal Y}^{I_1}_{K_1}\cdots {\cal Y}^{I_J}_{K_J}
{\bar\varphi^{j_J}_{K_1}\varphi_{K_1 j_1}\bar\varphi^{j_1}_{K_2}\varphi_{K_2 j_2}\cdots 
\bar\varphi^{j_{J-1}}_{K_{J}}\varphi_{K_{J} j_J}
\over |x-x_{K_1}|^2\, |x-x_{K_2}|^2\,\cdots\, |x-x_{K_J}|^2}\cr\cr
&&
\end{eqnarray}
Thus, at the leading order in the large $N$ expansion we have
\begin{eqnarray}
\langle {\cal O}(S^I(x))\rangle_\varphi&=&
\left({\lambda\over 4\pi^2}\right)^J\sum_{K_1,\cdots,K_J=1}^Q t_{K_1}\cdots t_{K_J}
{\cal Y}^{I_1}_{K_1}\cdots {\cal Y}^{I_J}_{K_J}\cr\cr
&&\times {(M^{-1})_{K_1K_2}(M^{-1})_{K_2K_3}\cdots (M^{-1})_{K_J K_1}\over 
|x-x_{K_1}|^2\, |x-x_{K_2}|^2\,\cdots\, |x-x_{K_J}|^2}
\end{eqnarray}
If we now define
\begin{equation}
\Phi^I(x)\equiv {\lambda\over 4\pi^2}{\rm diag}\left({t_1{\cal Y}^I_1\over |x-x_1|^2},
{t_2{\cal Y}^I_2\over |x-x_2|^2},\cdots,{t_Q{\cal Y}^I_Q\over |x-x_Q|^2}\right)(M^{-1})
\end{equation}
we have
\begin{equation}
\langle {\cal O}(S^I(x))\rangle_\varphi={\rm Tr}_Q (\Phi^{I_1}(x)\Phi^{I_2}(x)\cdots\Phi^{I_J}(x))
\end{equation}
It is again possible to rewrite this as an overlap between a matrix product state
\begin{equation}
|\Phi\rangle=\sum_{I_1,I_2,\cdots I_J=1}^6 {\rm Tr}_Q (\Phi^{I_1}(x)\Phi^{I_2}(x)\cdots\Phi^{I_J}(x))
|\phi^{I_1}\cdots\phi^{I_J}\rangle
\end{equation}
and a state corresponding to ${\cal O}$
\begin{equation}
|{\cal O}\rangle =|\phi^{I_1}\cdots\phi^{I_J}\rangle
\end{equation}
as
\begin{equation}
\langle {\cal O}(S^I(x))\rangle_\varphi=\langle \Phi |{\cal O}\rangle
\end{equation}
We will see that this is again a useful change in notation.
 
\subsection{Simple tests}

The simplest test we can consider is a two point function of dual giant gravitons.
This case corresponds to ${\cal O}=1$ and $K=2$.
The exact answer for the correlators we are considering is
\begin{equation}
\langle \chi_{(J)}(x_1)\chi_{(J)}^\dagger (x_2)\rangle ={(N+J-1)!\over (N-1)!}
\left({g_{YM}^2 \over 4\pi^2}\right)^J {1\over |x_1-x_2|^{2J}}
\end{equation}
where $J$ is order $N$.
We would like to show that this answer can be reproduced using the effective theory.
The effective theory computation entails evaluating
\begin{eqnarray}
\sum_{i_1,i_2=1}^\infty t_1^{i_1}t_2^{i_2}
\left\langle \chi_{(i_1)}(x_1)\chi_{(i_2)}(x_2)\right\rangle
= \int [d\rho] e^{-{8\pi^2 N\over\lambda}{\rm Tr}(\rho^2)
-N {\rm Tr}\log\left[\delta^K_J+2\sqrt{t_1 t_2\over x_{12}^2}\rho_{KJ}\right]}
\end{eqnarray}
%
Since $\rho$ is hermitian with vanishing diagonal elements we have
\begin{equation}
\rho=\left[
\begin{matrix}
0      &\rho^*\\
\rho  &0
\end{matrix}\right]
\end{equation}
Plugging this into the ``effective action'' we  get
\begin{equation}
S_{\rm eff}={16\pi^2 N\over\lambda}|\rho|^2
+N \log\left[ 1-4{t_1 t_2\over x_{12}^2}|\rho|^2\right]
\end{equation}
Thus, the integral we need to perform is
\begin{equation}
\sum_{i_1,i_2=1}^\infty t_1^{i_1}t_2^{i_2}\left\langle \chi_{(i_1)}(x_1)\chi_{(i_2)}(x_2)\right\rangle
=\int [d\rho] {e^{-{16\pi^2 N\over\lambda}|\rho|^2}\over (1-4{t_1 t_2\over x_{12}^2}|\rho|^2)^N}
\end{equation}
Now, as we explained above, because we are dealing with a formal expression, we should expand the integrand and 
pick up a specific coefficient of a $(t_1t_2)^n$ monomial.
Only after doing this should we do the integral.
The formula we need is
\begin{equation}
\left( {1\over 1-x}\right)^N=\sum_{J=0}^\infty {(N+J-1)!\over (N-1)! J!}x^J
\end{equation}
It is now simple to find
\begin{equation}
\sum_{i_1,i_2=1}^\infty t_1^{i_1}t_2^{i_2}\left\langle \chi_{(i_1)}(x_1)\chi_{(i_2)}(x_2)\right\rangle
=\sum_{J=0}^{\infty}\left( 4{t_1 t_2\over x_{12}^2}\right)^J{(N+J-1)!\over (N-1)! J!}
\int [d\rho] e^{-{16\pi^2 N\over\lambda}|\rho|^2}
|\rho|^{2J}
\end{equation}
After a suitable change of variables the above integral can be expressed in terms of the $\Gamma$ function.
The result is
\begin{equation}
\sum_{i_1,i_2=1}^\infty t_1^{i_1}t_2^{i_2}\left\langle \chi_{(i_1)}(x_1)\chi_{(i_2)}(x_2)\right\rangle
=\sum_{J=0}^\infty (t_1t_2)^J {(N+J-1)!\over (N-1)!}\left({g_{YM}^2 \over 4\pi^2}\right)^J {1\over |x_1-x_2|^{2J}}
\end{equation}
There are obvious generalizations of this check that we will not pursue here.

\subsection{Graph Duality}

Our goal in this subsection is to show that the graph duality that underlies the $\rho$ theory derived for the giant graviton
correlation functions also plays a role for $\rho$ theory derived for the dual giant graviton correlation functions.
We will do this by again studying the graphs of the $\rho$ theory and understanding how they are related to the graphs of the 
original matrix model. 

For simplicity, again consider $Q=2$.
The effective action for the dual giant gravitons is
\begin{eqnarray}
S_{\rm eff}&=&{16\pi^2 N\over\lambda}|\rho|^2+N\log \left(1-4{t_1 t_2\over |x_{12}|^2}|\rho|^2\right)\cr
&=&{16\pi^2 N\over\lambda}|\rho|^2-N{4t_1t_2\over x_{12}^2}|\rho|^2
-{N\over 2}\left({4t_1t_2\over x_{12}^2}\right)^2|\rho|^4
-{N\over 3}\left({4t_1t_2\over x_{12}^2}\right)^3|\rho|^6
+\cdots
\end{eqnarray}
From this effective action we can see that the propagator for the $\rho$ field is ${\lambda\over 16\pi^2 N}$ which
is exactly what we had for the giants.
There are also $2n$ point vertices with $n=1,2,3,...$.
The most important difference as compared to the giant graviton effective theory is that all interaction vertices here
are positive.
This is exactly what must happen: recall the form of the Schur polynomials
\begin{eqnarray}
\chi_{\tiny\yng(1,1,1)}(Z)&=&{1\over 3!}\left({\rm Tr}(Z)^3-3{\rm Tr}(Z){\rm Tr}(Z^2)+2{\rm Tr}(Z^3)\right)\cr
\chi_{\tiny\yng(3)}(Z)&=&{1\over 3!}\left({\rm Tr}(Z)^3+3{\rm Tr}(Z){\rm Tr}(Z^2)+2{\rm Tr}(Z^3)\right)
\end{eqnarray}

\begin{figure}[h]%
\begin{center}
\includegraphics[width=1\columnwidth]{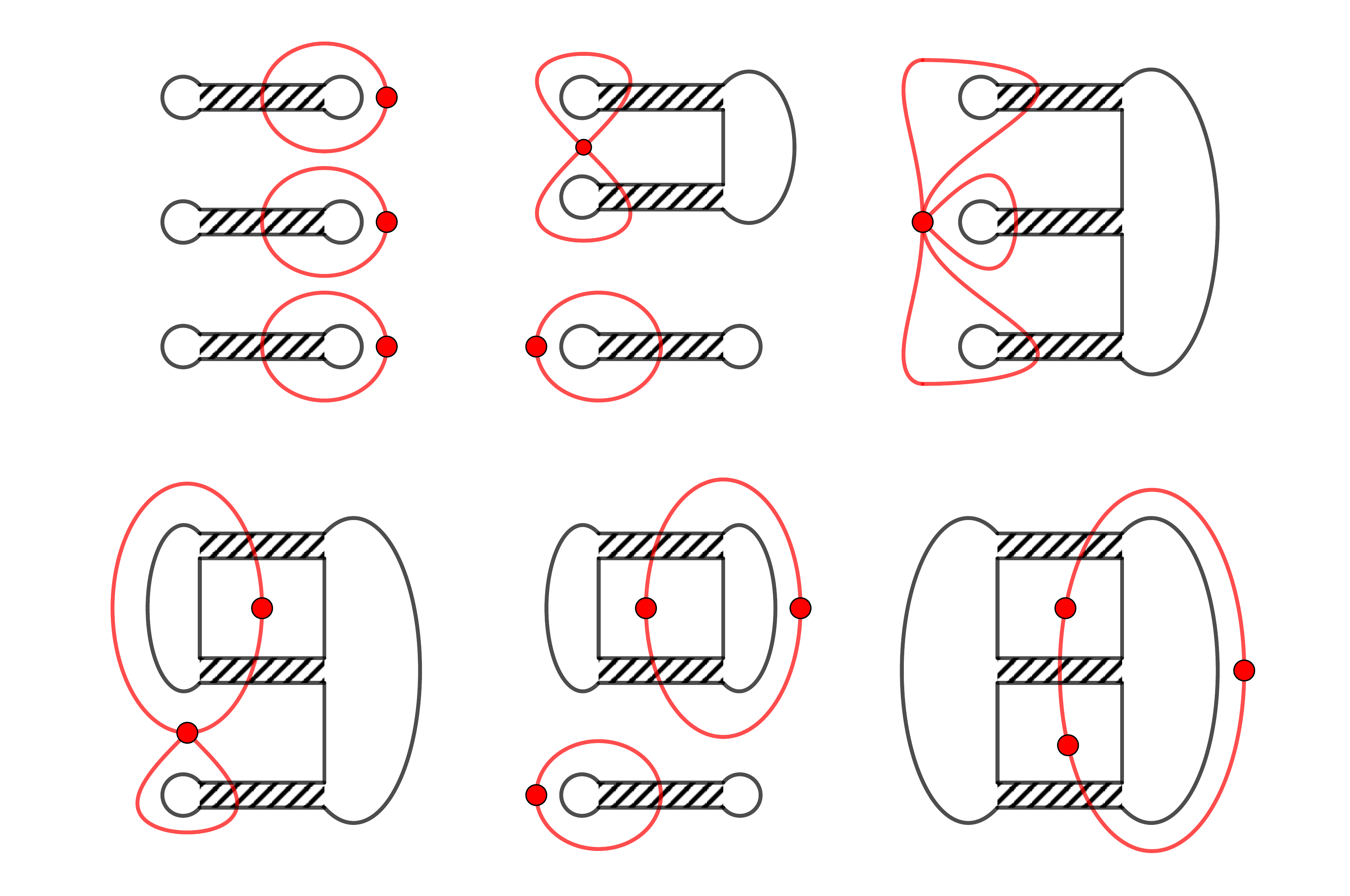}%
\caption{Ribbon graphs (in black) and $\rho$ graphs (in red) for $n=3$ fields.}%
\label{FDiagnis3}%
\end{center}
\end{figure}

To compute a giant graviton correlator we will have both positive and negative signs - and this is coded in detail to the signs of
vertices of the $\rho$ theory of the giants.
There are positive and negative signs because the character appearing in the Schur polynomial can be both positive and 
negative.
To compute a dual giant graviton correlator we will have only positive signs - and this is coded into the positivity of all
vertices of the $\rho$ theory of the dual giants.
Only positive signs appear because the character appearing in the Schur polynomial is always equal to 1.
In Figure \ref{FDiagnis3} we show the types of ribbon graphs and the corresponding $\rho$ graphs that contribute for 
$n=3$ boxes.
The detailed expressions for the two different diagrams agree perfectly.
The $N$ dependence depends on the number of vertices and propagators (each vertex gives an $N$ and each
propagator a $1/N$).
The coefficient depends in detail on things like symmetry factors, which all work out correctly.
This again establishes a duality at the levels of graphs: there is one $\rho$ graph for each ribbon graph.

\section{Correlation functions of two giant gravitons and a single trace operator}\label{Sec:GiantCorrelators}

In this section we will reconsider the computation of correlation functions of two giant gravitons and a single unprotected
single trace operator.
We will consider a single trace operator that belongs to the $SU(2)$ sector.
Our main goal is to test if the selection rules present for maximal giant gravitons  are respected by correlators
involving sub-maximal giants. 
We find that the selection rules are respected only for the maximal giants and operators close to the maximal giants.

Our starting point is the formula
\begin{equation}
\langle {\cal O}(S^I(x))\rangle_\chi=-{\rm Tr}_Q (\Phi^{I_1}(x)\Phi^{I_2}(x)\cdots\Phi^{I_J}(x))
\end{equation}
We consider $Q=2$ giants, located at $x_1$ and $x_2$. The first one is constructed using the matrix
\begin{equation}
{\cal Z}(x^\mu)=Z(x^\mu)+ Z^\dagger(x^\mu)+ Y(x^\mu)-Y^\dagger(x^\mu) \label{ZMat}
\end{equation}
and the second one is constructed using its hermitian conjugate.
We build our single trace operator ${\cal O}$ using only $Z$ and $Y$ fields.
The two point functions we are using have the standard normalization
\begin{equation} 
\langle Z^i_j (x)(Z^\dagger)^k_l (y)\rangle ={g_{YM}^2\over 4\pi^2}{1\over |x-y|^2}\delta^i_l\delta^k_j=
\langle Y^i_j (x)(Y^\dagger)^k_l (y)\rangle
\end{equation}
We will consider the case that our operator ${\cal O}$ is located at $x_0^\mu=(0,0,0,0)$ so that
\begin{equation} 
\langle Z^i_j (x_0){\cal Z}^k_l (y)\rangle =\langle Z^i_j (x_0)({\cal Z}^\dagger)^k_l (y)\rangle={g_{YM}^2\over 4\pi^2}{1\over |y|^2}\delta^i_l\delta^k_j
\end{equation}
\begin{equation} 
\langle Y^i_j (x_0){\cal Z}^k_l (y)\rangle =-\langle Y^i_j (x_0)({\cal Z}^\dagger)^k_l (y)\rangle=-{g_{YM}^2 \over 4\pi^2}{1\over |y|^2}\delta^i_l\delta^k_j
\end{equation}
For an operator ${\cal O}={\rm Tr}(Z^{n_1}Y^{n_2}Z^{n_3}\cdots)$ we have
\begin{equation}
\langle {\cal O}\rangle_\chi=-{\rm Tr}_Q (M_Z^{n_1}M_Y^{n_2}M_Z^{n_3}\cdots)
\end{equation}
where
\begin{equation}
M_Z=-\left[\begin{matrix} {t_1 \over x_1^2}  &0\cr 0 & {t_2 \over x_2^2} \end{matrix}\right]{\lambda\over 4\pi^2} M^{-1}
\end{equation}

\begin{equation}
M_Y=\left[\begin{matrix}{- t_1\over x_1^2} &0\cr 0 &{ t_2\over x_2^2}\end{matrix}\right]
{\lambda\over 4\pi^2} M^{-1}
\end{equation}


\begin{equation}
M^{-1}=
{1\over 1+32  {t_1 t_2 \over x_{12}^2} |\rho|^2}
\left[
\begin{matrix} 1 &&4 \sqrt{{2 t_1 t_2 \over x_{12}^2}}\rho\cr -4 \sqrt{{2 t_1 t_2 \over x_{12}^2}}\rho^* &&1\end{matrix}
\right]
\end{equation}
Based on experience with the planar limit we only expect integrability at large $N$, so we will test the selection rules at
large $N$.
Thus, we will use these expressions at the saddle point where
\begin{equation}
|\rho|^2={\lambda\over 16\pi^2}-{x_{12}^2\over 32t_1 t_2}
\end{equation}

Imagine that we are computing a correlation function of the form
\begin{eqnarray}
\left\langle \chi_{(1^{K})}({\cal Z}(x_1))\chi_{(1^{K})}({\cal Z}^{\dagger}(x_2))\,{\cal O}(0)\right\rangle
\end{eqnarray}
The single trace operator is a trace of a product of $L$ fields with $L$ even.
Since there are no self contractions allowed, this operator has a definite spatial dependence
\begin{eqnarray}
\left\langle \chi_{(1^{K})}({\cal Z}(x_1))\chi_{(1^{K})}({\cal Z}^{\dagger}(x_2))\,{\cal O}(0)\right\rangle\propto
{1\over x_1^L}{1\over x_2^L}{1\over x_{12}^{2K-L}}
\end{eqnarray}
This spatial dependence must be reproduced by the generating function
\begin{eqnarray}
\sum_{i_1,i_2}^N t_1^{i_1}t_2^{i_2}\left\langle \chi_{(1^{i_1})}({\cal Z}(x_1))\chi_{(1^{i_2})}({\cal Z}^{\dagger}(x_2))\,{\cal O}(0)\right\rangle
=\sum_{R=0}^N e^{-R}\left( {t_1 t_2 \lambda\over 4\pi^2}\right)^R\langle {\cal O}\rangle_\chi
\end{eqnarray}
Now, it is clear that $\langle {\cal O}\rangle_\chi$ is a function of ${t_1\over x_1^2}$, ${t_2\over x_2^2}$ and
${x_{12}^2 \over t_1 t_2}$.
To obtain the correct spatial dependence, we need to pick the monomial of the form
\begin{equation}
 \left({t_1\over x_1^2}\right)^{L\over 2}\left({t_2\over x_2^2}\right)^{L\over 2}
\left({x_{12}^2 \over t_1 t_2}\right)^{Q}
\end{equation}
from $\langle {\cal O}\rangle_\chi$, and the term with $R=K-{L\over 2}+Q$ from the sum.
We have two constraints: we must have $0\le R\le N$ and by explicitly expanding  $\langle {\cal O}\rangle_\chi$
we find that $Q\ge {L\over 2}$.
We can easily find the coefficient of the term with $Q={L\over 2}$
\begin{equation}
{\rm Tr}(M_Z^{n_1}M_YM_Z^{n_2-n_1}M_Y\cdots )=\left({\lambda\over 32 \pi^4}\right)^{L\over 2}
{x_{12}^L\over x_1^L x_2^L}{\rm Tr}(\sigma_2^{n_1}\sigma_1\sigma_2^{n_2-n_1}\sigma_1\cdots )
\label{niceidentity}
\end{equation}
The importance of this term follows because in certain circumstances, it is the only term that contributes.
Indeed, for the maximal giant which has $K=N$ we have $R=N-{L\over 2}+Q$. 
The inequalities given above become $N-{L\over 2}+Q\le N$ and $Q\ge {L\over 2}$ which in fact implies that
$Q={L\over 2}$.
When $K<N$ extra terms will contribute, weighted with powers of ${N-K\over N}$.
When $K$ is close to $N$ so that we are very close to the maximal giant, this factor goes to zero at large $N$ so that
only the term with $Q={L\over 2}$ needs to be included.
In these cases, (\ref{niceidentity}) together with the results of \cite{deLeeuw:2015hxa} immediately imply that the
selection rules of \cite{Jiang:2019xdz} are respected, suggesting that integrability is present.
However, as soon as ${N-K\over N}$ is of order $1$, the extra terms will contribute.
We have verified numerically that in this case the selection rules are not obeyed, strongly suggesting that sub-maximal
giant graviton correlators do not enjoy integrability.

In conclusion, our results suggest that the maximal and close to maximal giants lead to an integrable problem, but
sub-maximal giants don't.

\section{Correlation functions of two dual giant gravitons and a single trace operator}\label{Sec:DualGiantCorrelators}

We now consider the correlation functions in the dual giant graviton case.  
Following the discussion for giant gravitons, we consider $Q = 2$ dual giant gravitons located at $x_1$ and $x_2$ respectively and built using the matrix (\ref{ZMat}).  The two-point correlation functions are as above so that, for an operator ${\cal O}={\rm Tr}(Z^{n_1}Y^{n_2}Z^{n_3}\cdots)$ we again have
\begin{equation}
\langle {\cal O}\rangle_\chi={\rm Tr}_Q (M_Z^{n_1}M_Y^{n_2}M_Z^{n_3}\cdots)
\end{equation}
where
\begin{equation}
M_Z=-\left[\begin{matrix} {t_1 \over x_1^2}  &0\cr 0 & {t_2 \over x_2^2} \end{matrix}\right]{\lambda\over 4\pi^2} M^{-1}
\end{equation}

\begin{equation}
M_Y=\left[\begin{matrix}{- t_1\over x_1^2} &0\cr 0 &{ t_2\over x_2^2}\end{matrix}\right]
{\lambda\over 4\pi^2} M^{-1}
\end{equation}
with the only change being the form of the matrix $M$. 
This difference is a result of the fact that $\rho$ is now a hermitian matrix.  We have
\begin{equation}
M=\left[
\begin{matrix} 1 && 4\sqrt{{2 t_1 t_2 \over x_{12}^2}}\rho\cr 4\sqrt{{2 t_1 t_2 \over x_{12}^2}}\rho^* &&1\end{matrix}
\right]
\end{equation}

\begin{equation}
M^{-1}=
{1\over 1-32{t_1t_2 \over x_{12}^2}|\rho|^2}
\left[
\begin{matrix} 1 &&-4\sqrt{{2 t_1 t_2 \over x_{12}^2}}\rho\cr -4\sqrt{{2 t_1 t_2 \over x_{12}^2}}\rho^* &1\end{matrix}
\right].  
\end{equation}

Using these results we again find that once again there are a number of terms from the expansion of $\langle{\cal O}\rangle_\chi$
that contribute. The term that respects the selection rules never dominates over the terms that spoil it, so that for dual
giants integrability does not appear to be present. 
This difference between giant gravitons and dual giant gravitons can be traced back to the fact that there is a maximum
size for the giant gravitons, whereas the dual giant gravitons can be arbitrarily large.

It is interesting to note that the dynamics of an open string attached to a maximal giant graviton is 
integrable \cite{Hofman:2007xp}.
For open strings attached to giant gravitons that are almost maximal, the dynamics is integrable at the leading order 
in the large $N$ expansion. 
Open strings attached to either less than maximal giant gravitons or to dual giant gravitons do not lead to
integrable dynamics \cite{Koch:2015pga}.

\section{Bound states of Giants and of dual giants}\label{Sec:BoundStates}

Giant gravitons correspond to Schur polynomials labeled with Young diagrams that have a single column.
Dual giant gravitons correspond to Schur polynomials labeled with Young diagrams that have a single row.
In general, a Schur polynomial will be labeled by a Young diagram that has multiple rows and columns.
This operator is dual to a bound state of giant gravitons and dual giant gravitons.
There are formulas\cite{Lederman} that express any general Schur polynomial as a linear combination of products of Schur
polynomials with a single column or as a linear combination of products of Schur polynomials with a single row.
Using these formulas, in this section, we develop formulas that are applicable to these bound states.

The relevant formulas are nicely illustrated with a few examples.
Consider the two giant graviton bound state given by
\begin{equation}
\chi_{\tiny\yng(2,2,1,1,1)}
\end{equation}
Both columns should be of order $N$ - this example is just to illustrate the idea.
We can separate the above giant graviton bound state into ``two pieces'' by separating the two columns, namely
\begin{equation}
\chi_{\tiny\yng(1,1,1,1,1)}\qquad\qquad \chi_{\tiny\yng(1,1)}
\end{equation}
Use these operators as the diagonals of a matrix, ordered so that the length of the column does not increase down the diagonal.
To populate the complete matrix, start from the diagonal element and add a box when you move right and remove a box when
you move left.
The original Schur polynomial is given by the determinant of this matrix.
For our example
\begin{equation}
\chi_{\tiny\yng(2,2,1,1,1)}=\det\left[
\begin{matrix}
\chi_{\tiny\yng(1,1,1,1,1)} & \chi_{\tiny\yng(1,1,1,1,1,1)}\cr
\chi_{\tiny\yng(1)} &\chi_{\tiny\yng(1,1)}
\end{matrix}
\right]
=\chi_{\tiny\yng(1,1,1,1,1)}\chi_{\tiny\yng(1,1)}  - \chi_{\tiny\yng(1,1,1,1,1,1)}\chi_{\tiny\yng(1)} 
\label{eq:BS_X52}
\end{equation}
This is easy to verify using the products
\begin{equation}
\chi_{\tiny\yng(1,1,1,1,1)}\chi_{\tiny\yng(1,1)}=
\chi_{\tiny\yng(2,2,1,1,1)}+\chi_{\tiny\yng(2,1,1,1,1,1)}+\chi_{\tiny\yng(1,1,1,1,1,1,1)}
\end{equation}
\begin{equation}
\chi_{\tiny\yng(1,1,1,1,1,1)}\chi_{\tiny\yng(1)}=
\chi_{\tiny\yng(2,1,1,1,1,1)}+\chi_{\tiny\yng(1,1,1,1,1,1,1)}
\end{equation}
Here is a slightly more complicated example
\begin{equation}
\chi_{\tiny\yng(3,2,1,1,1)}=\det\left[
\begin{matrix}
\chi_{\tiny\yng(1,1,1,1,1)} & \chi_{\tiny\yng(1,1,1,1,1,1)} & \chi_{\tiny\yng(1,1,1,1,1,1,1)}\cr
\chi_{\tiny\yng(1)} &\chi_{\tiny\yng(1,1)} &\chi_{\tiny\yng(1,1,1)}\cr
0 &1 &\chi_{\tiny\yng(1)}
\end{matrix}
\right]
\end{equation}
There is an equally easy rule for the Schur polynomials corresponding to dual giant gravitons. Here is an example
\begin{equation}
\chi_{\tiny\yng(4,2)}=\det\left[
\begin{matrix}
\chi_{\tiny\yng(4)} &\chi_{\tiny\yng(5)}\cr
\chi_{\tiny\yng(1)} &\chi_{\tiny\yng(2)}
\end{matrix}
\right]
\end{equation}

Using these rules we are able to generalize our results to bound states of giant gravitons and dual giant gravitons. 
Using equation (\ref{GenRep}),  we can express a Schur polynomial dual to a giant graviton as
\begin{equation}
\chi_{(1^{p})}=\frac{1}{p!}\left(-\frac{d}{dt}\right)^{p}\int d\bar{\chi}^{N}\cdots d\bar{\chi}^{1}\int d\chi^{1}\cdots d\chi^{N}\,e^{-t\bar{\chi}^{i}Z_{i}^{j}\chi_{j}-\bar{\chi}^{i}\chi_{i}}\Bigg|_{t=0}\label{eq:BS_chi1p}
\end{equation}
 For convenience, we define the derivative operator
\begin{equation}
\partial_{t}^{(p)}\equiv\frac{1}{p!}\left(-\frac{d}{dt}\right)^{p}\Bigg|_{t=0}
\end{equation}
so that (\ref{eq:BS_chi1p}) can be rewritten as
\begin{equation}
\chi_{\left(1^{p}\right)}\left(Z\right)=\partial_{t}^{(p)}\int d\bar{\chi}^{N}\cdots d\bar{\chi}^{1}\int d\chi^{1}\cdots d\chi^{N}\,e^{-t\bar{\chi}^{i}Z_{i}^{j}\chi_{j}-\bar{\chi}^{i}\chi_{i}}
\end{equation}

For a bound state of giant gravitons, given by a Schur polynomial labeled by Young diagram $R$ with column
lengths $c_{1}\geq c_{2}\geq\cdots\geq c_{W}$, we define a $W\times W$ matrix $\mathcal{D}^{(R)}$
\begin{equation}
\left(\mathcal{D}^{(R)}\right)_{KL}=\partial_{t_{K}}^{(c_{K}+L-K)}\qquad\qquad1\leq K,L\leq W
\end{equation}
The Schur polynomial is given by
\begin{equation}
\chi_{R}\left(Z\right)=\det\left(\mathcal{D}^{(R)}\right)\int\prod_{K=1}^{W}\left[d\bar{\chi}_{K}d\chi_{K}\right]\,e^{-t_{K}\bar{\chi}_{K}^{i}Z_{i}^{j}\chi_{Kj}-\bar{\chi}_{K}^{i}\chi_{Ki}}
\end{equation}

In analogy to the bound state of giant gravitons, from (\ref{DGntsProject}),  we can express a dual giant graviton as 
\begin{equation}
\chi_{(k)}=\bar{\partial}_{t}^{(k)}\int d^{N}\bar{\varphi}\int d^{N}\varphi\,e^{-\bar{\varphi}^{i}\varphi_{i}+t\bar{\varphi}^{i}Z_{i}^{j}\varphi_{j}}
\end{equation}
where the derivative operator 
\begin{equation}
\bar{\partial}_{t}^{(k)}=\frac{1}{k!}\left(\frac{d}{dt}\right)^{k}\Big|_{t=0}
\end{equation}
takes the $k$-th order derivative with respect to $t$ and evaluates the result at $t=0$. 

For a bound state of dual giants, labeled by Young diagram $R$ with row lengths $r_{1}\geq r_{2}\geq\cdots\geq r_{H}$, 
define a $H\times H$ matrix $\bar{\mathcal{D}}^{(R)}$ as
\begin{equation}
\left(\bar{\mathcal{D}}^{(R)}\right)_{KL}=\bar{\partial}_{t_{K}}^{(r_{K}+L-K)}\qquad\qquad1\leq K,L\leq H
\end{equation}
The bound state of dual giants is given by
\begin{equation}
\chi_{R}\left(Z\right)=\det\left(\bar{\mathcal{D}}^{(R)}\right)\int\prod_{K=1}^{H}\left[d\bar{\varphi}_{K}d\varphi_{K}\right]\,e^{-\bar{\varphi}_{K}^{i}\varphi_{Ki}+t_{K}\bar{\varphi}_{K}^{i}Z_{i}^{j}\varphi_{Kj}}
\end{equation}

Before leaving this section, we describe another approach to the bound state description which maybe a simpler description
in some settings.
To motivate the discussion, consider the bound state of giants as (\ref{eq:BS_X52}). 
As we have explained above, using the generating function of previous sections, we need to extract the coefficient of 
$t_{1}^{5}t_{2}^{2}$ for the first term and $t_{1}^{6}t_{2}$ for the second term. 
This is cumbersome in practice and can be avoided as we now explain. 
To combine the two terms, multiply the generating function with a factor
\begin{equation}
1-\frac{t_{2}}{t_{1}}=\det\begin{bmatrix}1 & t_{1}^{-1}\\
t_{2} & 1
\end{bmatrix}.
\end{equation}
With above factor, the coefficient of $t_{1}^{5}t_{2}^{2}$ term of the generating function represents the 
bound state ((\ref{eq:BS_X52})). 
The generalization to bound states of $W$ giants is obvious. 
The additional factor is now the determinant of a $W\times W$ matrix
\begin{equation}
V_{W}=\begin{pmatrix}1 & t_{1}^{-1} & t_{1}^{-2} & \cdots & t_{1}^{-W+1}\\
t_{2} & 1 & t_{2}^{-1} & \cdots & t_{2}^{-W+2}\\
t_{3}^{2} & t_{3} & 1 & \cdots & \vdots\\
\vdots & \vdots & \vdots & \ddots & \vdots\\
t_{W}^{W-1} & t_{W}^{W-2} & t_{W}^{W-3} & \cdots & 1
\end{pmatrix},\label{eq:BSN_VW}
\end{equation}
which can also be expressed as the product of a diagonal matrix with the Vandermonde matrix
\begin{equation}
V_{W}=\begin{pmatrix}1\\
 & t_{2}\\
 &  & t_{3}^{2}\\
 &  &  & \ddots\\
 &  &  &  & t_{W}^{W-1}
\end{pmatrix}\begin{pmatrix}1 & t_{1}^{-1} & t_{1}^{-2} & \cdots & t_{1}^{-W+1}\\
1 & t_{2}^{-1} & t_{2}^{-2} & \cdots & t_{2}^{-W+1}\\
1 & t_{3}^{-1} & t_{3}^{-2} & \cdots & \vdots\\
\vdots & \vdots & \vdots & \ddots & \vdots\\
1 & t_{W}^{-1} & t_{W}^{-2} & \cdots & t_{W}^{-W+1}
\end{pmatrix}.
\end{equation}
The determinant is therefore 
\begin{equation}
\det\left(V_{W}\right)=\left(\prod_{i=1}^{W}t_{i}^{i-1}\right)\prod_{1\leq i<j\leq W}\left(t_{j}^{-1}-t_{i}^{-1}\right)=\prod_{1\leq i<j\leq W}\left(1-\frac{t_{j}}{t_{i}}\right).
\end{equation}
The generating function of bound states of $W$ giants is given by
\begin{equation}
\sum_{c_{1},c_{2},\cdots c_{W}\leq N}t_{1}^{c_{1}}t_{2}^{c_{2}}\cdots t_{W}^{c_{W}}\chi_{\left(1^{c_{1}}1^{c_{2}}\cdots1^{c_{W}}\right)}\left(Z\right)=\det\left(V_{W}\right)\int\prod_{K=1}^{W}\left[d\bar{\chi}_{K}d\chi_{K}\right]\,e^{-t_{K}\bar{\chi}_{K}^{i}Z_{i}^{j}\chi_{Kj}-\bar{\chi}_{K}^{i}\chi_{Ki}}.\label{eq:BSN_GeneratingFun}
\end{equation}

A remark on the use  of (\ref{eq:BSN_GeneratingFun}) is in order. 
A well-define bound state is associated with a Young diagram, whose column lengths satisfy 
$N\geq c_{1}\geq c_{2}\geq\cdots\geq c_{W}$.
On the r.h.s of (\ref{eq:BSN_GeneratingFun}), we do not have a convenient way to impose this constraint. 
Therefore (\ref{eq:BSN_GeneratingFun}) contains many terms do not correspond to well-defined bound states,
that is, terms that do not satisfy $c_{1}\geq c_{2}\geq\cdots\geq c_{W}$.
Thus, when performing concrete computation with this generating function, some care needs to be taken when
extracting terms that correspond to well-defined bound states.

Similarly, for bound states of $H$ dual giants, we have 
\begin{equation}
\sum_{r_{1},r_{2},\dots r_{H}\geq1}t_{1}^{r_{1}}t_{2}^{r_{2}}\cdots t_{H}^{r_{H}}\chi_{\left(r_{1},r_{2},\dots,r_{H}\right)}\left(Z\right)=\det\left(V_{H}\right)\int\prod_{K=1}^{H}\left[d\bar{\varphi}_{K}d\varphi_{K}\right]\,e^{-\bar{\varphi}_{K}^{i}\varphi_{Ki}+t_{K}\bar{\varphi}_{K}^{i}Z_{i}^{j}\varphi_{Kj}},
\end{equation}
 where the matrix $V_{H}$ has the same form as (\ref{eq:BSN_VW})
\begin{equation}
V_{H}=\begin{pmatrix}1 & t_{1}^{-1} & t_{1}^{-2} & \cdots & t_{1}^{-H+1}\\
t_{2} & 1 & t_{2}^{-1} & \cdots & t_{2}^{-H+2}\\
t_{3}^{2} & t_{3} & 1 & \cdots & \vdots\\
\vdots & \vdots & \vdots & \ddots & \vdots\\
t_{H}^{H-1} & t_{H}^{H-2} & t_{H}^{H-3} & \cdots & 1
\end{pmatrix}.
\end{equation}

To distinguish these two approaches in what follows, we refer to the first as the derivative approach and to the second
as the polynomial approach. 

\subsection{Large $N$ Effective Theory: Giant Bound States}\label{sec:GBS}

In this section we will follow the derivative approach.
Consider a three-point function $\left\langle \chi_{R_{1}}\left(\mathcal{Z}_{1}\right)\chi_{R_{2}}\left(\mathcal{Z}_{2}\right)\mathcal{O}\right\rangle $, where $\mathcal{O}$ is a normal ordered single trace operator. 
The Young diagram $R_{1}$ has $W_1$ columns with lengths $c_{1}\geq c_{2}\geq\cdots\geq c_{W_{1}}$ and the Young
diagram $R_{2}$ has $W_2$ columns with lengths $c_{W_{1}+1}\geq c_{W_{1}+2}\geq\cdots\geq c_{W}$, where
$W=W_1+W_2$. 
Without loss of generality, assume $W_{1}\geq W_{2}$.
Use $t_{1},t_{2},\cdots,t_{W_{1}}$ for the variables of the generating function used in the construction of
$\chi_{R_1}\left(\mathcal{Z}_1\right)$, where $t_{K}$ is the variable for the $K$-th giant of the bound state.
Similarly, $t_{W_1+1},\dots,t_{W}$ are variables of the generating function for $\chi_{R_{2}}\left(\mathcal{Z}_{2}\right)$ 
and $t_{W_{1}+K}$ is for the $K$-th giant of the bound state. 
Denote $\mathcal{D}^{(R_{1})}$ the derivative operator acting on variables of $\chi_{R_{1}}\left(\mathcal{Z}_{1}\right)$
and $\mathcal{D}^{(R_{2})}$ acting on those of $\chi_{R_{2}}\left(\mathcal{Z}_{2}\right)$.

With these conventions, we write
\begin{multline}
\left\langle \chi_{R_{1}}\left(\mathcal{Z}_{1}\right)\chi_{R_{2}}\left(\mathcal{Z}_{2}\right)\mathcal{O}\right\rangle =\det\left(\mathcal{D}^{(R_{1})}\right)\det\left(\mathcal{D}^{(R_{2})}\right)\int\left[d\phi^{I}\right]\prod_{K=1}^{W}\left[d\bar{\chi}_{K}d\chi_{K}\right]\\
\,e^{-\frac{1}{g_{YM}^{2}}\int d^{4}x\,\left[\frac{1}{2}{\rm Tr}\left(\partial_{\mu}\phi^{I}\partial^{\mu}\phi^{I}\right)+g_{YM}^{2}\sum_{K=1}^{W}\delta(x-x_{\left\lceil K/W_{1}\right\rceil })\bar{\chi}_{K}^{i}\left(\delta_{i}^{j}+t_{K}\left(\mathcal{Z}_{\left\lceil K/W_{1}\right\rceil }\right)_{i}^{j}\right)\chi_{Ki}\right]}\mathcal{O}\left(\phi^{I}\right)
\end{multline}
where $\left\lceil \cdot\right\rceil$  is the ceiling function so that
\begin{equation}
\mathcal{Z}_{\left\lceil K/W_{1}\right\rceil }=\begin{cases}
\mathcal{Z}_{1} & 1\leq K\leq W_{1}\\
\mathcal{Z}_{2} & W_{1}+1\leq K\leq W
\end{cases},
\end{equation}
and the measure of $\phi^{I}$ is normalized as in (\ref{NormPhi}).
Following the analysis developed above, complete the square to obtain
\begin{multline}
-\int d^{4}x\,\left[\frac{1}{2}{\rm Tr}\left(\partial_{\mu}\phi^{I}\partial^{\mu}\phi^{I}\right)+g_{YM}^{2}\sum_{K=1}^{W}\delta(x-x_{\left\lceil K/W_{1}\right\rceil})\bar{\chi}_{K}^{i}\left(\delta_{i}^{j}+t_{K}\left(\mathcal{Z}_{\left\lceil K/W_{1}\right\rceil }\right)_{i}^{j}\right)\chi_{Kj}\right]\\
=\int d^{4}x\,\left[\frac{1}{2}{\rm Tr}\left[\partial_{\mu}\left(\phi^{I}-S^{I}\right)\partial^{\mu}\left(\phi^{I}-S^{I}\right)-S^{I}\partial^{\mu}\partial_{\mu}S^{I}\right]-g_{YM}^{2}\sum_{K=1}^{W}\delta(x-x_{\left\lceil K/W_{1}\right\rceil})\bar{\chi}_{K}^{i}\chi_{Ki}\right]
\end{multline}
where 
\begin{equation}
\left(S^{I}\left(x\right)\right)_{j}^{i}=-\frac{g_{YM}^{2}}{4\pi^{2}}\left[\sum_{K=1}^{W_{1}}\frac{t_{K}\left(\mathcal{Y}_{1}\right)_{K}^{I}\bar{\chi}_{K}^{i}\chi_{Kj}}{\left|x-x_{K}\right|}+\sum_{K=W_{1}+1}^{W}\frac{t_{K}\left(\mathcal{Y}_{2}\right)_{K}^{I}\bar{\chi}_{K}^{i}\chi_{Kj}}{\left|x-x_{K}\right|}\right]
\end{equation}
Integrating over the $\phi^{I}$ fields and performing the HS transformation, we arrive at
\begin{multline}
\left\langle \chi_{R_{1}}\left(\mathcal{Z}_{1}\right)\chi_{R_{2}}\left(\mathcal{Z}_{2}\right)\mathcal{O}\right\rangle =\det\left(\mathcal{D}^{(R_{1})}\right)\det\left(\mathcal{D}^{(R_{2})}\right)\\
\int\left[d\rho\right]\int\prod_{K=1}^{2W}\left[d\bar{\chi}_{K}d\chi_{K}\right]\,e^{\frac{8\pi^{2}}{g_{YM}^{2}}{\rm Tr}\rho^{2}+2\sum_{K,J}^{\prime}\sqrt{\frac{t_{K}t_{J}\vec{\mathcal{Y}}_{1}\cdot\vec{\mathcal{Y}}_{2}}{x_{12}^{2}}}\rho_{KJ}\bar{\chi}_{K}^{a}\chi_{Ja}-\sum_{K=1}^{W}\bar{\chi}_{K}^{a}\chi_{Ka}}\mathcal{O}\left(S^{I}\right)
\end{multline}
where $\Sigma_{K,J}^{\prime}$ sums over the indices obeying $\left\lceil K/W_1\right\rceil \neq\left\lceil J/W_1\right\rceil$.
The anti-Hermitian matrix variable $\rho$ takes the form
\begin{equation}
\rho=\begin{pmatrix}0 & A\\
-A^{\dagger} & 0
\end{pmatrix},\quad A=\begin{pmatrix}A_{11} & A_{12} & \cdots & A_{1W_{2}}\\
A_{21} & A_{22} & \cdots & A_{2W_{2}}\\
\vdots & \vdots & \vdots & \vdots\\
A_{W_{1}1} & A_{W_{1}2} & \cdots & A_{W_{1}W_{2}}
\end{pmatrix}\label{eq:BS_rho}
\end{equation}
where $A$ is a $W_1\times W_2$ complex matrix.
Integrate over the fermion fields to find
\begin{multline}
\left\langle \chi_{R_{1}}\left(\mathcal{Z}_{1}\right)\chi_{R_{2}}
\left(\mathcal{Z}_{2}\right)\mathcal{O}\right\rangle \\
=\det\left(\mathcal{D}^{(R_{1})}\right)\det\left(\mathcal{D}^{(R_{2})}\right)
\int\left[d\rho\right]\int\,e^{\frac{8\pi^{2}}{g_{YM}^{2}}{\rm Tr}\rho^{2}
+N{\rm Tr}\ln\left[\delta_{JK}-2\sqrt{\frac{t_{K}t_{J}\vec{\mathcal{Y}}_1
\cdot\vec{\mathcal{Y}}_{2}}{x_{12}^{2}}}\rho_{KJ}\right]}
\left\langle \mathcal{O}\left(S^{I}\right)\right\rangle _{\chi}\label{eq:BS_2ptRho}
\end{multline}
where $\left\langle \mathcal{O}^{I}\left(S^{I}\right)\right\rangle _{\chi}$
is defined by Wick contracting all pairs of $\chi$, $\bar{\chi}$
according to Wick's theorem with the basic contraction given by 
\begin{equation}
\left\langle \bar{\chi}_{J}^{a}\chi_{Kb}\right\rangle =\delta_{b}^{a}M_{JK}^{-1}\qquad\qquad
M_{JK}^{-1}=\delta_{JK}-2\sqrt{\frac{t_{K}t_{J}\vec{\mathcal{Y}}_{1}\cdot
\vec{\mathcal{Y}}_{2}}{x_{12}^{2}}}\rho_{KJ}
\end{equation}

We can also express integral in (\ref{eq:BS_2ptRho}) in terms of the matrix $A$ defined in (\ref{eq:BS_rho}). 
Towards this end define the diagonal matrices 
\begin{equation}
V_{1}=\mathrm{diag}\left(\sqrt{t_{1}},\dots,\sqrt{t_{W_{1}}}\right)\quad V_{2}=\mathrm{diag}\left(\sqrt{t_{W_{1}+1}},\dots,\sqrt{t_{W}}\right)
\end{equation}
so that the matrix $M$ can be written as 
\begin{equation}
M=\begin{pmatrix}I_{W_{1}} & 2\sqrt{\frac{\vec{\mathcal{Y}}_1
\cdot\vec{\mathcal{Y}}_{2}}{x_{12}^{2}}}V_{1}AV_{2}\\
-2\sqrt{\frac{\vec{\mathcal{Y}}_{1}\cdot\vec{\mathcal{Y}}_{2}}{x_{12}^{2}}}V_{2}A^{\dagger}V_{1} & I_{W_{2}}
\end{pmatrix}
\end{equation}
where $I_{W_{1}}$ is a $W_{1}\times W_{1}$ identity matrix. 
Its determinant is given by
\begin{align}
\det M & =\det\left(I_{W_{1}}+4\frac{\vec{\mathcal{Y}}_{1}\cdot\vec{\mathcal{Y}}_{2}}{x_{12}^{2}}V_{1}AV_{2}V_{2}A^{\dagger}V_{1}\right)\equiv\det M_{1}
\end{align}
where we have used the identity 
\begin{equation}
\det\begin{pmatrix}A & B\\
C & D
\end{pmatrix}=\det\left(D\right)\det\left(A-BD^{-1}C\right)
\end{equation}
The equation (\ref{eq:BS_2ptRho}) can be written as
\begin{equation}
\left\langle \chi_{R_{1}}\left(\mathcal{Z}_{1}\right)\chi_{R_{2}}\left(\mathcal{Z}_{2}\right)
\mathcal{O}\right\rangle =\det\left(\mathcal{D}^{(R_{1})}\right)
\det\left(\mathcal{D}^{(R_{2})}\right)\int\left[dA\right]\int
\,e^{-\frac{16\pi^{2}}{g_{YM}^{2}}{\rm Tr} A^{\dagger}A
+N{\rm Tr}\ln M_{1}}\left\langle \mathcal{O}\left(S^{I}\right)\right\rangle _{\chi}
\label{eq:BS_chichiO}
\end{equation}
with the measure normalized as follows
\begin{equation}
\int\left[dA\right]\int\,e^{-\frac{16\pi^{2}}{g_{YM}^{2}}{\rm Tr} A^{\dagger}A}=1
\end{equation}
Using the identity 
\begin{equation}
\begin{pmatrix}A & B\\
C & D
\end{pmatrix}^{-1}=\begin{pmatrix}\left(A-BD^{-1}C\right)^{-1} & -\left(A-BD^{-1}C\right)^{-1}BD^{-1}\\
-D^{-1}C\left(A-BD^{-1}C\right)^{-1} & D^{-1}+D^{-1}C\left(A-BD^{-1}C\right)^{-1}BD^{-1}
\end{pmatrix}
\end{equation}
we can also express $M^{-1}$ in terms of $M_{1}$
\begin{equation}
M^{-1}=\begin{pmatrix}M_{1}^{-1} & -2\sqrt{\frac{\vec{\mathcal{Y}}_{1}\cdot\vec{\mathcal{Y}}_{2}}{x_{12}^{2}}}M_{1}^{-1}V_{1}AV_{2}\\
2\sqrt{\frac{\vec{\mathcal{Y}}_{1}\cdot\vec{\mathcal{Y}}_{2}}{x_{12}^{2}}}V_{2}A^{\dagger}V_{1}M_{1}^{-1} & I_{W_{2}}-4\frac{\vec{\mathcal{Y}}_{1}\cdot\vec{\mathcal{Y}}_{2}}{x_{12}^{2}}V_{2}A^{\dagger}V_{1}M_{1}^{-1}V_{1}AV_{2}
\end{pmatrix}.
\end{equation}

We have verified that the formulas of this section can be used to reproduce the result for the two-point function of
Schur polynomials with two columns. The exact result is known
\begin{equation}
\left\langle \chi_{(1^{J_{1}}1^{J_{2}})}\left(Z\right)\chi_{(1^{J_{1}}1^{J_{2}})}\left(Z^{\dagger}\right)\right\rangle =\frac{N!}{\left(N-J_{1}\right)!}\frac{\left(N+1\right)!}{\left(N+1-J_{2}\right)!}\delta_{J_{1},J_{3}}\delta_{J_{2},J_{4}}.
\end{equation}
Our generating function agrees perfectly with this result.
Since there are many details and these results are not crucial for the arguments that follow, we have collected them 
in Appendix \ref{TwoGiantBound}.

\subsection{Large $N$ Effective Theory: Dual Giant Bound States}

In this section we will again follow the derivative approach.
For bound states of dual giants, again consider a three-point function $\left\langle \chi_{R_{1}}\left(\mathcal{Z}_{1}\right)\chi_{R_{2}}\left(\mathcal{Z}_{2}\right)\mathcal{O}\right\rangle $,
where $\mathcal{O}$ is a normal ordered single trace operator.
The Young diagram $R_{1}$ has $H_{1}$ columns with row lengths $r_{1}\geq r_{2}\geq\cdots\geq c_{H_{1}}$
and the Young diagram $R_{2}$ has $H_{2}$ rows with row lengths
$r_{H_{1}+1}\geq c_{H_{1}+2}\geq\cdots\geq c_{H}$, where $H=H_{1}+H_{2}$.
Without loss of generality assume $H_{1}\geq H_{2}$. Use $t_{1},t_{2},\cdots,t_{H_{1}}$
for the variables of the generating function for $\chi_{R_{1}}\left(\mathcal{Z}_{1}\right)$,
where $t_{K}$ is the variable for the $K$-th dual giant of the bound
state. Similarly, $t_{H_{1}+1},\dots,t_{H}$ are variables of the
generating function for $\chi_{R_{2}}\left(\mathcal{Z}_{2}\right)$
and $t_{H_{1}+K}$ is for the $K$-th dual giant of the bound state.
Denote the derivative operator acting on the variables of 
$\chi_{R_{1}}\left(\mathcal{Z}_{1}\right)$ by $\bar{\mathcal{D}}^{(R_{1})}$.
Similarly $\bar{\mathcal{D}}^{(R_{2})}$
acts on the variables of $\chi_{R_{2}}\left(\mathcal{Z}_{2}\right)$.

We can write
\begin{multline}
\left\langle \chi_{R_{1}}\left(\mathcal{Z}_{1}\right)\chi_{R_{2}}\left(\mathcal{Z}_{2}\right)\mathcal{O}\right\rangle =\det\left(\bar{\mathcal{D}}^{(R_{1})}\right)\det\left(\bar{\mathcal{D}}^{(R_{2})}\right)\int\left[d\phi^{I}\right]\prod_{K=1}^{H}\left[d\bar{\varphi}_{K}d\varphi_{K}\right] \\
\,e^{-\frac{1}{g_{YM}^{2}}\int d^{4}x\,\left[\frac{1}{2}{\rm Tr}\left(\partial_{\mu}\phi^{I}\partial^{\mu}\phi^{I}\right)+g_{YM}^{2}\sum_{K=1}^{W}\delta(x-x_{\left\lceil K/H_{1}\right\rceil })\bar{\varphi}_{K}^{i}\left(\delta_{i}^{j}-t_{K}\left(\mathcal{Z}_{\left\lceil K/H_{1}\right\rceil }\right)_{i}^{j}\right)\varphi_{Ki}\right]}\mathcal{O}\left(\phi^{I}\right)
\end{multline}
where the measure $[d \phi^{I}]$ is normalized as usual.
After completing the square we find
\begin{multline}
-\int d^{4}x\,\left[\frac{1}{2}{\rm Tr}\left(\partial_{\mu}\phi^{I}\partial^{\mu}\phi^{I}\right)+g_{YM}^{2}\sum_{K=1}^{W}\delta(x-x_{\left\lceil K/H_{1}\right\rceil })\bar{\varphi}_{K}^{i}\left(\delta_{i}^{j}-t_{K}\left(\mathcal{Z}_{\left\lceil K/H_{1}\right\rceil }\right)_{i}^{j}\right)\varphi_{Kj}\right]\\
=\int d^{4}x\,\left[\frac{1}{2}{\rm Tr}\left[\partial_{\mu}\left(\phi^{I}-S^{I}\right)\partial^{\mu}\left(\phi^{I}-S^{I}\right)-S^{I}\partial^{\mu}\partial_{\mu}S^{I}\right]-g_{YM}^{2}\sum_{K=1}^{H}\delta(x-x_{\left\lceil K/H_{1}\right\rceil })\bar{\varphi}_{K}^{i}\varphi_{Ki}\right]
\end{multline}
where 
\begin{equation}
\left(S^{I}\left(x\right)\right)_{j}^{i}=\frac{g_{YM}^{2}}{4\pi^{2}}\left[\sum_{K=1}^{H_{1}}\frac{t_{K}\left(\mathcal{Y}_{1}\right)_{K}^{I}\bar{\varphi}_{K}^{i}\varphi_{Kj}}{\left|x-x_{1}\right|^{2}}+\sum_{K=H_{1}+1}^{H}\frac{t_{K}\left(\mathcal{Y}_{2}\right)_{K}^{I}\bar{\varphi}_{K}^{i}\varphi_{Kj}}{\left|x-x_{2}\right|^{2}}\right]
\end{equation}
Integrating over the $\phi^{I}$ fields and performing a H-S transformation, we find
\begin{multline}
\left\langle \chi_{R_{1}}\left(\mathcal{Z}_{1}\right)\chi_{R_{2}}\left(\mathcal{Z}_{2}\right)\mathcal{O}\right\rangle =\det\left(\bar{\mathcal{D}}^{(R_{1})}\right)\det\left(\bar{\mathcal{D}}^{(R_{2})}\right) \\
\int\left[d\rho\right]\int\prod_{K=1}^{H}\left[d\bar{\varphi}_{K}d\varphi_{K}\right]\,
e^{-\frac{8\pi^{2}}{g_{YM}^{2}}{\rm Tr}\rho^{2}-2\sum_{K,J}^{\prime}
\sqrt{\frac{t_{K}t_{J}\vec{\mathcal{Y}}_{1}\cdot\vec{\mathcal{Y}}_{2}}{x_{12}^{2}}}\rho_{KJ}
\bar{\varphi}_{K}^{a}\varphi_{Ja}-\sum_{K=1}^{W}\bar{\varphi}_{K}^{a}\varphi_{Ka}}\mathcal{O}
\left(S^{I}\right)
\end{multline}
where $\Sigma_{K,J}^{\prime}$ sums over the indices obeying $\left\lceil K/H_1\right\rceil \neq\left\lceil J/H_1\right\rceil$.
The Hermitian matrix variable $\rho$ takes the form 
\begin{equation}
\rho=\begin{pmatrix}0 & A\\
A^{\dagger} & 0
\end{pmatrix},\quad A=\begin{pmatrix}A_{11} & A_{12} & \cdots & A_{1H_{2}}\\
A_{21} & A_{22} & \cdots & A_{2H_{2}}\\
\vdots & \vdots & \vdots & \vdots\\
A_{H_{1}1} & A_{H_{1}2} & \cdots & A_{H_{1}H_{2}}
\end{pmatrix}\label{eq:BS_rho_dual}
\end{equation}
where $A$ is a $H_{1}\times H_{2}$ complex matrix.
Finally, integrate over the $\varphi$ fields to arrive at
\begin{multline}
\left\langle \chi_{R_{1}}\left(\mathcal{Z}_{1}\right)\chi_{R_{2}}\left(\mathcal{Z}_{2}\right)\mathcal{O}\right\rangle \\
=\det\left(\bar{\mathcal{D}}^{(R_{1})}\right)
\det\left(\bar{\mathcal{D}}^{(R_{2})}\right)\int\left[d\rho\right]\int\,e^{-\frac{8\pi^{2}}{g_{YM}^{2}}{\rm Tr}\rho^{2}
-N{\rm Tr}\ln\left[\delta_{JK}+2\sqrt{\frac{t_{K}t_{J}\vec{\mathcal{Y}}_{1}\cdot\vec{\mathcal{Y}}_{2}}{x_{12}^{2}}}\rho_{KJ}\right]}
\left\langle \mathcal{O}\left(S^{I}\right)\right\rangle _{\varphi}
\label{eq:BS_2ptRho_dual}
\end{multline}
where $\left\langle \mathcal{O}^{I}\left(S^{I}\right)\right\rangle _{\chi}$
is defined by Wick contracting all pairs of $\varphi$, $\bar{\varphi}$
according to Wick's theorem, with the basic contraction given by
\begin{equation}
\left\langle \bar{\varphi}_{J}^{a}\varphi_{Kb}\right\rangle =\delta_{b}^{a}M_{JK}^{-1}
\end{equation}
where
\begin{equation}
M_{JK}^{-1}=\delta_{JK}+2\sqrt{\frac{t_{K}t_{J}\vec{\mathcal{Y}}_{1}\cdot\vec{\mathcal{Y}}_{2}}{x_{12}^{2}}}\rho_{KJ}
\end{equation}

It is again useful to express integral in (\ref{eq:BS_2ptRho_dual}) in terms of the matrix $A$ defined in (\ref{eq:BS_rho_dual}).
Towards this end, define the diagonal matrices
\begin{equation}
V_{1}=\mathrm{diag}\left(\sqrt{t_{1}},\dots,\sqrt{t_{H_{1}}}\right),\quad V_{2}=\mathrm{diag}\left(\sqrt{t_{H_{1}+1}},\dots,\sqrt{t_{H}}\right)
\end{equation}
The matrix $M$ is now
\begin{equation}
M=\begin{pmatrix}I_{H_{1}} & 2\sqrt{\frac{\vec{\mathcal{Y}}_{1}
\cdot\vec{\mathcal{Y}}_{2}}{x_{12}^{2}}}V_{1}AV_{2}\\
2\sqrt{\frac{\vec{\mathcal{Y}}_{1}\cdot\vec{\mathcal{Y}}_{2}}{x_{12}^{2}}}V_{2}
A^{\dagger}V_{1} & I_{H_{2}}
\end{pmatrix}
\end{equation}
Its determinant is given by
\begin{equation}
\det M=\det\left(I_{H_{1}}-4\frac{\vec{\mathcal{Y}}_{1}\cdot\vec{\mathcal{Y}}_{2}}{x_{12}^{2}}V_{1}AV_{2}V_{2}A^{\dagger}V_{1}\right)
\end{equation}
We define 
\begin{equation}
M_{1}\equiv I_{H_{1}}-4\frac{\vec{\mathcal{Y}}_{1}\cdot\vec{\mathcal{Y}}_{2}}{x_{12}^{2}}V_{1}AV_{2}V_{2}A^{\dagger}V_{1}
\end{equation}
which allows for the compact expression
\begin{equation}
M^{-1}=\begin{pmatrix}M_{1}^{-1} & -2\sqrt{\frac{\vec{\mathcal{Y}}_{1}\cdot\vec{\mathcal{Y}}_{2}}{x_{12}^{2}}}M_{1}^{-1}V_{1}AV_{2}\\
-2\sqrt{\frac{\vec{\mathcal{Y}}_{1}\cdot\vec{\mathcal{Y}}_{2}}{x_{12}^{2}}}V_{2}A^{\dagger}V_{1}M_{1}^{-1} & I_{H_{2}}+4\frac{\vec{\mathcal{Y}}_{1}\cdot\vec{\mathcal{Y}}_{2}}{x_{12}^{2}}V_{2}A^{\dagger}V_{1}M_{1}^{-1}V_{1}AV_{2}
\end{pmatrix}
\end{equation}
 The integral (\ref{eq:BS_2ptRho_dual}) becomes 
\begin{multline}
\left\langle \chi_{R_{1}}\left(\mathcal{Z}_{1}\right)\chi_{R_{2}}\left(\mathcal{Z}_{2}\right)\mathcal{O}\right\rangle \\
=\det\left(\bar{\mathcal{D}}^{(R_{1})}\right)\det\left(\bar{\mathcal{D}}^{(R_{2})}\right)
\int\left[dA\right]\int\,e^{-\frac{16\pi^{2}}{g_{YM}^{2}}{\rm Tr} A^{\dagger}A-N{\rm Tr}\ln M_{1}}
\left\langle \mathcal{O}\left(S^{I}\right)\right\rangle _{\varphi} \label{eq:BS_2ptA_dual}
\end{multline}
 where the measure is normalized as 
\begin{equation}
\int\left[dA\right]\int\,e^{-\frac{16\pi^{2}}{g_{YM}^{2}}{\rm Tr} A^{\dagger}A}=1
\end{equation}

\section{Giants and Dual Giants constructed using more than one matrix}\label{Sec:RestrictedSchur}

The Schur polynomials provide a basis for the local observables of a matrix field theory constructed using a single matrix.
There is a generalization to the case that we build observables using two matrices, known as the restricted Schur 
polynomial \cite{Bhattacharyya:2008rb,Bhattacharyya:2008xy}.
The restricted Schur polynomial is again constructed by taking the trace of a projector-like object
\begin{equation}
\chi_{R,(r,s)\alpha\beta}(Z,Y)={\rm Tr}(P_{R,(r,s)\alpha\beta}\, Z^{\otimes n}\otimes Y^{\otimes m})
\end{equation}
The $P_{R,(r,s)\alpha\beta}$ used in the construction of the restricted Schur polynomials are intertwining maps.
These maps can be thought of as matrices defined in the carrier space of representation $R$ of $S_{n+m}$.
$R$ is a Young diagram with $n+m$ boxes, written as $R\vdash n+m$.
$S_{n+m}$ permutes the indices of the $n+m$ matrices used to define the restricted Schur polynomial operator.
We then restrict to the $S_n\times S_m$ subgroup which permutes $Y$ indices (or $Z$ indices) but does not
permute $Y$ with $Z$ indices.
The representation of the subgroup can be labeled by a pair of Young diagrams $(r,s)$ where $r\vdash n$ records
the representation for the $Z$s and $s\vdash m$ the representation for the $Y$s.
Since the representation $(r,s)$ can be subduced more than once from $R$, we need a multiplicity label $\alpha$ to keep
track of the copy of $(r,s)$ that we consider.
We then have matrices $\Gamma_{(r,s)}^{(\alpha)}(\sigma)$ with $\sigma\in S_n\times S_m$ representing the subgroup.
The intertwining maps map between these representations
\begin{equation}
\Gamma_{(r,s)}^{(\alpha)}(\sigma)P_{R,(r,s)\alpha\beta}=P_{R,(r,s)\alpha\beta}\Gamma_{(r,s)}^{(\beta)}(\sigma)
\end{equation}
There is a dramatic simplification for giants (where $R$ is a single column of $m+n$ boxes) and for dual giants
(where $R$ is a single row with $n+m$ boxes).
In these cases $R$ is one dimensional and it restricts to a single representation of the subgroup, given by taking
$r$ to be a single row (column) with $n$ boxes and taking $s$ to be a single row (column) with $m$ boxes for a giant
(or dual giant).
Thus, we don't need the multiplicity indices and the intertwining maps are particularly simple - they are just the projectors
$P_R$ themselves
\begin{equation}
\chi_{(1^{n+m}),((1^n),(1^m))}(Z,Y)={\rm Tr}(P_{(1^{n+m})}\, Z^{\otimes n}\otimes Y^{\otimes m})
\end{equation}
\begin{equation}
\chi_{(n+m),((n),(m))}(Z,Y)={\rm Tr}(P_{(n+m)}\, Z^{\otimes n}\otimes Y^{\otimes m})
\end{equation}
Repeating the analysis of previous sections we find
\begin{equation}
\sum_{k_1=0}^N\sum_{k_2=0}^{N-k_1} t_Z^{k_1}t_Y^{k_2}\chi_{(1^{k_1+k_2}),((1^{k_1}),(1^{k_2}))}(Z,Y)
=\int d\bar\chi^N\cdots d\bar\chi^1 \int d\chi_1\cdots d\chi_N \,\, 
e^{-t_Z\bar\chi^{i} Z^{j}_{i}\chi_{j}-t_Y\bar\chi^{i} Y^{j}_{i}\chi_{j}-\bar\chi^{i} \chi_{i}}
\end{equation}
\begin{equation}
\sum_{k_1=0}^\infty \sum_{k_2=0}^\infty t_Z^{k_1}t_Y^{k_2}\chi_{(k_1+k_2),((k_1),(k_2))}(Z,Y)
=\int d^N \bar\varphi \int d^N\varphi\,\, 
e^{t_Z\bar\varphi^{i} Z^{j}_{i}\varphi_{j}+t_Y\bar\varphi^{i} Y^{j}_{i}\varphi_{j}
-\bar\varphi^{i} \varphi_{i}}
\end{equation}
The generalization to more than two matrices is now obvious.

\subsection{Large $N$ Effective Theory: Giants}

Again start with the free theory.
As above, we have giants at position $x_A$ for $A=1,2,...,Q$.
Imagine that the restricted Schur polynomials are built from two matrices ${\cal Y}$ and ${\cal Z}$. 
We can again be general and consider the generating function
\begin{eqnarray}
&&\!\!\!\!\!\!\!\!\!\!\!\!\!\!\!\sum_{n_1,m_1,\cdots n_Q,m_Q=1}^N 
t_{Z1}^{n_1}t_{Y1}^{m_1}\cdots t_{ZQ}^{n_Q}t_{YQ}^{m_Q}
\left\langle \chi_{(1^{n_1+m_1}),((1^{n_1}),(1^{m_1}))}(x_1)\cdots 
\chi_{(1^{n_Q+m_Q}),((1^{n_Q}),(1^{m_Q}))}(x_Q)\,{\cal O}\right\rangle\cr\cr
&=&\int [d\phi^I] \int \prod_{K=1}^Q
[d\bar\chi_K d\chi_K]\cr
&&e^{-{1\over g_{YM}^2}\int d^4 x\left( \sum_{I=1}^6 {1\over 2}
{\rm Tr}(\partial_\mu \phi^I\partial^\mu \phi^I)
+g_{YM}^2\sum_{K=1}^Q \delta (x-x_K)\bar \chi_K^i(\delta^j_i+t_{ZK} {\cal Z}^j_i
+ t_{YK} {\cal Y}^j_i)\chi_{Kj}\right)}
{\cal O}(\phi^I)\cr
&&
\end{eqnarray}
where ${\cal O}$ is a general single trace operator and the measure is normalized as before.
${\cal O}$ is again chosen normal ordered so that contractions of fields in ${\cal O}$ vanish.
We want to integrate over the $(\phi^I)^i_j$ field, which is easily performed because the (free) ``action'' above is
quadratic in the $\phi^I$ fields, after the integral identity for the Schur polynomials has been employed.
Completing the square as usual we find
\begin{eqnarray}
&&-\int d^4 x\left( {1\over 2}{\rm Tr}(\partial_\mu \phi^I \partial^\mu \phi^I)
+g_{YM}^2\sum_{K=1}^Q \delta (x-x_K)\bar \chi_K^i(\delta^j_i+t_{ZK} {\cal Z}^j_i + t_{YK} {\cal Y}^j_i )\chi_{Kj}\right)\cr
&=&-\int d^4 x \left(-{1\over 2} {\rm Tr}\left[(\phi^I-S^I)\partial^\mu\partial_\mu (\phi^I-S^I) 
- S^I\partial^\mu\partial_\mu S^I\right]
+g_{YM}^2\sum_{K=1}^Q \delta (x-x_K)\bar \chi_K^i\chi_{Ki}\right)\cr
&&
\end{eqnarray}
where
\begin{equation}
(S^I(x))^i_j=-{g_{YM}^2\over 4\pi^2}\sum_{K=1}^Q{ \left( t_{YK} {\cal Y}_K^I + t_{ZK} {\cal Z}_K^I \right)\bar\chi_K^i \chi_{Kj}\over |x-x_K|^2}
\end{equation}
Note that the solution is very similar to what we had when we considered giants build from a single matrix: 
$t_{K} {\cal Z}_K^I$ before becomes $t_{YK} {\cal Y}_K^I + t_{ZK} {\cal Z}_K^I$.  
The integration over $\phi$ proceeds as before with this change in the value for $S$.  
For the H-S transformation introduce the matrix, $\rho$.  
The answer after integrating over the fermions, is
\begin{eqnarray}
&&\!\!\!\!\!\!\!\!\!\!\!\!\!\!\!\sum_{n_1,m_1,n_2,m_2,\cdots n_Q,m_Q=1}^N 
t_{Z1}^{n_1}t_{Y1}^{m_1}\cdots t_{ZQ}^{n_Q}t_{YQ}^{m_Q}
\left\langle \chi_{(1^{n_1+m_1}),((1^{n_1}),(1^{m_1}))}(x_1)\cdots 
\chi_{(1^{n_Q+m_Q}),((1^{n_Q}),(1^{m_Q}))}(x_Q)\,{\cal O}\right\rangle\cr\cr
&=& \int [d\rho] e^{{8\pi^2 N\over\lambda}{\rm Tr}(\rho^2)+N {\rm Tr}\log\left[\delta_{JK}-2
\sqrt{{ \left(  (t_{KY} \vec{\cal Y}_K + t_{KZ} \vec{\cal Z}_K) \cdot  (t_{JY} \vec{\cal Y}_J + t_{JZ} \vec{\cal Z}_J) \over x_{KJ}^2 \right)}}\rho_{KJ}\right]}
\langle {\cal O}(S^I(x))\rangle_\chi\label{FnlEffAct}
\end{eqnarray}
As a check of the above result consider the two-point function of giant gravitons.  
For restricted Schur polynomials the two-point function is given by
\begin{eqnarray}
&&\langle \chi_{R, (r,s) \alpha \beta}(Z,Y) \chi_{T,(t,u)\mu\nu}(Z,Y)^\dag \rangle\cr
&&\qquad\qquad\qquad = \frac{f_R \textnormal{hooks}_R}{\textnormal{hooks}_r \textnormal{hooks}_s} \
\left( {g_{YM}^2 \over 4\pi ^2}\right)^{|R|}{1\over |x_{12}|^{2|R|}}
 \delta_{R T} \delta_{r t} \delta_{su} \delta_{\alpha \mu} \delta_{\beta \nu} \label{2Point}
\end{eqnarray}
where $|R|$ is the number of boxes in $R$.
In our present setup $R$ is a single column of $J_1+J_2$ boxes, $r$ is a single column of $J_1$ boxes and $s$ is a single column of $J_2$ boxes.  
The multiplicity labels $\alpha$, $\beta$, $\mu$ and $\nu$ may be dropped since there is a single way to subduce the representations.  
The two-point function is
\begin{eqnarray}
&&\langle \chi_{1^{J_1 + J_2}, (1^{J_1}, 1^{J_2})}(Z, Y) \chi_{1^{J_1 + J_2}, (1^{J_1}, 1^{J_2})}(Z, Y)^\dag\rangle\cr
&&\qquad\qquad\qquad
= \frac{N!}{(N-(J_1 + J_2))!} \frac{(J_1 + J_2)!}{J_1 ! J_2 !} \left( {\lambda\over 4\pi ^2}\right)^{J_1 + J_2}
{1\over |x_{12}|^{2J_1 + J_2}} 
\end{eqnarray}
To recover this result take ${\cal \vec{Y}}$ and ${\cal \vec{Z}}$ to be orthonormal and take ${\cal O} = 1$ so that the matrix inside the logarithm in (\ref{FnlEffAct}) becomes
\begin{equation}
M = \left( \begin{array}{cc} 1  &  -2 \sqrt{\frac{t_{1Y} t_{2Y} + t_{1Z} t_{2Z}}{x_{12}^2}} \rho  \nonumber  \\  2 \sqrt{\frac{t_{1Y} t_{2Y} + t_{1Z} t_{2Z}}{x_{12}^2}} \rho^{*} & 1     \end{array}  \right)
\end{equation}
The integral we need to compute is given by
\begin{eqnarray}
& & \frac{16 \pi N}{\lambda} \int_0^\infty r dr \int_0^{2 \pi} d\phi e^{-\frac{16 \pi^2 N}{\lambda} r^2 - N \log\left( 1 + 4 \frac{t_{1Y} t_{2Y} + t_{1Z} t_{2Z} }{x_{12}^2}r^2 \right)}    \nonumber \\
&& = \int_0^\infty d\tilde{r} \ e^{-\tilde{r} - N \log\left( 1 +  \frac{ g_{YM}^2 (t_{1Y} t_{2Y} + t_{1Z} t_{2Z}) }{4 \pi^2 x_{12}^2}\tilde{r} \right)} 
\end{eqnarray}
where $\rho = r e^{i\phi}$ and the normalisation condition on the $\rho$ integral provides the constant in front.  
Expand the integrand and perform the integral analytically, as before.  
The final answer is 
\begin{eqnarray}
& & \sum_{J = 0}^N \frac{N!}{(N-J)!}\left( \frac{g_{YM}^2 (t_{1Y} t_{2Y} + t_{1Z} t_{2Z}) }
{4 \pi^2 x_{12}^2} \right)^J \nonumber \\
& = & \sum_{J = 0}^N \sum_{K=0}^J \left(\frac{ g_{YM}^2 }{4 \pi^2 x_{12}^2}\right)^J \frac{N!}{(N-J)!} 
\left( \begin{array}{c} J \\ K \end{array} \right) (t_{1Y} t_{2Y})^{K} (t_{1Z} t_{2Z})^{J-K}
\end{eqnarray}
For $J_1$ $Z$-fields and $J_2$ $Y$-fields identify $J = J_1 + J_2$ and $K = J_2$ which yields the factor
\begin{equation}
 \frac{N!}{(N-(J_1 + J_2))^2} \frac{(J_1 + J_2)!}{J_1 ! J_2 !} \left(\frac{ g_{YM}^2 }{4 \pi^2 x_{12}^2}\right)^{J_1 + J_2} (t_{1Y} t_{2Y})^{J_2} (t_{1Z} t_{2Z})^{J_1}
\end{equation}
This is the expected answer (\ref{2Point}).

\subsection{Large $N$ Effective Theory: Maximal Giants}

We again start with the free theory.
As above, we have giants at position $x_A$ for $A=1,2,...,Q$.
Consider restricted Schur polynomials built from two matrices ${\cal Z}$ and $X$ or ${\cal Z}$ and $X^\dagger$ 
and consider the generating function
\begin{eqnarray}
&&\sum_{m_1,m_2,\cdots m_Q=1}^N 
t^{m_1}\cdots t^{m_Q}
\left\langle \chi_{(1^{N}),((1^{N-m_1}),(1^{m_1}))}(x_1)\cdots 
\chi_{(1^{N}),((1^{N-m_Q}),(1^{m_Q}))}(x_Q)\,{\cal O}\right\rangle\cr\cr
&=&\int [d\phi^I] \int \prod_{K=1}^Q
[d\bar\chi_K d\chi_K]\cr
&&\qquad\qquad e^{-{1\over g_{YM}^2}\int d^4 x\left( \sum_{I=1}^6 {1\over 2}
{\rm Tr}(\partial_\mu \phi^I\partial^\mu \phi^I)
+g_{YM}^2\sum_{K=1}^Q \delta (x-x_K)\bar \chi_K^i({\cal Z}^j_{Ki}
+ t_{K} X_{Ki}^j)\chi_{Kj}\right)}
{\cal O}(\phi^I)\cr
&&
\end{eqnarray}
where ${\cal O}$ is again a general single trace operator and the measure is normalized as before.
${\cal O}$ is chosen to be normal ordered so that contractions of fields in ${\cal O}$ vanish and it is
constructed using only $Z$ and $Y$ fields.
${\cal Z}$ is defined as above.
We can choose $X_K$ to be either $X$ or $X^\dagger$.
The procedure is by now routine: integrate over the $(\phi^I)^i_j$ fields by completing the square.
We find
\begin{eqnarray}
&&-\int d^4 x\left( {1\over 2}{\rm Tr}(\partial_\mu \phi^I \partial^\mu \phi^I)
+g_{YM}^2\sum_{K=1}^Q \delta (x-x_K)\bar \chi_K^i({\cal Z}^j_{Ki} + t_K X^j_{Ki} )\chi_{Kj}\right)\cr
&=&-\int d^4 x \left({1\over 2} {\rm Tr}\left[(\phi^I+S^I)\partial^\mu\partial_\mu (\phi^I+S^I) 
- S^I\partial^\mu\partial_\mu S^I\right]\right)
\end{eqnarray}
where
\begin{equation}
(S^I(x))^i_j=-{g_{YM}^2\over 4\pi^2}\sum_{K=1}^Q{ \left( {\cal Z}_K^I 
+ t_K X_K^I \right)\bar\chi_K^i \chi_{Kj}\over |x-x_K|^2}.  
\end{equation}
Now do the H-S transformation and integrate over the fermions to get
\begin{eqnarray}
&&\sum_{m_1,m_2,\cdots m_Q=1}^N 
t_X^{m_1}\cdots t_X^{m_Q}
\left\langle \chi_{(1^{N}),((1^{N-m_1}),(1^{m_1}))}(x_1)\cdots 
\chi_{(1^{N}),((1^{N-m_Q}),(1^{m_Q}))}(x_Q)\,{\cal O}\right\rangle\cr\cr
&=& \int [d\rho] e^{{8\pi^2 N\over\lambda}{\rm Tr}(\rho^2)+N {\rm Tr}\log\left[-2
\sqrt{{ \left(  (\vec{\cal Z}_K + t_K X_K) \cdot  ( \vec{\cal Z}_J + t_J X_J) \over x_{KJ}^2 \right)}}\rho_{KJ}\right]}
\langle {\cal O}^I(S^I(x))\rangle_\chi\label{FinalSEff}
\end{eqnarray}

Consider the case that a non-trivial ${\cal O}$ is included.
Our starting point is again the formula
\begin{equation}
\langle {\cal O}(S^I(x))\rangle_\chi=-{\rm Tr}_Q (\Phi^{I_1}(x)\Phi^{I_2}(x)\cdots\Phi^{I_J}(x))
\end{equation}
where the $\Phi^I$s are all $Y$'s and $Z$'s so that ${\cal O}$ belongs to the $SU(2)$ sector.
The simplest case is again $Q=2$ giants, located at $(0,x_1,0,0)$ and $(0,x_2,0,0)$ and they are both given by maximal
restricted Schur polynomials constructed using the matrices ${\cal Z}$ and we choose $X_1=X$ and $X_2=X^\dagger$.
Operator ${\cal O}$ is located at $x_0$.
For an operator ${\cal O}={\rm Tr}(Z^{n_1}Y^{n_2}Z^{n_3}\cdots)$ we will have
\begin{equation}
\langle {\cal O}\rangle_\chi=-{\rm Tr}_Q (M_Z^{n_1}M_Y^{n_2}M_Z^{n_3}\cdots)
\end{equation}
where
\begin{equation}
M_Z=-\left[\begin{matrix} \kappa^2 &0\cr 0 &\kappa^2\end{matrix}\right]{\lambda\over 4\pi^2} M^{-1}
\end{equation}

\begin{equation}
M_Y=-\left[\begin{matrix}-{\kappa x_1\over x_{01}^2} &0\cr 0 &-{\kappa x_2\over x_{02}^2}\end{matrix}\right]
{\lambda\over 4\pi^2} M^{-1}
\end{equation}

\begin{equation}
M=\left[
\begin{matrix} 0 &&-2\sqrt{\kappa^2+{t_1t_2\over |x_{12}|^2}}\rho\cr 2\sqrt{\kappa^2+{t_1t_2\over |x_{12}|^2}}\rho^* &&0\end{matrix}
\right]
\end{equation}

\begin{equation}
M^{-1}=
{1\over 4\left(\kappa^2 +{t_1t_2\over |x_{12}|^2}|\rho|^2\right)}
\left[
\begin{matrix} 0 &&2\sqrt{\kappa^2+{t_1t_2\over |x_{12}|^2}}\rho\cr 
-2\sqrt{\kappa^2+{t_1t_2\over |x_{12}|^2}}\rho^* &&0\end{matrix}
\right]
\end{equation}

%
Up to an overall scaling which trivially comes out of ${\cal O}$ these are the expressions of \cite{Jiang:2019xdz} so 
that we have non-trivial evidence that the maximal restricted Schur leads to an integrable problem.

\subsection{Large $N$ Effective Theory: Dual Giants}

The relevant generating function is
\begin{eqnarray}
&&\sum_{n_1,m_1,n_2,m_2,\cdots n_Q,m_Q=1}^\infty
t_{Z1}^{n_1}t_{Y1}^{m_1}\cdots t_{ZQ}^{n_Q}t_{YQ}^{m_Q}
\left\langle \chi_{(n_1+m_1),((n_1),(m_1))}(x_1)\cdots 
\chi_{(n_Q+m_Q),((n_Q),(m_Q))}(x_Q)\,{\cal O}\right\rangle\cr\cr
&=&\int [d\phi^I] \int \prod_{K=1}^Q[d\bar\varphi_K d\varphi_K]\cr
&& \qquad \qquad e^{-{1\over g_{YM}^2}\int d^4 x\left( \sum_{I=1}^6 {1\over 2}{\rm Tr}(\partial_\mu \phi^I\partial^\mu \phi^I)
+g_{YM}^2\sum_{K=1}^Q \delta (x-x_K)\bar \varphi_K^i(\delta^j_i-t_{ZK} {\cal Z}^j_i-t_{YK} {\cal Y}^j_i)
\varphi_{Kj}\right)}
{\cal O}^I(\phi^I)\cr\cr
&&
\end{eqnarray}
where ${\cal O}$ is again a normal ordered but otherwise general single trace operator and the measure is 
normalized as above.
Completing the square as usual, we find
\begin{eqnarray}
&&-\int d^4 x\left( {1\over 2}{\rm Tr}(\partial_\mu \phi^I \partial^\mu \phi^I)
+g_{YM}^2\sum_{K=1}^Q \delta (x-x_K)\bar \varphi_K^i (\delta^j_i-t_{ZK} {\cal Z}^j_i - t_{YK} {\cal Y}^j_i )\varphi_{Kj}\right)\cr
&=&\int d^4 x \left({1\over 2} {\rm Tr}\left[(\phi^I-S^I)\partial^\mu\partial_\mu (\phi^I-S^I) 
- S^I\partial^\mu\partial_\mu S^I\right]
-g_{YM}^2\sum_{K=1}^Q \delta (x-x_K)\bar \varphi_K^i\varphi_{Ki}\right)\cr
&&
\end{eqnarray}
where
\begin{equation}
(S^I(x))^i_j=-{g_{YM}^2\over 4\pi^2}\sum_{K=1}^Q{ \left( t_{YK} {\cal Y}_K^I + t_{ZK} {\cal Z}_K^I \right)\bar\varphi_K^i \phi_{Kj}\over |x-x_K|^2}.  
\end{equation}
After integrating over $\phi$ and trading $\varphi$ for $\rho$, with the H-S transformation, we find
\begin{eqnarray}
&&\sum_{n_1,m_1,n_2,m_2,\cdots n_Q,m_Q=1}^\infty 
t_{Z1}^{n_1}t_{Y1}^{m_1}\cdots t_{ZQ}^{n_Q}t_{YQ}^{m_Q}
\left\langle \chi_{(n_1+m_1),((n_1),(m_1))}(x_1)\cdots 
\chi_{(n_Q+m_Q),((n_Q),(m_Q))}(x_Q)\,{\cal O}\right\rangle\cr\cr
&=& \int [d\rho] e^{-{8\pi^2 N\over\lambda}{\rm Tr}(\rho^2)-N {\rm Tr}\log\left[\delta_{JK}+2 
\sqrt{{ \left(  (t_{KY} \vec{\cal Y}_K + t_{KZ} \vec{\cal Z}_K) \cdot  (t_{JY} \vec{\cal Y}_J + t_{JZ} \vec{\cal Z}_J) \over x_{KJ}^2 \right)}}\rho_{KJ}\right]}
\langle {\cal O}(S^I(x))\rangle_\chi\label{twodualeffaction}
\end{eqnarray}
As a simple test of this result, we will reproduce the two-point correlator of dual giants.
Choose ${\cal \vec{Y}}$ and ${\cal \vec{Z}}$ to be orthonormal and take ${\cal O} = 1$. 
We then obtain
\begin{equation}
M = \left( \begin{array}{cc} 1  &  2 \sqrt{\frac{t_{1Y} t_{2Y} + t_{1Z} t_{2Z}}{x_{12}^2}} \rho  \nonumber  \\  2 \sqrt{\frac{t_{1Y} t_{2Y} + t_{1Z} t_{2Z}}{x_{12}^2}} \rho^{*} & 1     \end{array}  \right)
\end{equation}
The key computation is the following integration
\begin{eqnarray}
& & \frac{1}{Z_{\rho}} \int [d\rho] e^{-\frac{8 \pi^2 N}{\lambda} {\rm Tr}(\rho^2 )- N \log\left( 1 - 4 \frac{t_{1Y} t_{2Y} + t_{1Z} t_{2Z} }{x_{12}^2} |\rho|^2 \right)}  = \frac{16 \pi N}{\lambda} \int [d\rho] \frac{e^{-\frac{16 \pi^2 N}{\lambda} |\rho|^2 } }{  \left( 1 - 4 \frac{t_{1Y} t_{2Y} + t_{1Z} t_{2Z} }{x_{12}^2} |\rho|^2 \right)^{N} }  \cr
&=& \frac{16 \pi N}{\lambda} \sum_{J=0}^{\infty} \sum_{K=0}^{J} \frac{4^{J}}{x_{12}^{2J}} \frac{(N+J-1)!}{(N-1)!J!} \left( \begin{array}{c} J \\ K \end{array} \right) (t_{1Y} t_{2Y})^{K} (t_{1Z} t_{2Z})^{J-K}  \int [d\rho] e^{-\frac{16 \pi^2 N}{\lambda} |\rho|^2 } |\rho|^{2J} \cr
&&
\end{eqnarray}
The coefficient of the term with $J=J_1 +J_2$ and $K= J_2 $ reproduces the expected result
\begin{equation}
G_2 = \frac{(N+J_1 + J_2 -1)!}{(N-1)!} \frac{(J_1 + J_2 )!}{J_1 ! J_2 !} 
\left(\frac{g_{YM}^2}{4\pi^2} \right)^J \frac{1}{|x_1 - x_2 |^{2(J_1 + J_2 )} }
\end{equation}

Finally consider the correlator between the dual giants and a single trace operator ${\cal O}$.
We need to compute the following effective one-point function
\begin{equation}
\langle {\cal O}\rangle_\chi=-{\rm Tr}_Q (M_Z^{n_1}M_Y^{n_2}M_Z^{n_3}\cdots)
\end{equation}
for an operator ${\cal O}={\rm Tr}(Z^{n_1}Y^{n_2}Z^{n_3}\cdots)$.
Working as we did above, we find
\begin{equation}
M^{-1}=
{\cal P}
\left[
\begin{matrix} 1 &&-2\sqrt{t_{1 z_1} t_{2 z_2} \kappa^2 +{t_{1 x_1 }t_{2 x_2}\over |x_{12}|^2}}\rho\cr 
-2\sqrt{t_{1 z_1} t_{2 z_2} \kappa^2 +{t_{1 x_1 }t_{2 x_2}\over |x_{12}|^2}}\rho^* &&1 \end{matrix}
\right]
\end{equation}
where
\begin{equation}
{\cal P}={1\over 1-4\left(t_{1 z_1} t_{2 z_2} \kappa^2 +{t_{1 x_1 }t_{2 x_2}\over |x_{12}|^2}\right) |\rho|^2}
\end{equation}
and
%
\begin{equation}
M_Z=-\left[\begin{matrix} \kappa^2 &0\cr 0 &\kappa^2\end{matrix}\right]{\lambda\over 4\pi^2} M^{-1}
\end{equation}

\begin{equation}
M_Y=-\left[\begin{matrix}-{\kappa x_1\over x_{01}^2} &0\cr 0 &-{\kappa x_2\over x_{02}^2}\end{matrix}\right]
{\lambda\over 4\pi^2} M^{-1}
\end{equation}

It is clear that the form of $M^{-1}$ differs from what we found for the Schur polynomials.
However, $M_{Y}$ and $M_{Z}$ are the same, which reflects the fact that the extra matrix $X_{1,2}$ is orthogonal to 
${\cal Z}$.

We are now in a position to test the selection rules. 
We again find that there is a term which respects the selection rules and other terms which spoil them.
There is never a limit in which the term that respects integrability dominates, so that in the case of dual giant gravitons
integrability does not seem to play any role.

\section{Conclusions}\label{Sec:Conclusions}

Recently\cite{Jiang:2019xdz} a formalism for the computation of correlators of two determinant operators with one single trace 
was developed.
This formalism is remarkably powerful and in fact leads to a proposal for the all orders in $\lambda$ answer, at large $N$.
A key ingredient has been integrability which manifested itself in selection rules enjoyed by the correlator.
By interpreting the effective field theory constructed in \cite{Jiang:2019xdz} in terms of projection operators that implement
symmetries under permuting indices of fields used to construct the determinant, we have managed to generalize to include
less than maximal giant gravitons and dual giant gravitons constructed using one or more than one fields.
The selection rules that signal integrability are only obeyed for maximal and close to maximal giant gravitons.

The effective field theory obtained in \cite{Jiang:2019xdz} was interpreted in terms of large $N$ ``collective fields'' which 
furnish the open string field theory on the giant gravitons. 
We have interpreted the same effective field theory simply as a construction to implement certain symmetries present
in the problem.
To really prove that this is recovering the open string field theory on the giant gravitons, there are a number of things
one might try to demonstrate.
Lets focus, for simplicity on the free Yang-Mills theory, which corresponds to the tensionless limit of the string.
First, the $\rho$ matrix that is being integrated over is a $K\times K$ matrix for the $K$-giant graviton problem, matching
what we would expect.
We expect a $\rho$ matrix for each direction traverse to the giant graviton's worldvolume, i.e. we expect
a total of 6 $\rho$ matrices.
However, there is a single $\rho$ matrix.
Further, the $\rho$ matrix does not appear to be a field - its more like a constant matrix.
The dependence on the field theory coordinates was fixed when we integrated over the $\phi^I$ fields.
This entailed solving the $\phi^I$ equation of motion with a source, so it looks very much like the $\rho$ field is on-shell.
We usually perform the path integral over all possible configurations so this is a little unusual.
The equation of motion of the $\rho$ field is algebraic and not a differential equation so it looks much more like an auxilliary
field than a true dynamical field.
At low energies this open string field theory should reduce to a Yang-Mills theory.
It does not look clear how the algebraic equations of the effective field theory will manage to do this.
Clearly there are many more interesting avenues to pursue to clarify the dynamics of the $\rho$ field.

An attractive feature of the $\rho$ theory is that all $N$ dependence is explicit so it provides a natural starting
point for a systematic $1/N$ expansion.
Indeed, because a factor of $N$ multiplies the $\rho$ action, the loop counting parameter of the $\rho$ theory is
${1\over N}$.
The collective field theory of \cite{Jevicki:1979mb,Jevicki:1980zg} also makes all $N$ dependence explicit and it too 
has loop counting parameter equal to ${1\over N}$.
However, an important difference between collective field theory and the $\rho$ theory is that collective field theory
carries out a systematic change of variables, to gauge invariant variables, so that all degrees of freedom in the theory 
are accounted for.
There is a non-trivial Jacobian associated with this change of variables and it gives rise to highly non-trivial non-linear
interactions for the collective field.
It is often possible to repackage the large number of collective fields into higher dimensional 
fields\cite{Das:1990kaa,Das:2003vw,Koch:2010cy}.
Indeed, when applied to the original example of holography, the collective field theory is a local theory in one higher
dimension than the original matrix model\cite{Das:1990kaa}.
Can the $\rho$ theory be understood as a truncation of the collective field theory of \cite{Jevicki:1979mb,Jevicki:1980zg} 
in which we keep only the maximal giant graviton degrees of freedom?

It might also be interesting to understand if there is a connection between the $\rho$ theory and the tiny graviton 
theory \cite{SheikhJabbari:2004ik}.
The tiny graviton theory provides an effective description of giant gravitons with the angular momentum 
$N^{1\over 2}\ll J\ll N$ and has already been used \cite{Hirano:2018xmh}
to reproduce non-perturbative (in $1/N$) effects that are present in giant graviton correlation functions.

A promising future direction is to compute the correlators of two maximal restricted Schur operators and one non-protected
single trace operator. 
This corresponds to considering maximal giant gravitons with two angular momenta on the S$^5$.
We have already found some hints that integrability may be present, since the correlator computed in the free limit of the theory
respects the selection rules of \cite{Jiang:2019xdz}.
Finally, to get further insight into the $\rho$ theory, it maybe interesting to understand the relation of the effective
field theory derived for restricted Schur polynomials to the $SU(2)$ spin 
matrix theory\cite{Harmark:2014mpa,Harmark:2016cjq}, which was derived using the non-abelian Dirac-Born-Infeld 
action as a starting point.

{\vskip 0.5cm}

\noindent
\begin{centerline} 
{\bf Acknowledgements}
\end{centerline} 

We would like to thank Sanjaye Ramgoolam for useful discussions.
This work is supported by the South African Research Chairs Initiative of the Department of Science and Technology and 
National Research Foundation of South Africa as well as by funds received from the National Institute for Theoretical 
Physics (NITheP).

\begin{appendix}

\section{The Two Giant Graviton Boundstate}\label{TwoGiantBound}

In this Appendix we will first develop the polynomial approach to describe the giant graviton bound state and then we will test
its correctness by reproducing the two point function of Schur polynomials with two columns.
Our notation is spelled out in Section \ref{sec:GBS}.
Consider the three-point function 
$\left\langle \chi_{R_{1}}\left(\mathcal{Z}_{1}\right)\chi_{R_{2}}\left(\mathcal{Z}_{2}\right)\mathcal{O}\right\rangle $,
where $\mathcal{O}$ is a normal ordered single trace operator.
The generating function of two bound states with one normal ordered single trace operator is
\begin{eqnarray}
&&t_{1}^{c_{1}}t_{2}^{c_{2}}\cdots t_{W_{1}}^{c_{W_{1}}}t_{W_{1}+1}^{c_{W_{1}+1}}\cdots t_{W}^{c_{W}}\left\langle \chi_{(1^{c_{1}}\cdots1^{c_{W_{1}}})}\left(\mathcal{Z}_{1}\right)\chi_{(1^{c_{W_{1}+1}}\cdots1^{c_{W}})}\left(\mathcal{Z}_{2}\right)\mathcal{O}\right\rangle\cr
&&=\det\left(V^{(1)}\right)\det\left(V^{(2)}\right)
\int\left[d\phi^{I}\right]e^{-\frac{1}{g_{YM}^{2}}\int d^{4}x\,\left[\frac{1}{2}{\rm Tr}\left(\partial_{\mu}\phi^{I}\partial^{\mu}\phi^{I}\right)\right]}\cr
&&\prod_{K=1}^{W}\left[d\bar{\chi}_{K}d\chi_{K}\right]\,
e^{-\frac{1}{g_{YM}^{2}}\int d^{4}x\,\left[g_{YM}^{2}\sum_{K=1}^{W}\delta(x-x_{\left\lceil K/W_{1}\right\rceil})
\bar{\chi}_{K}^{i}\left(\delta_{i}^{j}+t_{K}\left(\mathcal{Z}_{\left\lceil K/W_{1}\right\rceil }\right)_{i}^{j}\right)
\chi_{Ki}\right]}\mathcal{O}\left(\phi^{I}\right)\cr
&&
\end{eqnarray}
 where $\left\lceil \cdot\right\rceil $ is the ceiling function
\begin{equation}
\mathcal{Z}_{\left\lceil K/W_{1}\right\rceil }=\begin{cases}
\mathcal{Z}_{1} & 1\leq K\leq W_{1}\\
\mathcal{Z}_{2} & W_{1}+1\leq K\leq W
\end{cases},\label{eq:BS_ZKW-1}
\end{equation}
and the measure of $\phi^{I}$ is normalized as in (\ref{NormPhi}).
Completing the square, we find
\begin{multline*}
-\int d^{4}x\,\left[\frac{1}{2}{\rm T}\left(\partial_{\mu}\phi^{I}\partial^{\mu}\phi^{I}\right)+g_{YM}^{2}\sum_{K=1}^{W}\delta(x-x_{\left\lceil K/W_{1}\right\rceil })\bar{\chi}_{K}^{i}\left(\delta_{i}^{j}+t_{K}\left(\mathcal{Z}_{\left\lceil K/W_{1}\right\rceil }\right)_{i}^{j}\right)\chi_{Kj}\right]\\
=\int d^{4}x\,\left[\frac{1}{2}{\rm T}\left[\partial_{\mu}\left(\phi^{I}-S^{I}\right)\partial^{\mu}\left(\phi^{I}-S^{I}\right)-S^{I}\partial^{\mu}\partial_{\mu}S^{I}\right]-g_{YM}^{2}\sum_{K=1}^{W}\delta(x-x_{\left\lceil K/W_{1}\right\rceil })\bar{\chi}_{K}^{i}\chi_{Ki}\right]
\end{multline*}
 where 
\[
\left(S^{I}\left(x\right)\right)_{j}^{i}=-\frac{g_{YM}^{2}}{4\pi^{2}}\left[\sum_{K=1}^{W_{1}}\frac{t_{K}\left(\mathcal{Y}_{1}\right)_{K}^{I}\bar{\chi}_{K}^{i}\chi_{Kj}}{\left|x-x_{1}\right|^{2}}+\sum_{K=W_{1}+1}^{W}\frac{t_{K}\left(\mathcal{Y}_{2}\right)_{K}^{I}\bar{\chi}_{K}^{i}\chi_{Kj}}{\left|x-x_{2}\right|^{2}}\right]
\]
Integrating over the $\phi^{I}$ fields and performing a H-S transformation, we arrive at
\begin{multline}
t_{1}^{c_{1}}t_{2}^{c_{2}}\cdots t_{W_{1}}^{c_{W_{1}}}t_{W_{1}+1}^{c_{W_{1}+1}}\cdots t_{W}^{c_{W}}\left\langle \chi_{(1^{c_{1}}\cdots1^{c_{W_{1}}})}\left(\mathcal{Z}_{1}\right)\chi_{(1^{c_{W_{1}+1}}\cdots1^{c_{W}})}\left(\mathcal{Z}_{2}\right)\mathcal{O}\right\rangle \\
=
\det\left(V^{(1)}\right)\det\left(V^{(2)}\right)\int\left[d\rho\right]\int\prod_{K=1}^{W}\left[d\bar{\chi}_{K}d\chi_{K}\right]\\
\,e^{\frac{8\pi^{2}}{g_{YM}^{2}}{\rm T}\rho^{2}+2\sum_{K,J}^{\prime}\sqrt{\frac{t_{K}t_{J}\vec{\mathcal{Y}}_{1}\cdot\vec{\mathcal{Y}}_{2}}{x_{12}^{2}}}\rho_{KJ}\bar{\chi}_{K}^{a}\chi_{Ja}-\sum_{K=1}^{W}\bar{\chi}_{K}^{a}\chi_{Ka}}\mathcal{O}^{I}\left(S^{I}\right)
\end{multline}
where $\Sigma_{K,J}^{\prime}$ sums over indices fulfilling
$\left\lceil K/W_{1}\right\rceil \neq\left\lceil J/W_{1}\right\rceil $.
The anti-Hermitian matrix variable $\rho$ takes the form
\begin{equation}
\rho=\begin{pmatrix}0 & A\\
-A^{\dagger} & 0
\end{pmatrix},\quad A=\begin{pmatrix}A_{11} & A_{12} & \cdots & A_{1W_{2}}\\
A_{21} & A_{22} & \cdots & A_{2W_{2}}\\
\vdots & \vdots & \vdots & \vdots\\
A_{W_{1}1} & A_{W_{1}2} & \cdots & A_{W_{1}W_{2}}
\end{pmatrix},\label{eq:NBS_rho}
\end{equation}
$A$ is a $W_{1}\times W_{2}$ complex matrix.
Integrate over the fermion fields to arrive at
\begin{multline}
t_{1}^{c_{1}}t_{2}^{c_{2}}\cdots t_{W_{1}}^{c_{W_{1}}}t_{W_{1}+1}^{c_{W_{1}+1}}\cdots t_{W}^{c_{W}}\left\langle \chi_{(1^{c_{1}}\cdots1^{c_{W_{1}}})}\left(\mathcal{Z}_{1}\right)\chi_{(1^{c_{W_{1}+1}}\cdots1^{c_{W}})}\left(\mathcal{Z}_{2}\right)\mathcal{O}\right\rangle \\
=\det\left(V^{(1)}\right)\det\left(V^{(2)}\right)\int\left[d\rho\right]\int\,e^{\frac{8\pi^{2}}{g_{YM}^{2}}{\rm T}\rho^{2}+N{\rm T}\ln\left[\delta_{JK}-2\sqrt{\frac{t_{K}t_{J}\vec{\mathcal{Y}}_{1}\cdot\vec{\mathcal{Y}}_{2}}{x_{12}^{2}}}\rho_{KJ}\right]}\left\langle \mathcal{O}\left(S^{I}\right)\right\rangle _{\chi},\label{eq:BSN_2ptRho}
\end{multline}
where $\left\langle \mathcal{O}\left(S^{I}\right)\right\rangle _{\chi}$ is defined by Wick contracting all pairs of $\chi$, 
$\bar{\chi}$ fields according to Wick's theorem with the basic contraction given by
\begin{equation}
\left\langle \bar{\chi}_{J}^{a}\chi_{Kb}\right\rangle =-\delta_{b}^{a}M_{JK}^{-1}
\qquad\qquad
M_{JK}=\delta_{JK}-2\sqrt{\frac{t_{K}t_{J}\vec{\mathcal{Y}}_{1}\cdot\vec{\mathcal{Y}}_{2}}{x_{12}^{2}}}\rho_{KJ}
\end{equation}

As a test, of this generating function, we will compute the two-point function of Schur polynomials with two columns. 
The exact result is
\begin{equation}
\left\langle \chi_{(1^{J_{1}}1^{J_{2}})}\left(Z\right)\chi_{(1^{J_{3}}1^{J_{4}})}\left(Z^{\dagger}\right)\right\rangle =\frac{N!}{\left(N-J_{1}\right)!}\frac{\left(N+1\right)!}{\left(N+1-J_{2}\right)!}\delta_{J_{1},J_{3}}\delta_{J_{2},J_{4}}
\end{equation}
Parameterize the matrix $A$ as
\begin{equation}
A=\begin{pmatrix}z_{1} & z_{2}\\
z_{3} & z_{4}
\end{pmatrix},\quad z_{i}=r_{i}e^{i\theta_{i}}
\end{equation}
We need to compute the integral 
\begin{equation}
\int\prod_{i=1}^{4}d\theta_{i}r_{i}dr_{i}\,e^{-\frac{16\pi^{2}N}{\lambda}\sum_{i=1}^{4}r_{i}^{2}}\left(\det M\right)^{N}
\end{equation}
Note that 
\begin{eqnarray}
\det M&=&1+\frac{16t_{1}t_{2}t_{3}t_{4}}{x_{12}^{4}}\left(r_{2}^{2}r_{3}^{2}+r_{1}^{2}r_{4}^{2}-2r_{1}r_{2}r_{3}r_{4}\cos\theta\right)\\
&+&\frac{4}{x_{12}^{2}}\left(t_{1}t_{3}r_{1}^{2}+t_{2}t_{3}r_{3}^{2}+t_{1}t_{4}r_{2}^{2}+t_{2}t_{4}r_{4}^{2}\right)
\end{eqnarray}
where $\theta=\theta_{1}+\theta_{4}-\theta_{2}-\theta_{3}$. 
Expand $\left(\det M\right)^{N}$ using multinomial coefficients
\begin{equation}
\binom{n_{1}+n_{2}+\cdots+n_{p}}{n_{1},n_{2},\dots,n_{p}}=\frac{\left(n_{1}+n_{2}+\cdots+n_{p}\right)!}{n_{1}!n_{2}!\cdots n_{p}!}
\end{equation}
to obtain
\begin{multline}
\left(\det M\right)^{N} =\sum_{\sum n_{i}+\sum k_{i}+l+\tilde{m}=N}\binom{N}{n_{1},n_{2},\tilde{m},k_{1},k_{2},k_{3},k_{4},l}\left(\frac{16t_{1}t_{2}t_{3}t_{4}}{x_{12}^{4}}\right)^{n+\tilde{m}}\left(\frac{4}{x_{12}^{2}}\right)^{k} \\
\left(r_{2}^{2}r_{3}^{2}\right)^{n_{1}}\left(r_{1}^{2}r_{4}^{2}\right)^{n_{2}}\left(-2r_{1}r_{2}r_{3}r_{4}\cos\theta\right)^{\tilde{m}}\left(t_{1}t_{3}r_{1}^{2}\right)^{k_{1}}\left(t_{1}t_{4}r_{2}^{2}\right)^{k_{2}}\left(t_{2}t_{3}r_{3}^{2}\right)^{k_{3}}\left(t_{2}t_{4}r_{4}^{2}\right)^{k_{4}} \label{eq:BSN_detMN}
\end{multline}
where $n=n_{1}+n_{2}$ and $k=k_{1}+k_{2}+k_{3}+k_{4}$. 
Integrate over $\theta_4$ to remove the $\cos\theta$ dependence in the expansion. 
The integral of $\cos^{\tilde{m}}\theta$ is nonzero only when $\tilde{m}$ is even
\begin{equation}
\int_{0}^{2\pi}d\theta_{4}\left(-2\cos\theta\right)^{2m}=2\pi\binom{2m}{m}=\binom{2m}{m}\int_{0}^{2\pi}d\theta_{4}
\end{equation}
We can therefore replace the $\left(-2\cos\theta\right)^{\tilde{m}}$ factor in $\left(\det M\right)^{N}$ with 
$\binom{2m}{m}$, $2m\equiv \tilde{m}$.
The integration over $\theta_4$ can be dropped as it will be canceled by a normalization factor. 
Next absorb $m$ into the multinomial coefficient using
\begin{equation}
\binom{N}{n_{1},n_{2},2m,k_{1},k_{2},k_{3},k_{4},l}\binom{2m}{m}=\binom{N}{n_{1},n_{2},m,m,k_{1},k_{2},k_{3},k_{4},l}
\end{equation}
to arrive at
\begin{multline}
\left(\det M\right)^{N} = \\
\binom{N}{n_{1},n_{2},m,m,k_{1},k_{2},k_{3},k_{4},l}\left(\frac{4}{x_{12}^{2}}\right)^{2n+4m+k}t_{1}^{n+2m+k_{12}}t_{2}^{n+2m+k_{34}}t_{3}^{n+2m+k_{13}}t_{4}^{n+2m+k_{24}} \\
 \times r_{1}^{2(n_{2}+k_{1})}r_{2}^{2(n_{1}+k_{2})}r_{3}^{2(n_{1}+k_{3})}r_{4}^{2(n_{2}+k_{4})}\left(r_{1}r_{2}r_{3}r_{4}\right)^{2m}
\end{multline}
The final step of the computation entails extracting specific powers of $t_{1}$, $t_{2}$, $t_{3}$, and $t_{4}$. 
With this final step in mind, we use new variables to simplify the powers of $t_{i}$.
The new variables are
\begin{equation}
J=2n+4m+k,\quad J_{1}=n+2m+k_{12},\quad J_{3}=n+2m+k_{13}
\end{equation}
Clearly, $J$ is the total number of boxes in the Young diagram of the bound state while $J_1$ and $J_3$ are the number of 
boxes in the first columns. 
This motivates the variables
\begin{align}
k_{2} & =J_{1}-n-2m-k_{1},\\
k_{3} & =J_{3}-n-2m-k_{1},\\
k_{4} & =J-J_{13}+k_{1},\quad J_{13}\equiv J_{1}+J_{3}
\end{align}
Then $\left(\det M\right)^{N}$ now becomes 
\begin{multline}
\left(\det M\right)^{N}=\binom{N}{n_{1},n_{2},m,m,k_{1},J_{1}-n-2m-k_{1},J_{3}-n-2m-k_{1},J-J_{13}+k_{1},l}\\
\times\left(\frac{4}{x_{12}^{2}}\right)^{J}t_{1}^{J_{1}}t_{2}^{J-J_{1}}t_{3}^{J_{3}}t_{4}^{J-J_{3}}r_{1}^{2(n_{2}+m+k_{1})}r_{2}^{2(J_{1}-n_{2}-m-k_{1})}r_{3}^{2(J_{3}-n_{2}-m-k_{1})}r_{4}^{2(J-J_{13}+n_{2}+m+k_{1})}
\end{multline}
To obtain nonzero results, the total number of boxes of two bound states must be equal. 
Define new indices, again to simplify the powers of $r_{i}$, as follows
\begin{equation}
h=n_{2}+m+k_{1},\quad j_{1}=n_{1}+m,\quad j_{2}=n_{2}+m,\quad h\geq n_{2}+m+J_{13}-J
\end{equation}
We can eliminate $n_{1}$, $n_{2}$, and $k_{1}$ to find
\begin{align}
\left(\det M\right)^{N} & =\binom{N}{j_{1}-m,j_{2}-m,m,m,h-j_{2},J_{1}-j_{1}-h,J_{3}-j_{1}-h,J-J_{13}+h-j_{2},l}\nonumber \\
 & \times\left(\frac{4}{x_{12}^{2}}\right)^{J}t_{1}^{J_{1}}t_{2}^{J-J_{1}}t_{3}^{J_{3}}t_{4}^{J-J_{3}}r_{1}^{2h}r_{2}^{2(J_{1}-h)}r_{3}^{2(J_{3}-h)}r_{4}^{2(J-J_{13}+h)}
\end{align}
The ranges of the new indices are determined by requiring that the lower indices of the multimonial coefficients are 
non-negative and less than or equal to $N$. The integration over $r_{i}$ can easily be performed to find 
\begin{multline}
\int\prod r_{i}dr_{i}\,e^{S_{\mathrm{eff}}}=\left(\frac{\lambda}{4\pi^{2}Nx_{12}^{2}}\right)^{J}t_{1}^{J_{1}}t_{2}^{J-J_{1}}t_{3}^{J_{3}}t_{4}^{J-J_{3}}h!\left(J_{1}-h\right)!\left(J_{3}-h\right)!\left(J-J_{13}+h\right)!\\
\times\binom{N}{j_{1}-m,j_{2}-m,m,m,h-j_{2},J_{1}-j_{1}-h,J_{3}-j_{1}-h,J-J_{13}+h-j_{2},l}
\end{multline}
where we have divided the above result by an appropriate normalization.
For convenience, denote
\begin{equation}
\int\prod r_{i}dr_{i}\,e^{S_{\mathrm{eff}}}\equiv\sum_{J,J_{1},J_{3}}\left(\frac{\lambda}{4\pi^{2}Nx_{12}^{2}}\right)^{J}t_{1}^{J_{1}}t_{2}^{J-J_{1}}t_{3}^{J_{3}}t_{4}^{J-J_{3}}F\left(J,J_{1},J_{3}\right)
\end{equation}
The multinomial coefficients in $F\left(J,J_{1},J_{3}\right)$ can
be factored into a product of binomial coefficients and factorials, as follows
\begin{multline}
F\left(J,J_{1},J_{3}\right)=\sum_{h,m,j_{1},j_{2}}h!\left(J_{1}-h\right)!\left(J_{3}-h\right)!\left(J-J_{13}+h\right)!\\
\times\binom{N}{j_{1}-m,j_{2}-m,m,m,h-j_{2},J_{1}-j_{1}-h,J_{3}-j_{1}-h,J-J_{13}+h-j_{2},l}\\
=\sum_{h,m,j_{1},j_{2}}\frac{N!\left(j_{1}!\right)^{2}\left(j_{2}!\right)^{2}}{\left(j_{1}-m\right)!\left(j_{2}-m\right)!\left(m!\right)^{2}\left(N+j_{1}+j_{2}-J\right)!}\binom{h}{j_{2}}\binom{J_{1}-h}{j_{1}}\binom{J_{3}-h}{j_{1}}\binom{J-J_{13}+h}{j_{2}}
\end{multline}
 Using the Chu-Vandermonde identity 
\begin{equation}
\sum_{j=0}^{k}\binom{m}{j}\binom{n-m}{k-j}=\binom{n}{k}
\end{equation}
 we have 
\begin{align}
\sum_{m}\frac{N!\left(j_{1}!\right)^{2}\left(j_{2}!\right)^{2}}{\left(j_{1}-m\right)!\left(j_{2}-m\right)!\left(m!\right)^{2}\left(N+j_{1}+j_{2}-J\right)!} & =\sum_{m}\binom{j_{1}}{m}\binom{j_{2}}{j_{2}-m}\frac{N!j_{1}!j_{2}!}{\left(N+j_{1}+j_{2}-J\right)!}\nonumber \\
  & =\frac{N!\left(j_{1}+j_{2}\right)!}{\left(N+j_{1}+j_{2}-J\right)!}
\end{align}
 Therefore 
\begin{align}
F\left(J,J_{1},J_{3}\right) & =\sum_{h,j,j_{1}}\frac{N!j!}{\left(N+j-J\right)!}\binom{h}{j-j_{1}}\binom{J_{1}-h}{j_{1}}\binom{J_{3}-h}{j_{1}}\binom{J-J_{13}+h}{j-j_{1}}\nonumber \\
 & \equiv\sum_{j}\frac{N!j!}{\left(N+j-J\right)!}G\left(J,J_{1},J_{3},j\right)\label{eq:BSN_FJJJ}
\end{align}
 where we have defined 
\begin{equation}
G\left(J,J_{1},J_{3},j\right)=\sum_{h=J_{13}-J,}^{\mathrm{min}\left(J_{1},J_{3}\right)}\sum_{j_{1}=0}^{j}\binom{h}{j-j_{1}}\binom{J_{1}-h}{j_{1}}\binom{J_{3}-h}{j_{1}}\binom{J-J_{13}+h}{j-j_{1}}
\end{equation}
With the help of explicit numerical evaluation, we have found an interesting formula for $G\left(J,J_{1},J_{3},j\right)$
\begin{align}
G\left(J,J_{1},J_{3},j\right) & =\sum_{i=0}^{J-j-\max\left(J_{1},J_{3}\right)}\binom{j+i}{i}\binom{J+1-j-i}{j},\quad\max\left(J_{1},J_{3}\right)+j\leq J\leq J_{1}+J_{3}\label{eq:BSN_GJJJ}\\
G\left(J,J_{1},J_{3},j\right) & =0,\quad J<\max\left(J_{1},J_{3}\right)+j
\end{align}
Using (\ref{eq:BSN_GJJJ}), we have 
\begin{align}
G\left(J,J_{1},J_{3},j\right) & =G\left(J,J_{1},J_{3}+1,j\right),\quad\text{if }J_{1}>J_{3},J\leq J_{1}+J_{3}
\label{eq:BSN_GFRelation}\\
G\left(J,J_{1},J_{3},j\right) & =G\left(J,J_{1}+1,J_{3},j\right)\quad\text{if }J_{1}<J_{3},J\leq J_{1}+J_{3}
\end{align}
and $F\left(J,J_{1},J_{3}\right)$ have the same properties. These relations will be used shortly.
Now, we need to multiply the path integral result with a factor 
\begin{equation}
\det\left(V^{(1)}\right)\det\left(V^{(2)}\right)=1-\frac{t_{2}}{t_{1}}-\frac{t_{4}}{t_{3}}+\frac{t_{2}t_{4}}{t_{1}t_{3}}
\label{eq:BSN_detV1V2}
\end{equation}
The action of each term of (\ref{eq:BSN_detV1V2}) on the path integral result is equivalent to changing the arguments of
$F\left(J,J_{1},J_{3}\right)$ by one. 
As an example, multiplying with $\frac{t_{2}}{t_{1}}$ we have 
\begin{align*}
\frac{t_{2}}{t_{1}}\int\prod r_{i}dr_{i}\,e^{-S_{\mathrm{eff}}} & =\sum_{J,J_{3}}\sum_{J_{1}\leq N}\left(\frac{\lambda}{4\pi^{2}Nx_{12}^{2}}\right)^{J}t_{1}^{J_{1}-1}t_{2}^{J-J_{1}+1}t_{3}^{J_{3}}t_{4}^{J-J_{3}}F\left(J,J_{1},J_{3}\right)\\
 & =\sum_{J,J_{3}}\sum_{J_{1}\leq N-1}\left(\frac{\lambda}{4\pi^{2}Nx_{12}^{2}}\right)^{J}t_{1}^{J_{1}}t_{2}^{J-J_{1}}t_{3}^{J_{3}}t_{4}^{J-J_{3}}F\left(J,J_{1}+1,J_{3}\right)
\end{align*}
Consequently we have 
\begin{multline}
\det\left(V^{(1)}\right)\det\left(V^{(2)}\right)\int\prod r_{i}dr_{i}\,e^{S_{\mathrm{eff}}}=\sum_{J,J_{1},J_{3}}\left(\frac{\lambda}{4\pi^{2}Nx_{12}^{2}}\right)^{J}\\
t_{1}^{J_{1}}t_{2}^{J-J_{1}}t_{3}^{J_{3}}t_{4}^{J-J_{3}}\Big[F\left(J,J_{1},J_{3}\right)-F\left(J,J_{1}+1,J_{3}\right)-F\left(J,J_{1},J_{3}+1\right)+F\left(J,J_{1}+1,J_{3}+1\right)\Big]\label{eq:BSN_ddPathIntegral}
\end{multline}
As mentioned above, equation (\ref{eq:BSN_ddPathIntegral}) contains many terms that do not correspond to 
well defined Young diagrams.
We need to impose the constraint $J\leq2\min\left(J_{1},J_{3}\right)$.
It is clear that terms on the r.h.s of (\ref{eq:BSN_ddPathIntegral}) with $J_1\neq J_3$ vanishes. 
Assume $J_1>J_3$ without loss of generality. 
Using (\ref{eq:BSN_GFRelation}), we see that $F\left(J,J_{1},J_{3}\right)=F\left(J,J_{1},J_{3}+1\right)$ and
$F\left(J,J_{1}+1,J_{3}\right)=F\left(J,J_{1}+1,J_{3}+1\right)$, from which it follows
\begin{equation}
\left\langle \chi_{(1^{J_{1}}1^{J-J_{1}})}\left(Z\right)\chi_{(1^{J_{3}}1^{J-J_{3}})}^{\dagger}\left(Z\right)\right\rangle =0,\quad\text{if }J_{1}\neq J_{3}
\end{equation}
Consider the case $J_1=J_3$. Using (\ref{eq:BSN_GFRelation}), we have
\begin{equation}
F\left(J,J_{1}+1,J_{3}\right)=F\left(J,J_{1}+1,J_{3}+1\right)=F\left(J,J_{1},J_{3}+1\right)
\end{equation}
and thus 
\begin{multline}
t_{1}^{J_{1}}t_{2}^{J-J_{1}}t_{3}^{J_{1}}t_{4}^{J-J_{1}}\left\langle \chi_{(1^{J_{1}}1^{J-J_{1}})}\left(Z\right)\chi_{(1^{J_{1}}1^{J-J_{1}})}^{\dagger}\left(Z\right)\right\rangle \\
=\sum_{J_{1}}\sum_{J\leq2J_{1}}\left(\frac{\lambda}{4\pi^{2}Nx_{12}^{2}}\right)^{J}t_{1}^{J_{1}}t_{2}^{J-J_{1}}t_{3}^{J_{1}}t_{4}^{J-J_{1}}\Big[F\left(J,J_{1},J_{1}\right)-F\left(J,J_{1}+1,J_{1}+1\right)\Big]
\end{multline}
The difference between two $F\left(J,J_{1},J_{1}\right)$ can be expressed in terms of difference between $G\left(J,J_{1},J_{1},j\right)$ functions
\begin{equation}
F\left(J,J_{1},J_{1}\right)-F\left(J,J_{1}+1,J_{1}+1\right)=\sum_{j}\frac{N!j!}{\left(N+j-J\right)!}\left[G\left(J,J_{1},J_{1},j\right)-G\left(J,J_{1}+1,J_{1}+1,j\right)\right]
\end{equation}
The only difference between $G\left(J,J_{1},J_{1},j\right)$ and $G\left(J,J_{1}+1,J_{1}+1,j\right)$ is that the former 
has one extra term. 
Thus, we have
\begin{align}
F\left(J,J_{1},J_{1}\right)-F\left(J,J_{1}+1,J_{1}+1\right) & =\sum_{j}\frac{N!j!}{\left(N+j-J\right)!}\binom{J-J_{1}}{J-j-J_{1}}\binom{J_{1}+1}{j}\nonumber \\
 & =\sum_{j}\frac{N!j!}{\left(N+j-J\right)!}\binom{J-J_{1}}{j}\binom{J_{1}+1}{j}
\end{align}
If we now let $J_{2}=J-J_{1}$, and use the formula 
\begin{equation}
\sum_{j=0}^{J_{2}}\frac{j!}{\left(N-J_{1}-J_{2}+j\right)!}\binom{J_{2}}{j}\binom{J_{1}+1}{j}=\frac{\left(N+1\right)!}{\left(N-J_{1}\right)!\left(N+1-J_{2}\right)!}\label{eq:BSN_Formula2}
\end{equation}
the demonstration is complete.
We have numerically verified (\ref{eq:BSN_Formula2}).

\end{appendix}

\end{document}